\documentclass[manuscript]{aastex61}

\newcommand\aastex{AAS\TeX}

\newcommand{\pd}{\partial}

\newcommand{\meanv}[1]{\overline{\bm #1}}

\def\Ro{\mbox{\rm Ro}}

\def\Brms{B_{\rm rms}}

\def\urms{u_{\rm rms}}

\def\apjl{ApJL}
\def\apjs{ApJ}
\def\aap{A\&A}

\def\Rs{R_{\odot}}
\def\prot{P_{\rm rot}}
\def\pcyc{P_{\rm cyc}}
\def\taud{\tau_{\rm dyn}}
\def\etat{\eta_{\rm t}}
\newcommand{\brac}[1]{\langle #1 \rangle}
\newcommand{\mean}[1]{\overline{#1}}
\def\rhk{\brac{R'_{HK}}}

\usepackage{natbib}
\usepackage{bm}
\usepackage{enumerate}
\usepackage{hyperref}
\usepackage{float}
\usepackage[latin1]{inputenc}
\usepackage{graphicx}
\usepackage{graphics}
\graphicspath{{eps/}{png/}}

\received{\today}
\submitjournal{ApJ}

\shorttitle{\aastex\ Stellar dynamos}
\shortauthors{Guerrero et al.}
\begin{document}

\title{What sets the magnetic field strength and cycle period in solar-type 
stars? }

\correspondingauthor{G. Guerrero}
\email{guerrero@fisica.ufmg.br}

\author{G. Guerrero}
\affiliation{Physics Department, Universidade Federal de Minas Gerais \\
Av. Antonio Carlos, 6627, Belo Horizonte, MG 31270-901, Brazil}

\author{B. Zaire}
\affiliation{Physics Department, Universidade Federal de Minas Gerais \\
Av. Antonio Carlos, 6627, Belo Horizonte, MG 31270-901, Brazil}

\author{P.~K. Smolarkiewicz}
\affiliation{European Centre for Medium-Range Weather Forecasts, Reading RG2 9AX, UK}

\author{E. M. de Gouveia Dal Pino}
\affiliation{Astronomy Department, Universidade de S\~{a}o Paulo, IAG-USP, 
Rua do Mat\~{a}o, 1226, S\~{a}o Paulo, SP, Brasil, 05508-090}

\author{A.~G. Kosovichev}
\affiliation{New Jersey Institute of technology, Newark, NJ 07103,USA}

\author{N. N. Mansour}
\affiliation{NASA, Ames Research Center, Moffett Field, Mountain View, CA 94040, USA}

\begin{abstract}
Two fundamental properties of stellar magnetic fields have been determined by observations 
for solar-like stars with different Rossby numbers ($\Ro$), namely, the magnetic field 
strength and the magnetic cycle period. The field strength exhibits two regimes: 1) for fast
rotation it is independent of $\Ro$, 2) for slow rotation it decays with $\Ro$ following 
a power law. For the magnetic cycle period two regimes of activity, the active and inactive 
branches, also have been identified. For both of them, the longer the rotation period, 
the longer the activity cycle. Using global dynamo simulations of solar like stars with 
Rossby numbers between $\sim0.4$ and $\sim2$, this paper explores the relevance of rotational 
shear layers in determining these observational properties. Our results, consistent with 
non-linear $\alpha^2\Omega$ dynamos, show that the total magnetic field  strength is 
independent of the rotation period.  Yet at surface levels, the origin of the magnetic 
field is determined by $\Ro$. While for $\Ro\lesssim1$  it is generated in the convection
zone, for $\Ro\gtrsim1$ strong toroidal fields are generated at the tachocline and  
rapidly emerge towards the surface. In agreement with the observations,  the magnetic cycle 
period increases with the rotational period.  However, a bifurcation is observed for 
$\Ro\sim1$, separating a regime where oscillatory dynamos operate mainly in the convection 
zone, from the regime where the tachocline has a predominant role. In the latter the 
cycles are believed to result from the periodic energy exchange between the dynamo and 
the magneto-shear instabilities developing in the tachocline and the radiative interior.
\end{abstract}


\section{Introduction} \label{sec:intro}

Modern observations have revealed the existence of large-scale magnetic
fields in most types of stars across the HR diagram. Among these are solar-type
stars, with convective envelopes and radiative cores, as well as fully convective
stars which are either in a pre-main sequence phase or represent main sequence 
M-type dwarfs. 
There is no doubt that the magnetic field is relevant in every phase of the
life of stars. It also plays a critical role in the evolution of planetary
discs and, ultimately, may define criteria for habitability \citep{dN+16}. 

For late-type and solar-like stars which have convective envelopes,  
large-scale magnetic fields, as well as different field topologies observed on 
their surface \citep{Petit+08,GDMHNHJ12}, are convincing evidences of a 
dynamo mechanism operating in the stellar interiors. 
The dynamo is the result of a complex system of electric currents induced by the  
differential rotation, in a processes known as the $\Omega$-effect;
and the helical turbulent convective motions and fields, producing the so-called 
$\alpha$-effect \citep{Pa55,SKR66}.  Furthermore, as it will be detailed below, 
observations show clear correlations of the magnetic field strength and the activity 
cycle period with the stellar Rossby number, $\Ro = \prot/\tau_c$, where $\prot$ is 
the period of rotation, and $\tau_c$ is the convective turnover time. These correlations 
provide information 
about the dynamo process which might help to decode its elusive details.

The relationship between the magnetic field strength and $\Ro$ has
two well defined regimes. They are evident in the stellar X-ray luminosity data, 
$L_{\rm X}$ \citep{pizzolato+03,Wright+11}, as well
as in direct measurements of the mean magnetic field, $\brac{B}$ \citep{Vidotto+14}. 
For $\Ro \gtrsim 0.1$, the magnetic activity shows a power law behavior,
$\brac{B} \propto \Ro^{-1.38} $\citep{Vidotto+14}. 
For $\Ro \lesssim 0.1$,  observations indicate a 
regime of activity independent of $\Ro$, which is often called the saturated phase. 
Recent observational results by \cite{Wright+16} point out that the two regimes
described above occur in both, fully and partially convective stars. 
These results question the canonical theory in which a rotational shear layer
at the interface between the radiative and the convection zones is fundamental. 
This layer is called tachocline.
 
A number of stars, specially of types F, G and K, exhibit chromospheric
variations consistent with cyclic magnetic activity \citep{Baliunas+95, SB99,BMM17}. 
The seminal studies of \citep{Noyes+84b,BST98} identified correlations between
the magnetic cycles and the stellar rotation.  Their results
suggested the existence of two main branches, dividing  active ($A$) from inactive 
($I$) stars \citep{SB99}. The $A$ and $I$ branches show positive dependences between 
the ratio $\prot/\pcyc$ (where $\pcyc $ is the magnetic cycle period) and either with 
$\Ro^{-1}$ or $\brac{R'_{HK}}$ (where $\brac{R'_{HK}}=F'_{HK}/F_{\rm bol}$, is the
mean fractional ${\rm Ca} ~_{\rm II} \; {\rm H}$ and ${\rm K}$ flux
relative to the stellar bolometric flux, $F_{\rm bol}$). 
The same branches, $A$ and $I$, are confirmed by \cite{BMM17} using recalibrated 
measurements and new data from the Kepler satellite.

\cite{BV07} compared the period of 
rotation with the magnetic cycle period ($P_{\rm rot}$~vs~$P_{\rm cyc}$) of the
Mount Wilson sample of stars and
found the same two branches, $A$ and $I$, having different positive slopes. 
\cite{BMM17} claimed that each of the trends found under this representation 
describes a family of lines for different values of the convective turnover time 
rather than an universal behavior.
The comparison of $\prot/\pcyc$~vs~$\brac{R'_{HK}}$ provides an universal 
trend given that all the quantities are observables and do not depend
on the unknown convective turnover time, $\tau_c$. 

The correlations between the magnetic field amplitude and the magnetic
cycle period with the Rossby number 
have challenged theoreticians and modellers over the last decades. 
For the scaling of the field strength with $\Ro$, 
explanations rely on the so called mean-field dynamo number 
$D = C_{\alpha}C_{\Omega}$, where $C_{\alpha}$ and $C_{\Omega}$ are non-dimensional 
quantities that compare the inductive effects of the turbulent $\alpha$-effect and 
the shear, against the 
dissipative effects of turbulence \citep{Noyes+84b}. Nevertheless, the scaling of 
the dynamo coefficients with the Rossby number is unknown, and the hypothesis based 
on the linear mean-field theory remains unproven \citep{Noyes+84b,SB99,BT15}.

Under the same linear dynamo theory, the activity cycle period is proportional 
to $D^{-1/2}$ \citep{Stix76,Noyes+84b}. 
Yet,  this is an incomplete approach since it does not 
consider the back reaction of the magnetic field on the flow. 
The mean-field simulations of \cite{PK16} that consider the dynamic
evolution of the $\alpha$-effect (i.e., a form of including the magnetic 
field back reaction on the flow) produce a magnetic cycle period which increases
with the period of rotation, while the magnetic field amplitude 
decreases with the increase of the rotational period.
These results are in agreement with the observations 
except for the saturated phase which was not considered in their model.
When the non-linearity is considered through a simple algebraic quenching, 
the opposite relation is obtained, i.e., $\pcyc$ decreases with the increase
of $\prot$.  Compared with other mean-field results, \cite{PK16} 
clearly demonstrate the importance of the non-linear processes occurring in the dynamo. 
For instance, flux transport mean-field simulations, in which  the cycle period 
is mainly determined by the meridional circulation and the buoyant rise of magnetic
flux tubes, result in correlations that are at odds with the observations 
\citep{JBB10,KKC14}. Similarly, the global numerical simulations of 
solar-like stars performed by \cite{SBCBdN17} and  \cite{Warnecke17} 
obtained  $P_{\rm cyc}$ 
decreasing with the increase of $P_{\rm rot}$.  More recently, 
\cite{Viviani+18} reported high resolution simulations of stars with rotation
between $1$ and $30$ times the solar rotation rate.  In the $\prot/\pcyc$ representation,
they found that the slow rotating cases, displaying anti-solar differential rotation, 
fall close to the $I$ branch but have a negative slope. Interestingly, the dynamo solutions
in the fast rotating cases are all non-axisymmetric and fall in a different branch
of activity for super-active stars. 
One common aspect of the global simulations above \citep{SBCBdN17,Warnecke17,Viviani+18}
is the absence of the radial shear layers which are well observed in the 
Sun and should exist also in stars with radiative zones. Thus, these models do not 
generate strong toroidal fields, neglecting the most important source of non-linearity. 
It is worth then exploring the influence of these regions in the dynamo mechanism and 
in the determination of stellar magnetic cycles.

In this paper we study the scaling of the magnetic field 
strength and the magnetic cycle period with the Rossby number in
global convective dynamo simulations including rotational shear layers.
The numerical model employed here is the same as described in \cite{GSDKM16a}. 
A detailed analysis of the angular momentum budget and the generation of torsional 
oscillations was presented in \cite{GSDKM16b}. In this paper we present an 
extensive series of simulations where the only varying parameter is the 
rotation rate of the reference frame, and therefore $\Ro$.
Our previous results have demonstrated that the presence of tacholines 
result in dynamos where the evolution of the plasma is governed in large 
extent by deep seated magnetic fields. Here we show how the scaling laws 
obtained in these dynamos exhibit similarities with the observations. 
The goal of this paper is to provide a theoretical analysis explaining 
the physics behind the resulting scaling laws. 

In the next section we describe the numerical model,  the results are 
described in \S \ref{s.results}. We discuss the implications
of our results for solar and stellar dynamos in \S~\ref{s.con}.   
Technical details of the analysis are presented in Appendix. 

\section{The model}
\label{s.model}

We consider a full spherical shell domain, $0\le \phi \le 2\pi$,
$0\le \theta \le \pi$, with the bottom boundary at $r_b=0.61\Rs$
and the top boundary at $r_t=0.96\Rs$. The simulations have a
grid resolution of $128 \times 64 \times64 $ points in longitude
($\phi$), latitude ($\theta$) and radius ($r$), respectively.

We solve a set of anelastic MHD equations in the following form:
\begin{equation}                                                                                               
{\bm \nabla}\cdot(\rho_s\bm u)=0, \label{equ:cont}
\end{equation}
\begin{equation}
        \frac{D \bm u}{Dt}+ 2{\bm \Omega} \times {\bm u} =  
    -{\bm \nabla}\left(\frac{p'}{\rho_s}\right) + {\bm g}\frac{\Theta'}
     {\Theta_s} + \frac{1}{\mu_0 \rho_s}({\bm B} \cdot \nabla) {\bm B} \;, \label{equ:mom} 
\end{equation}
\begin{equation}
        \frac{D \Theta'}{Dt} = -{\bm u}\cdot {\bm \nabla}\Theta_e -\frac{\Theta'}{\tau}\;, \label{equ:en} 
\end{equation}
\begin{equation}
 \frac{D {\bm B}}{Dt} = ({\bm B}\cdot \nabla) {\bm u} - {\bm B}(\nabla \cdot {\bm u})  \;,
 \label{equ:in} 
\end{equation}
\noindent
where $D/Dt = \pd/\pd t + \bm{u} \cdot {\bm \nabla}$ is the total
time derivative, ${\bm u}$ is the velocity field in a rotating
frame with angular velocity ${\bm \Omega}=(\Omega_r,\Omega_{\theta},\Omega_{\phi}) 
= \Omega_0(\cos\theta,-\sin\theta,0)$,
$p'$ is the pressure perturbation variable that accounts for both the gas
and magnetic pressure,
${\bm B}$ is the magnetic field, and $\Theta'$ is the potential temperature
perturbation with respect to an
ambient state $\Theta_e$ \citep[see][for comprehensive discussions]{GSKM13b,Cossette+17}.
Furthermore, $\rho_s$  and $\Theta_s$ are the density and potential temperature
of the reference state which is chosen to be isentropic (i.e., $\Theta_s={\rm const}$)
and in hydrostatic equilibrium;   ${\bm g}=GM/r^2 \bm{\hat{e}}_r$
is the gravity acceleration, $G$ and $M$ are the gravitational
constant and the stellar mass, respectively, and $\mu_0$ is the magnetic
permeability. The potential temperature, $\Theta$, is related to the specific
entropy: $s=c_p \ln\Theta+{\rm const}$.

The simulations were performed using the EULAG-MHD code\footnote{The code is available at 
the dedicated website: \url{http://www.astro.umontreal.ca/~paulchar/grps/eulag-mhd.html}},
a spin-off of the hydrodynamical model EULAG predominantly used in atmospheric 
and climate research \citep{PSW08}. The time evolution is calculated using 
a bespoke semi-implicit approach derivable from the trapezoidal-rule path integration 
of the prognostic equations (2)-(4). At the heart of the approach there is a 
non-oscillatory (viz. high resolution) forward-in-time Multidimensional Positive Definite
Advection Transport Algorithm (MPDATA) broadly documented in the literature 
(see \cite{S06} for an overview, and \cite{WKPS18} for recent advancements).
A comprehensive description of the MHD implementation is presented in \cite{SC13}. 

The truncation terms in MPDATA  evince
viscosity comparable to the explicit sub-grid scale (SGS) viscosity used in large-eddy 
simulation (LES) models \citep{ES02,DXS03,MSW06}. Thus, the results of 
MPDATA are often interpreted as implicit LES or ILES \citep{GMR07}. 
This implicit SGS approach has been fundamental to successfully reproduce the 
solar tachocline and deep seated magnetic dynamos with time scales compatible with the 
solar cycle 
\citep{GCS10,RCGS11,SC13,GSKM13b, GSDKM16a, GSDKM16b}. 

For the velocity field we use impermeable, stress-free conditions at the 
top and bottom surfaces of the shell; whereas the magnetic field is
assumed to be radial at these boundaries. 
Finally, for the thermal boundary condition, we consider zero radial derivative
of the radial convective flux of potential temperature 
perturbations at the bottom, and zero convective radial flux of the potential 
temperature perturbations at the top surface.
All simulations start from a random noise, centered about zero and 
the same for each experiment,  in the potential temperature perturbations, 
velocity and the magnetic field. For the vector fields the noise is divergence
free. All simulations are run until reaching statistically steady 
state, using a constant time step $\Delta t = 1800$ s. A list of the 
simulation runs used in the current paper is presented in Table~\ref{table.1}.

\section{Results}
\label{s.results}

\subsection{Large-scale flows and magnetic field}
\label{s.flows}

\begin{deluxetable*}{cccccccccccccc}
\tablenum{1}
\tablecaption{Simulation parameters and results \label{table.1}}
\tablewidth{0.8\columnwidth}
\tablehead{
\colhead{Model} & \colhead{$\prot$} & \colhead{$\brac{\urms}^{CZ}$} & \colhead{$\tau_c$} & \colhead{$\Ro$} &
\colhead{$\Ro_1$} & \colhead{$D^{\prime}_r$} & \colhead{$\pcyc$} & \colhead{$\brac{\mean{B}_{\phi}}^{TAC}$} &
\colhead{$\brac{\mean{B}_{\phi}}^{CZ}$} & \colhead{$\brac{\mean{B}_{\phi}}^{NSL}$} & 
\colhead{$\brac{\mean{B}_{p}}^{TAC}$} & \colhead{$\brac{\mean{B}_{p}}^{CZ}$} & 
\colhead{$\brac{\mean{B}_{p}}^{NSL}$} \\
\colhead{} &  \colhead{[days]} & \colhead{[m s$^{-1}$]} & \colhead{$10^5$ [s]} & \colhead{ } &
\colhead{ } & \colhead{$10^3$} & \colhead{[yr]} & \colhead{[T]} &
\colhead{[T]} & \colhead{[T]} & \colhead{[T]} & \colhead{[T]} & \colhead{[T]}
}
\startdata
RC07 &  7.0   & 35.18  &  2.71 &  0.36   & 0.26  & 2.93  &  3.0   &  0.095  &   0.063  &  0.072  &  0.241  &  0.035  &  0.024\\ 
RC14 &  14.0  & 37.69  &  3.41 &  0.56   & 0.60  & 3.74  & 9.6   &  0.167  &   0.128  &  0.152  &  0.483  &  0.045  &  0.039\\
RC18 &  18.0  & 38.99  &  2.83 &  0.87   & 0.80  & 1.84  & --    &  0.215  &   0.044  &  0.062  &  0.302  &  0.024  &  0.023\\
RC21 &  21.0  & 38.85  &  2.87 &  1.01   & 0.91  & 0.99  &30.1   &  0.246  &   0.037  &  0.062  &  0.240  &  0.022  &  0.024\\
RC24 &  24.0  & 38.83  &  2.96 &  1.12   & 1.02  & 0.35  & 16.3   &  0.286  &   0.058  &  0.088  &  0.110  &  0.040  &  0.041\\
RC28 &  28.0  & 38.90  &  3.11 &  1.24   & 1.19  & -0.27 &16.3   &  0.318  &   0.064  &  0.099  &  0.106  &  0.040  &  0.044\\
RC35 &  35.0  & 39.84  &  3.37 &  1.43   & 1.51  & -1.06 &19.1   &  0.409  &   0.066  &  0.087  &  0.111  &  0.034  &  0.037\\
RC42 &  42.0  & 40.70  &  3.64 &  1.59   & 1.87  & -1.54 &22.9   &  0.353  &   0.068  &  0.088  &  0.106  &  0.029  &  0.034\\
RC49 &  49.0  & 42.19  &  3.83 &  1.76   & 2.24  & -1.92 &26.2   &  0.456  &   0.075  &  0.090  &  0.106  &  0.025  &  0.028\\
RC56 &  56.0  & 43.83  &  3.95 &  1.95   & 2.69  & -2.18 & --    &  0.406  &   0.079  &  0.101  &  0.169  &  0.022  &  0.026\\
RC63 &  63.0  & 44.07  &  4.08 &  2.12   & 3.03  & -2.28 & --    &  0.340  &   0.079  &  0.089  &  0.149  &  0.021  &  0.023\\\hline
\enddata
\tablecomments{
The convective turnover time, $\tau_c$ is computed from the spectra of the non-axisymmetric velocity and 
magnetic fields as explained in Appendix~\ref{ap.A}. $\Ro = \prot/2 \pi \tau_c$ is the Rossby number
evaluated with $\tau_c$ in this table; $\Ro_1 = \prot/\tau_c^*$ is computed with $\tau_c^*$ estimated at one 
pressure height scale above the bottom of the convection zone, as in \cite{Noyes+84a}. 
The dynamo number, $D^{\prime}_r = C_{\alpha}^{\prime} C_{\Omega}^{\prime r}$ corresponds to the 
average over TAC in the polar region.
in the polar region. The primes in this definition mean that it is computed from polynomial fits to the 
dynamo coefficients. 
The period is computed by using the Fourier transform as explained in Appendix \ref{ap.C}. The quantities 
in angular brackets correspond to the time and volume average over the regions TAC, CZ and NSL (see the text).}
\end{deluxetable*}

\begin{figure*}[h]
\begin{center}
\includegraphics[width=0.3\columnwidth]{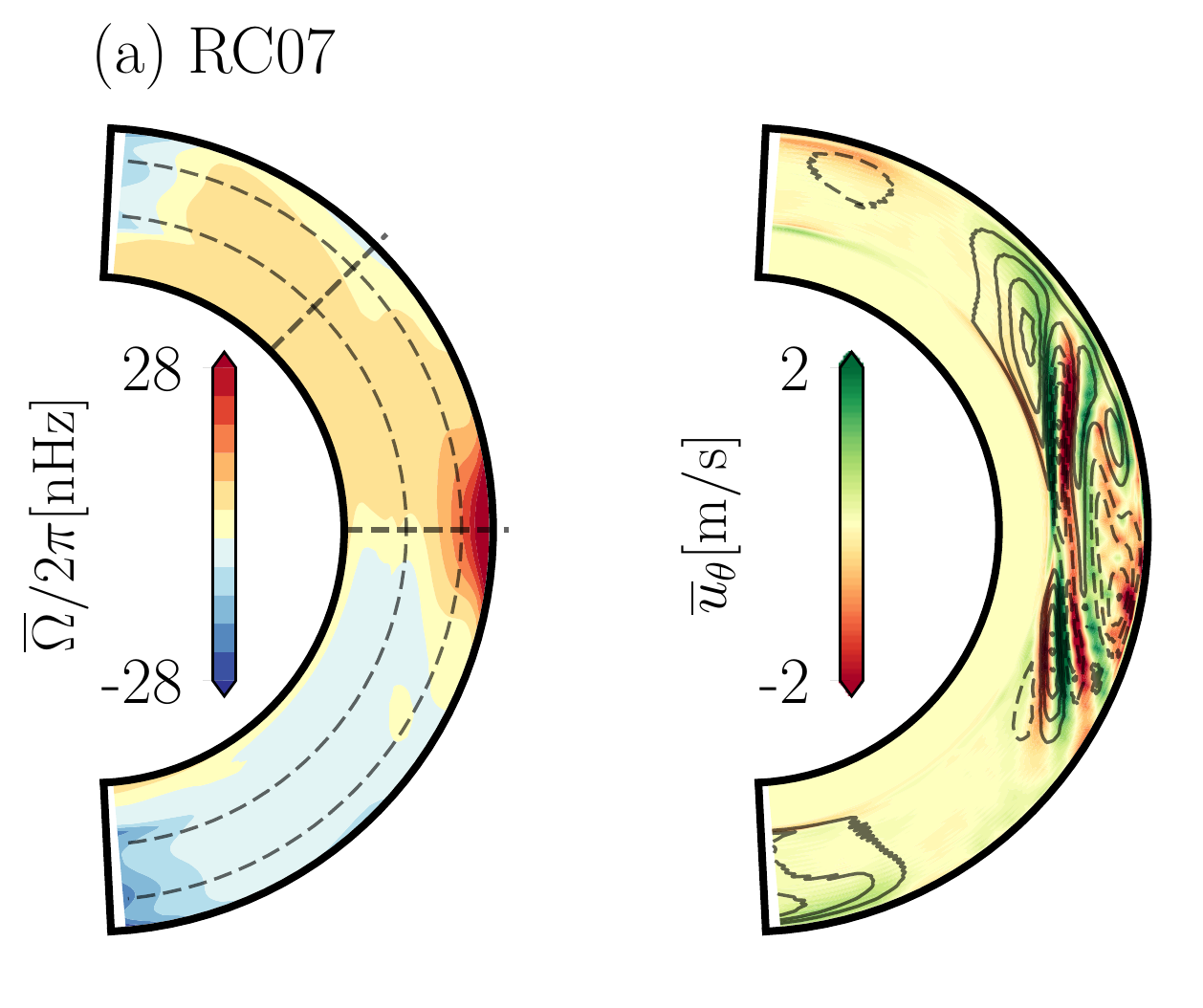} \hspace{0.2cm}
\includegraphics[width=0.3\columnwidth]{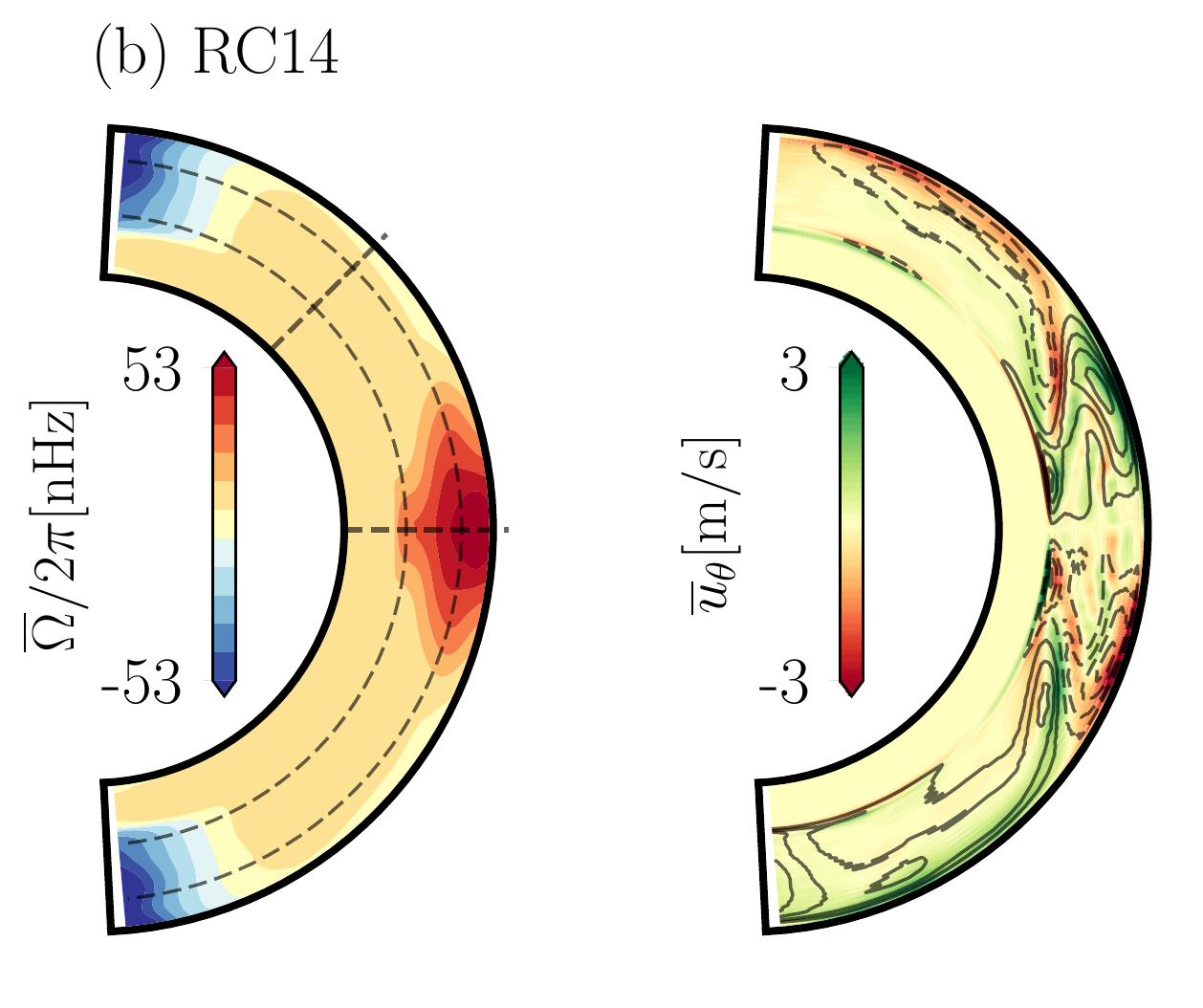}\hspace{0.2cm}
\includegraphics[width=0.3\columnwidth]{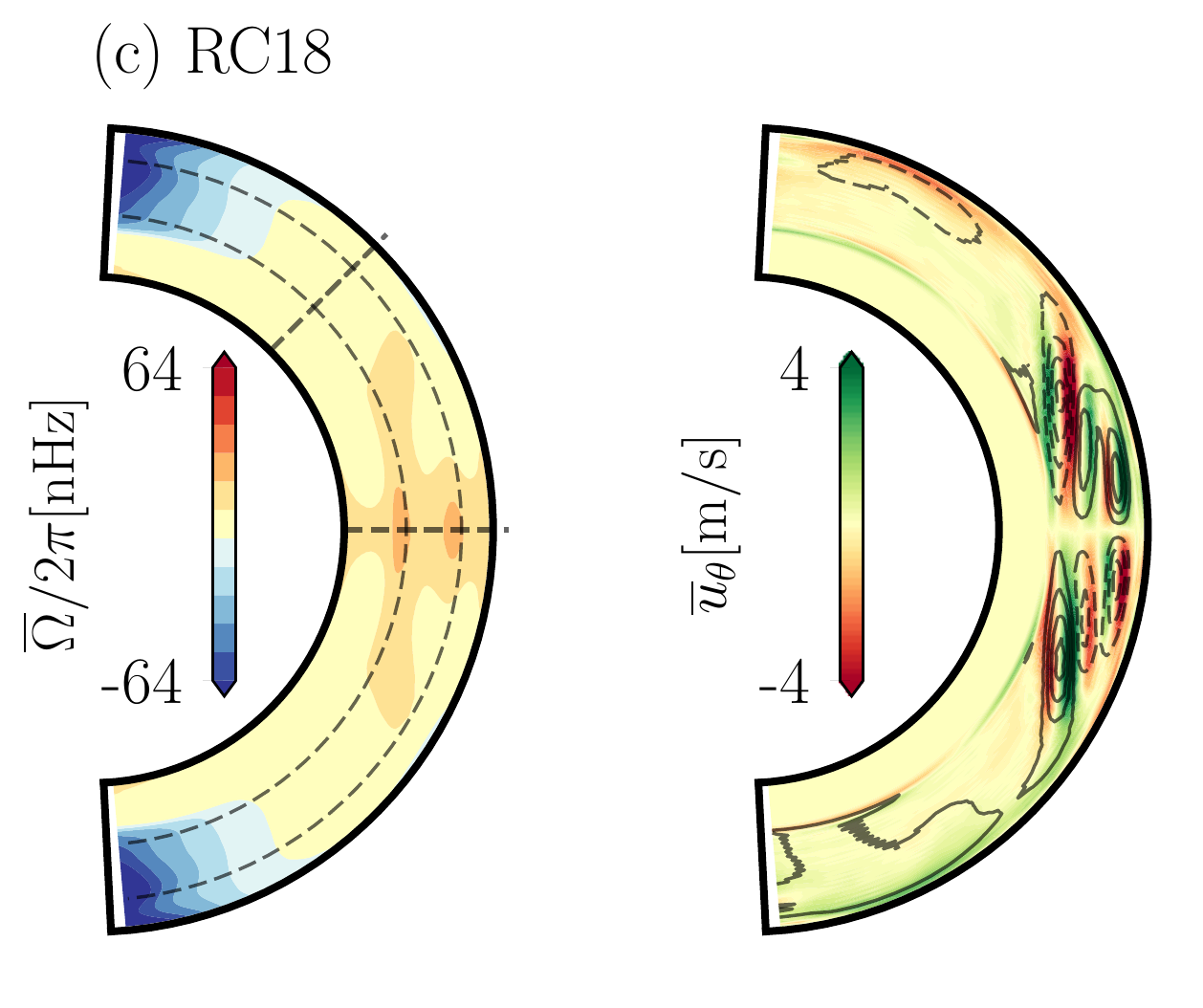}\\
\includegraphics[width=0.3\columnwidth]{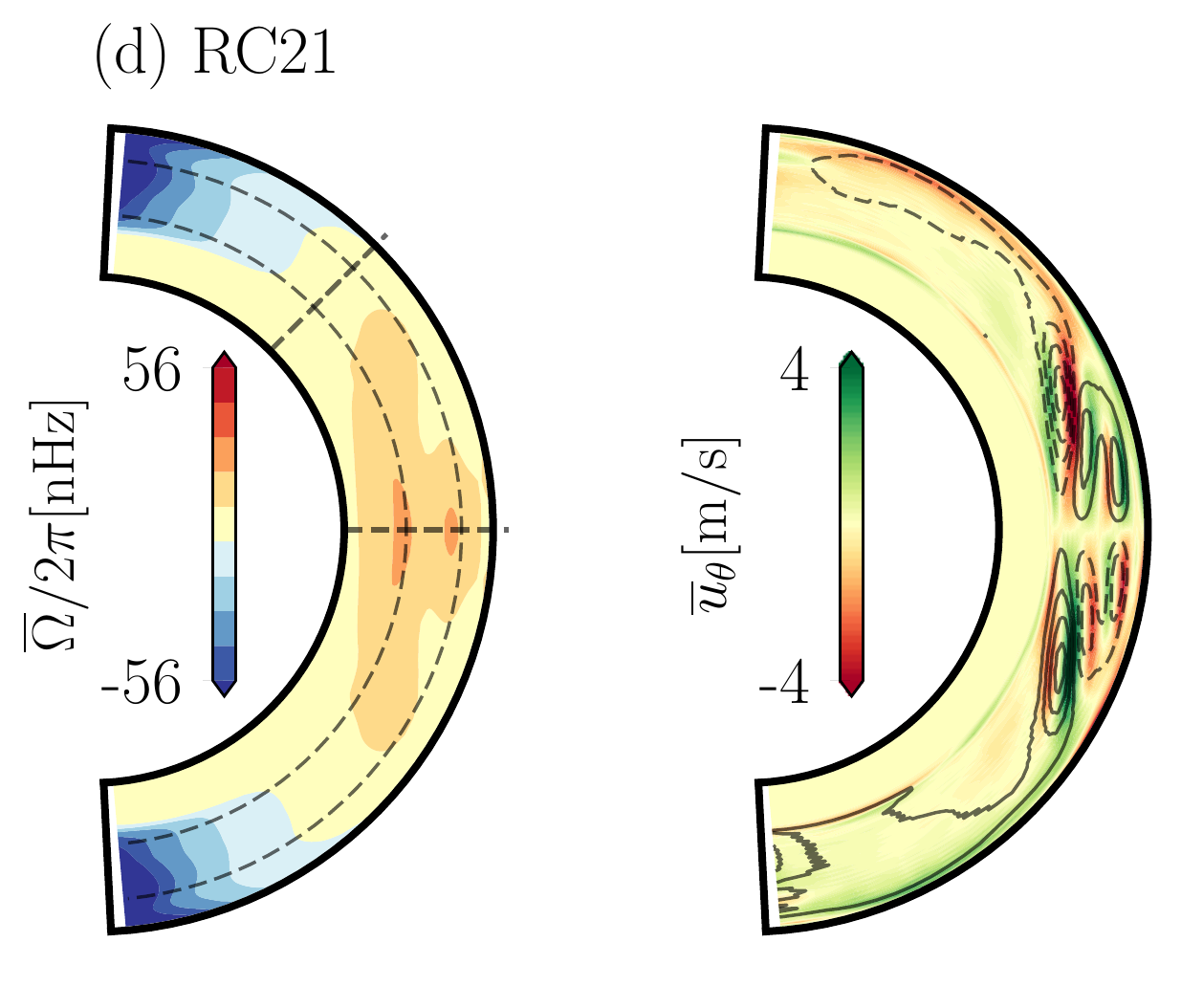} \hspace{0.2cm}
\includegraphics[width=0.3\columnwidth]{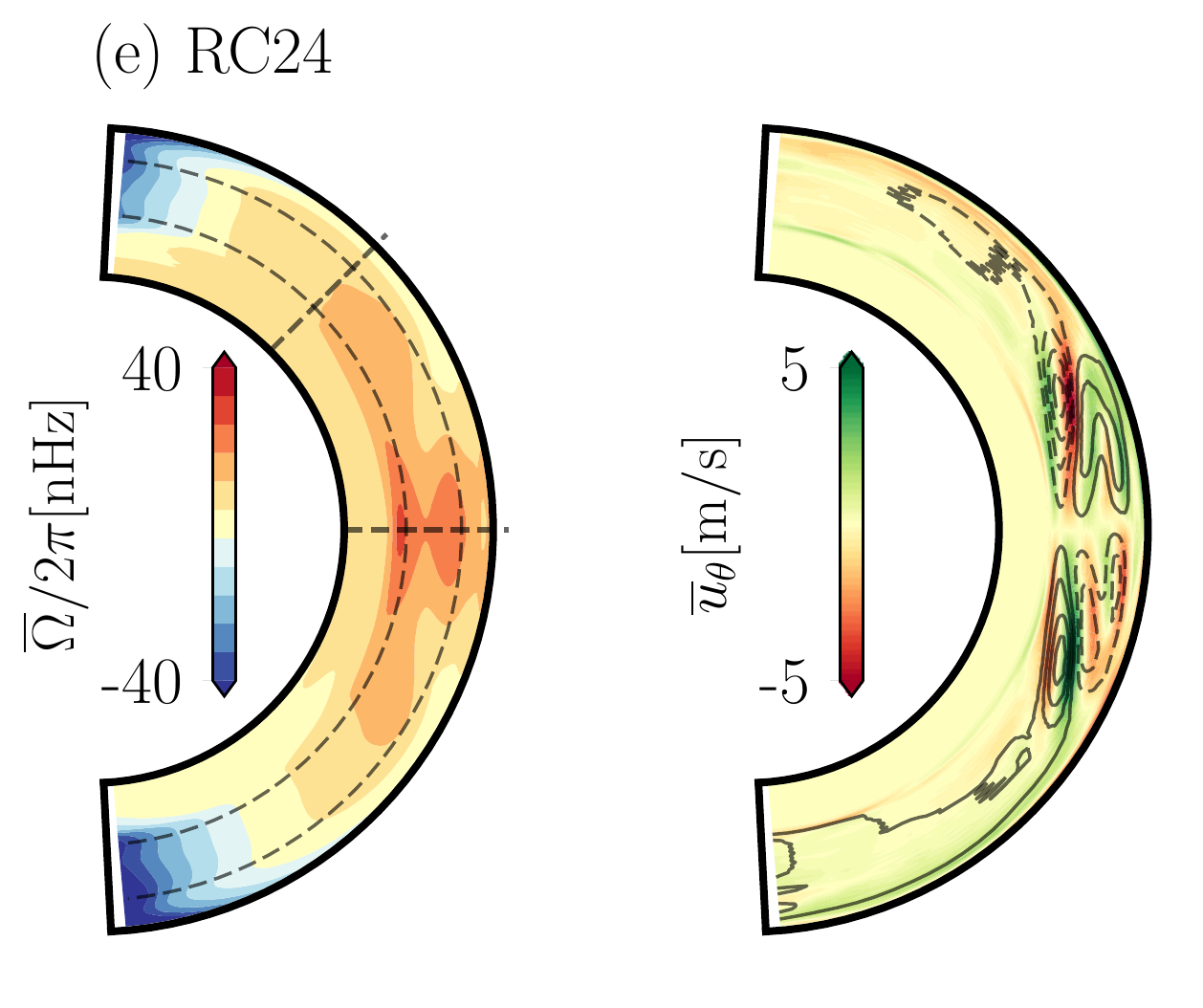} \hspace{0.2cm}
\includegraphics[width=0.3\columnwidth]{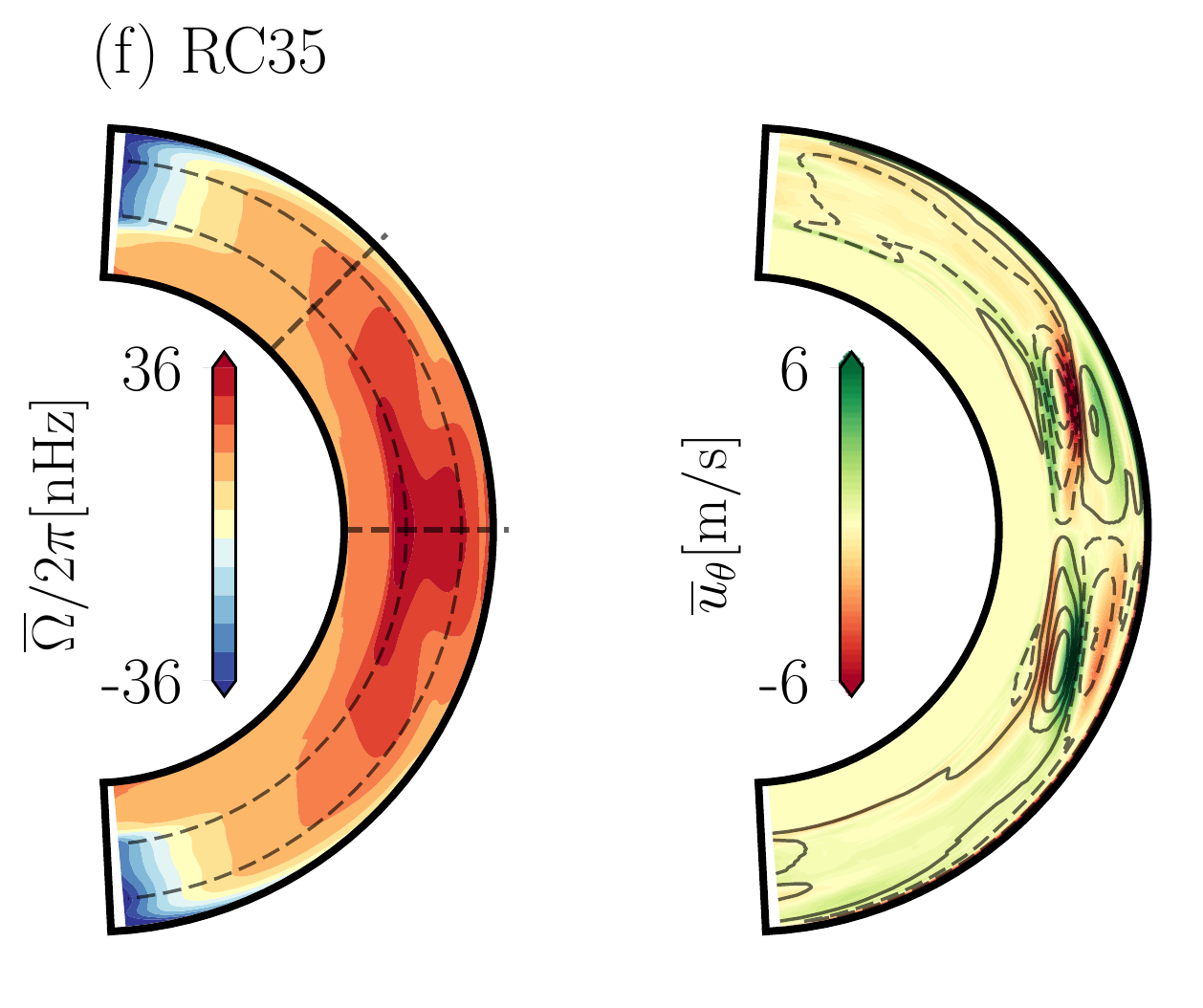}\\
\includegraphics[width=0.3\columnwidth]{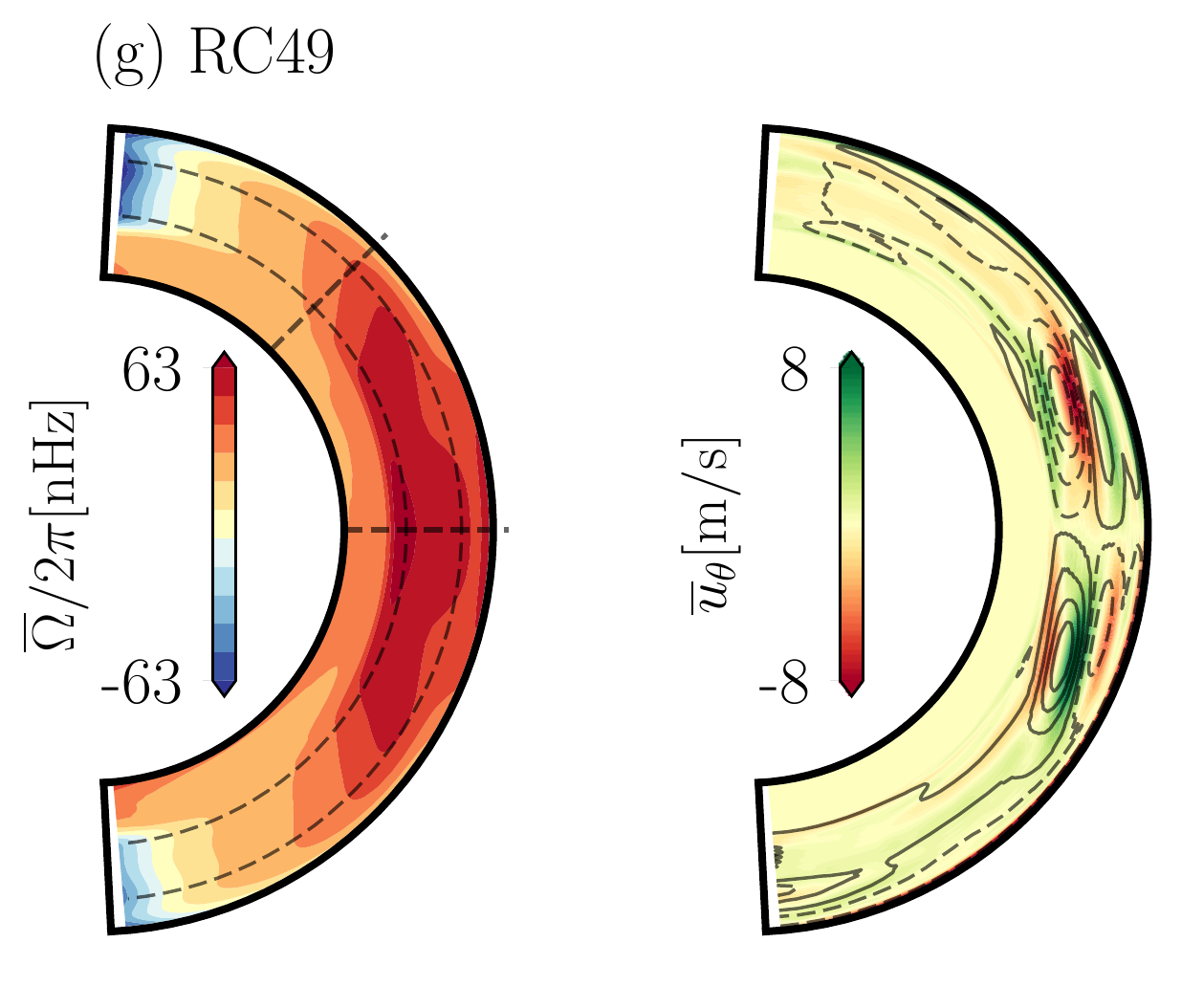} \hspace{0.2cm}
\includegraphics[width=0.3\columnwidth]{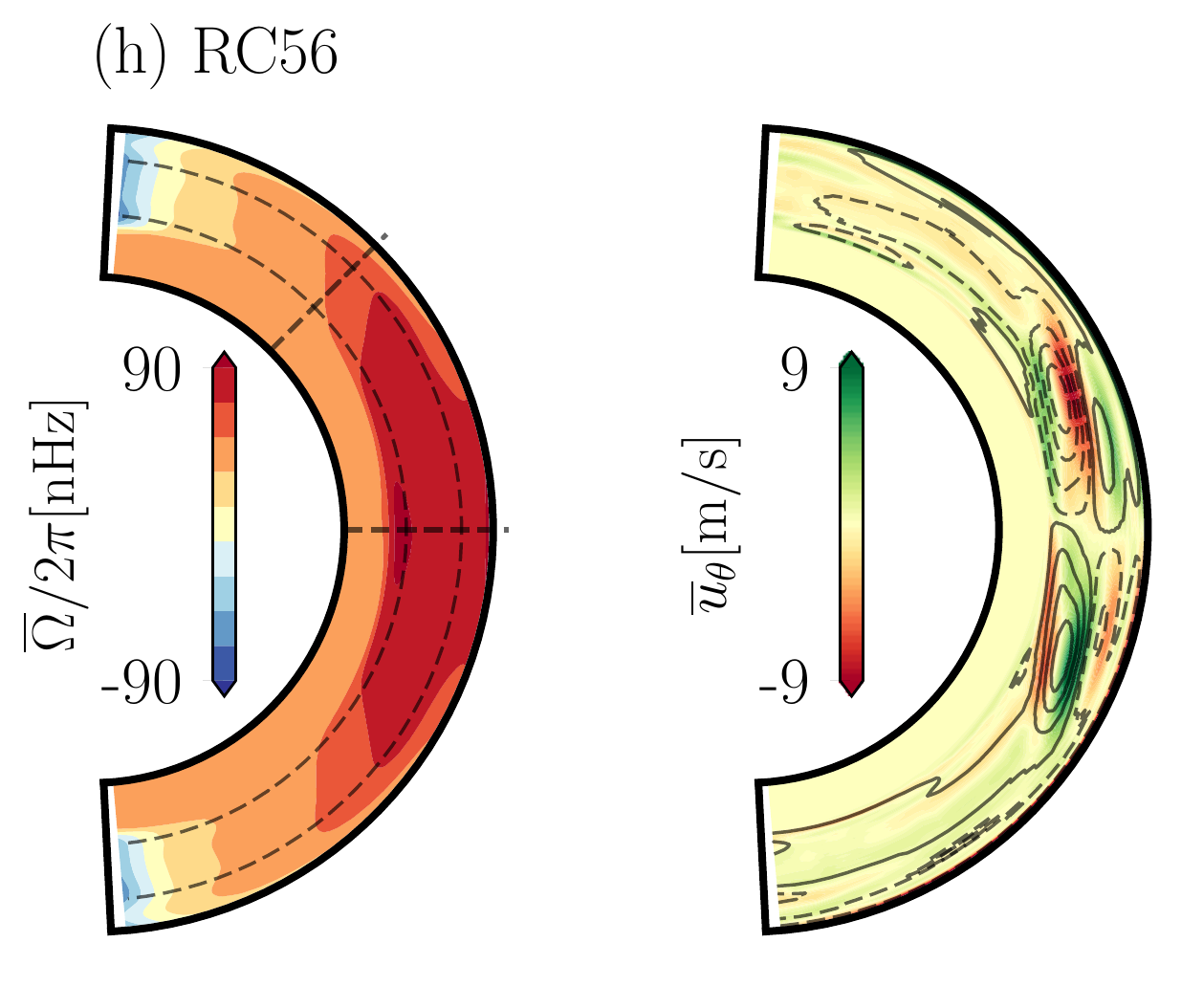} \hspace{0.2cm}
\includegraphics[width=0.3\columnwidth]{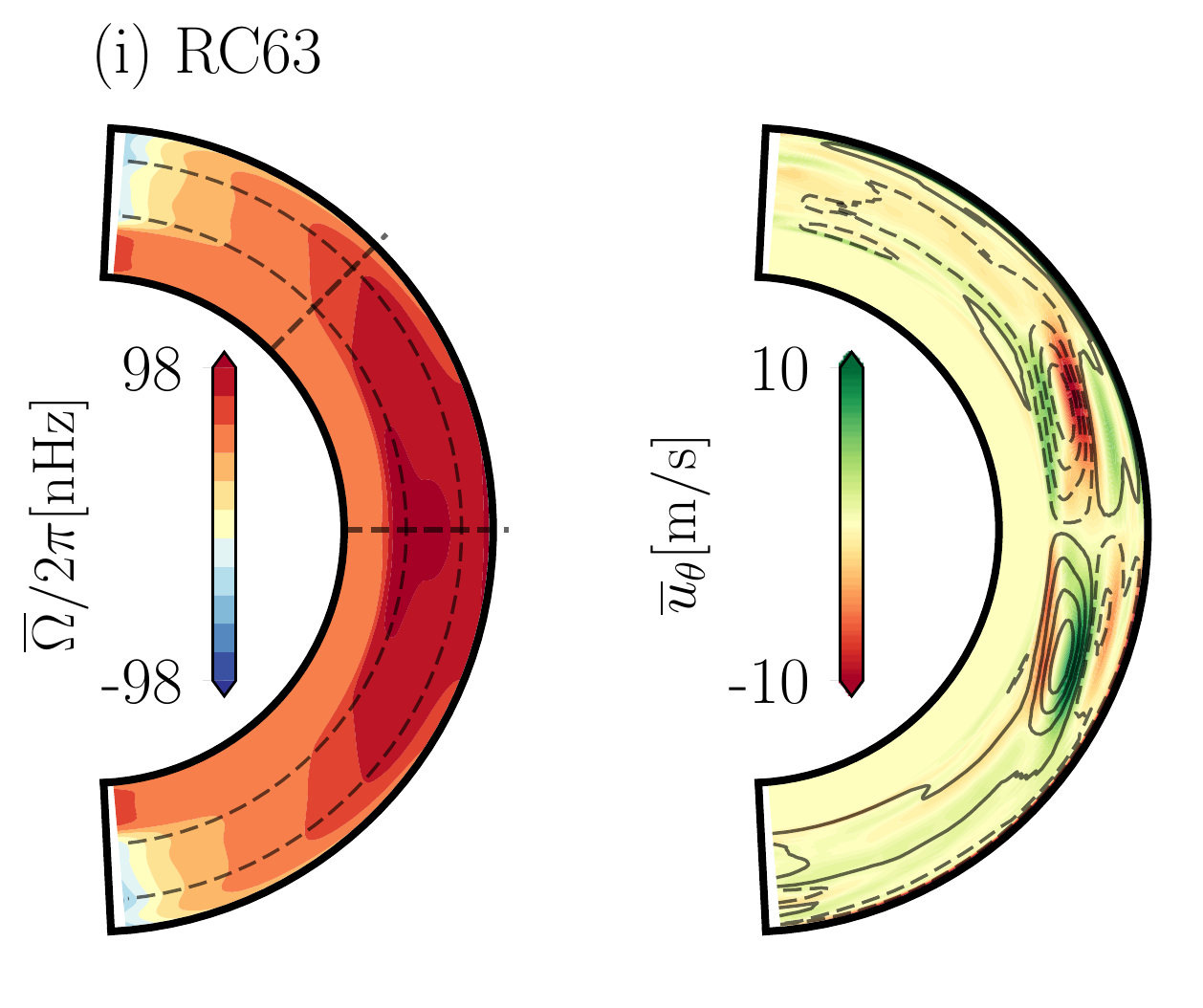}\\
\caption{Differetial rotation (left) and meridional circulation (right) of 
the models (a) RC07 to (i) RC63. The colors in the differential rotation
show iso-rotation contours in the rotating frame. 
The continuous lines divide the domain in the six different analysis
regions.  In the meridional circulation panels
the colored contours show the latitudinal velocity, $\mean{u}_{\theta}$. The
continuous (dashed) lines represent clockwise (counter-clockwise) circulation.}
\label{fig.dr} 
\end{center}
\end{figure*}

Figure \ref{fig.dr} shows the differential rotation (left panels of each model)  and 
meridional circulation (right panels) of some representative models from (a) RC07 to (i) RC63 
(the profiles of models RC28 and RC56 are not presented here since they appear in
\cite{GSDKM16a,GSDKM16b}).  
In the differential rotation profiles, colored contours depict the variations 
of the mean angular velocity, $\mean{\Omega}$, calculated as a 
temporal and azimuthal average, with respect to the rotating
frame. The results qualitatively show that the gradients of angular velocity
become prominent with the increase of the rotation period.  
Observational results also indicate that the latitudinal differential rotation
increases with the rotation period \citep[see Section 6.1 of ][and references
therein]{Lehtinen+16}.  The radial differential rotation for stars 
other than the Sun is evasive to observations. 
Nevertheless, it is commonly assumed that the shear is stronger for
rapidly rotating stars  
\citep[e.g.,][]{Noyes+84b,BT15}.
However, in our fast rotating simulations (RC07 - RC21) there is almost no
radial shear at the tachocline and in the near-surface layer.
As the rotation diminishes progressively from model~RC21 to 
model~RC63 the clearest gradients are observed at the tachocline as well as 
in the near-surface layers (a quantitative analysis is presented in 
\S\ref{sec.mfa}).

In the meridional circulation panels, the colored contours show the mean latitudinal
velocity ($\mean{u}_{\theta}$).  The contour lines show the stream function.
For the fast rotating models we observe a pattern of multiple slow convection 
cells, circulating over thin cylinders.  The number of cells decreases with
the increase of the rotation period. For instance, from model~RC28 to model~RC63  
only two cells are developed in each meridional quadrant. 
One is a broad counterclockwise cell going from $\sim 0.72 \Rs$ to
$\sim 0.86 \Rs$  and in all latitudes. The second one is clockwise. It is located 
above $\sim 0.86 \Rs$ and close to the equator.  The amplitude of $\mean{u}_{\theta}$
is monotonically increasing from the faster (RC07) to the slower (RC63) 
rotating models as can be noticed in the corresponding color bars.  

\begin{figure*}[h]
\begin{center}
\includegraphics[width=0.4\columnwidth]{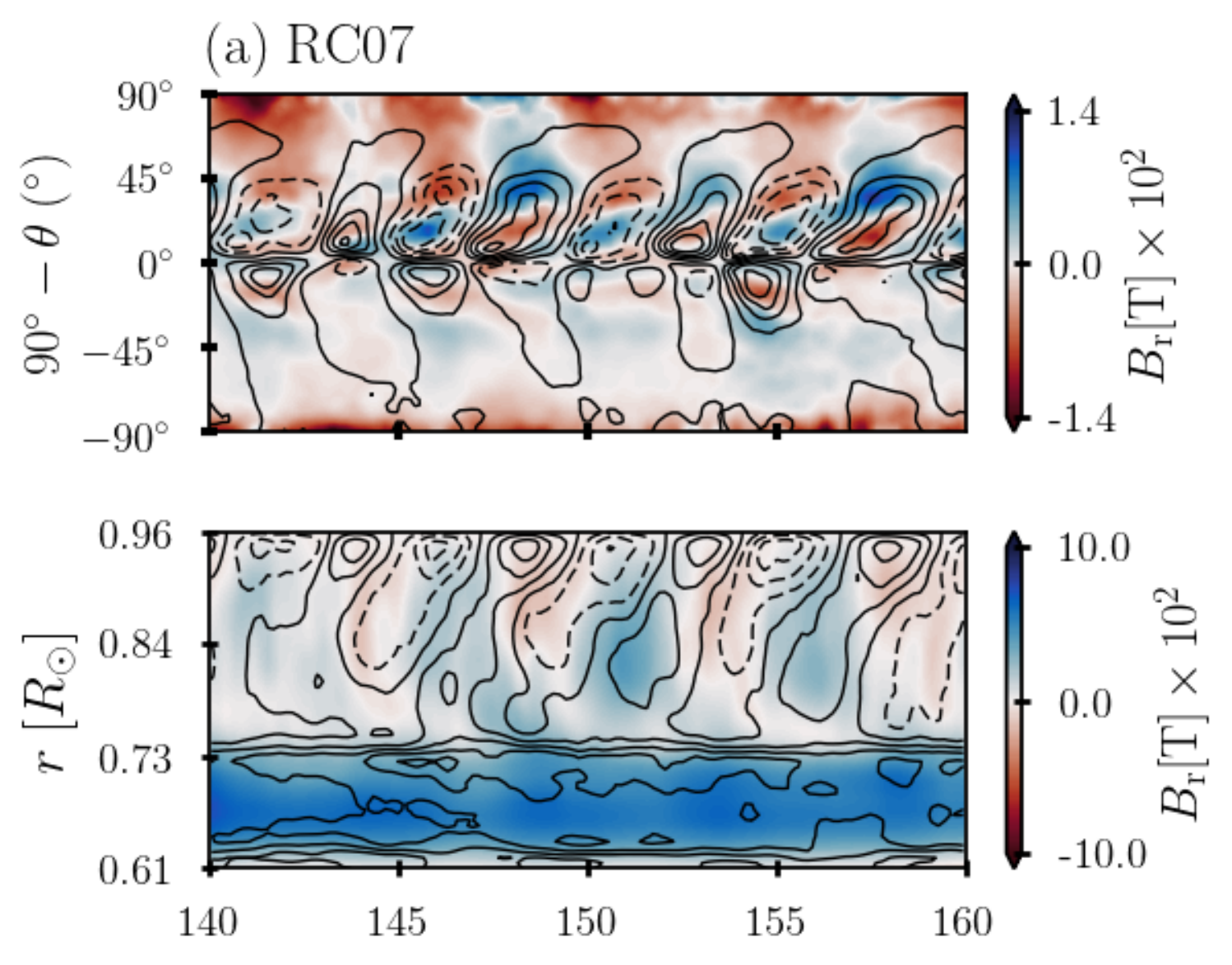}
\hspace{0.4cm}
\includegraphics[width=0.4\columnwidth]{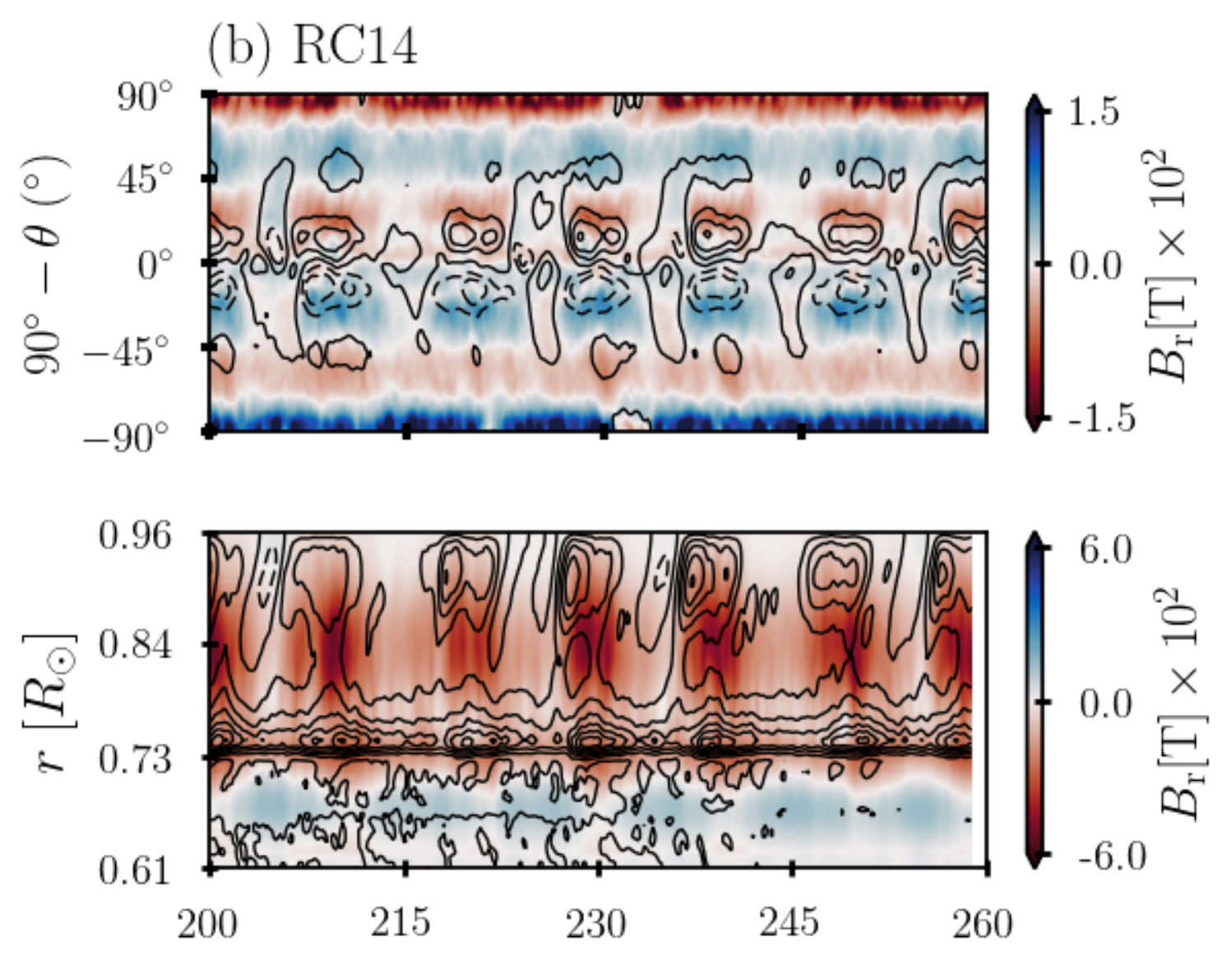}\\
\includegraphics[width=0.4\columnwidth]{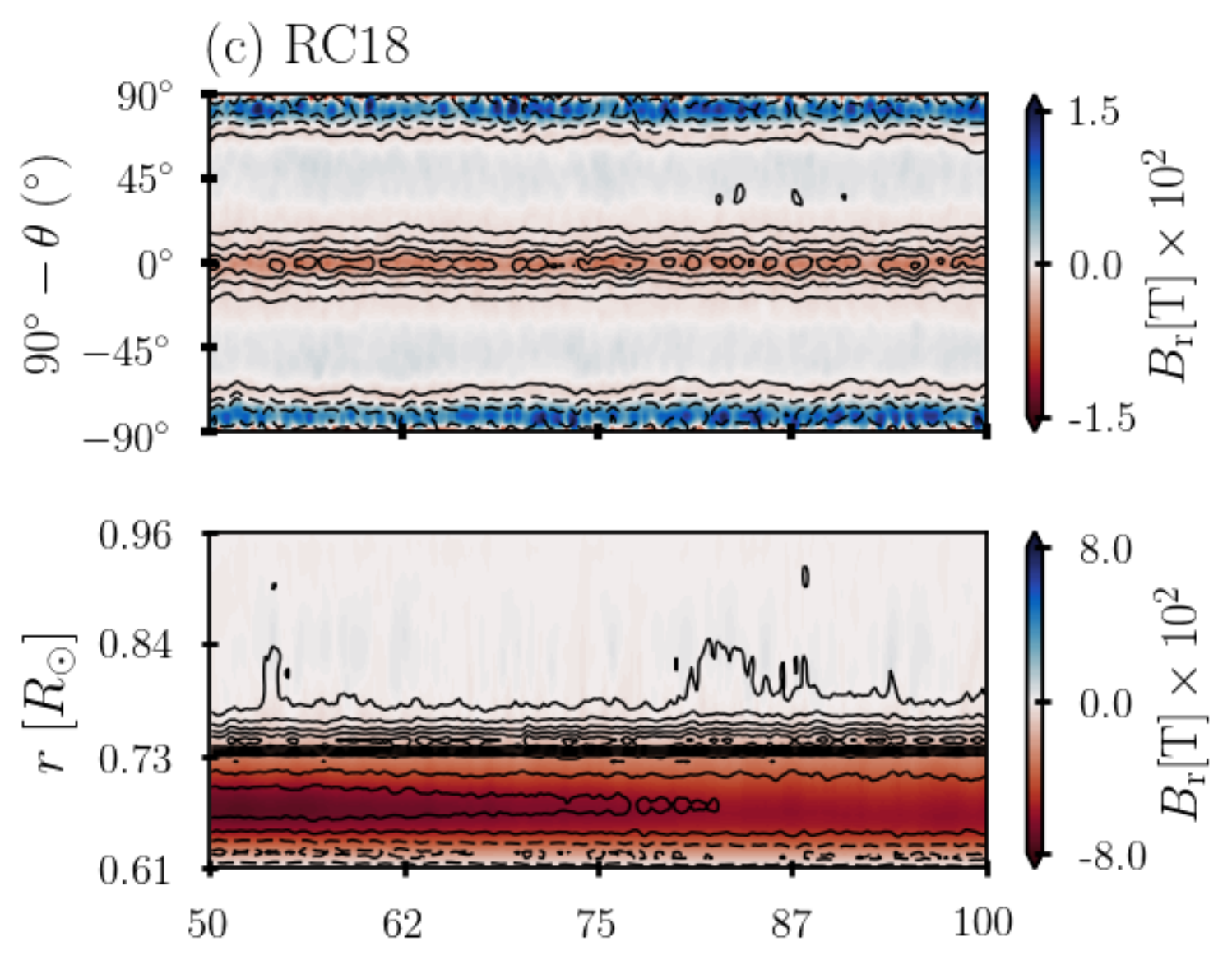}
\hspace{0.4cm}
\includegraphics[width=0.4\columnwidth]{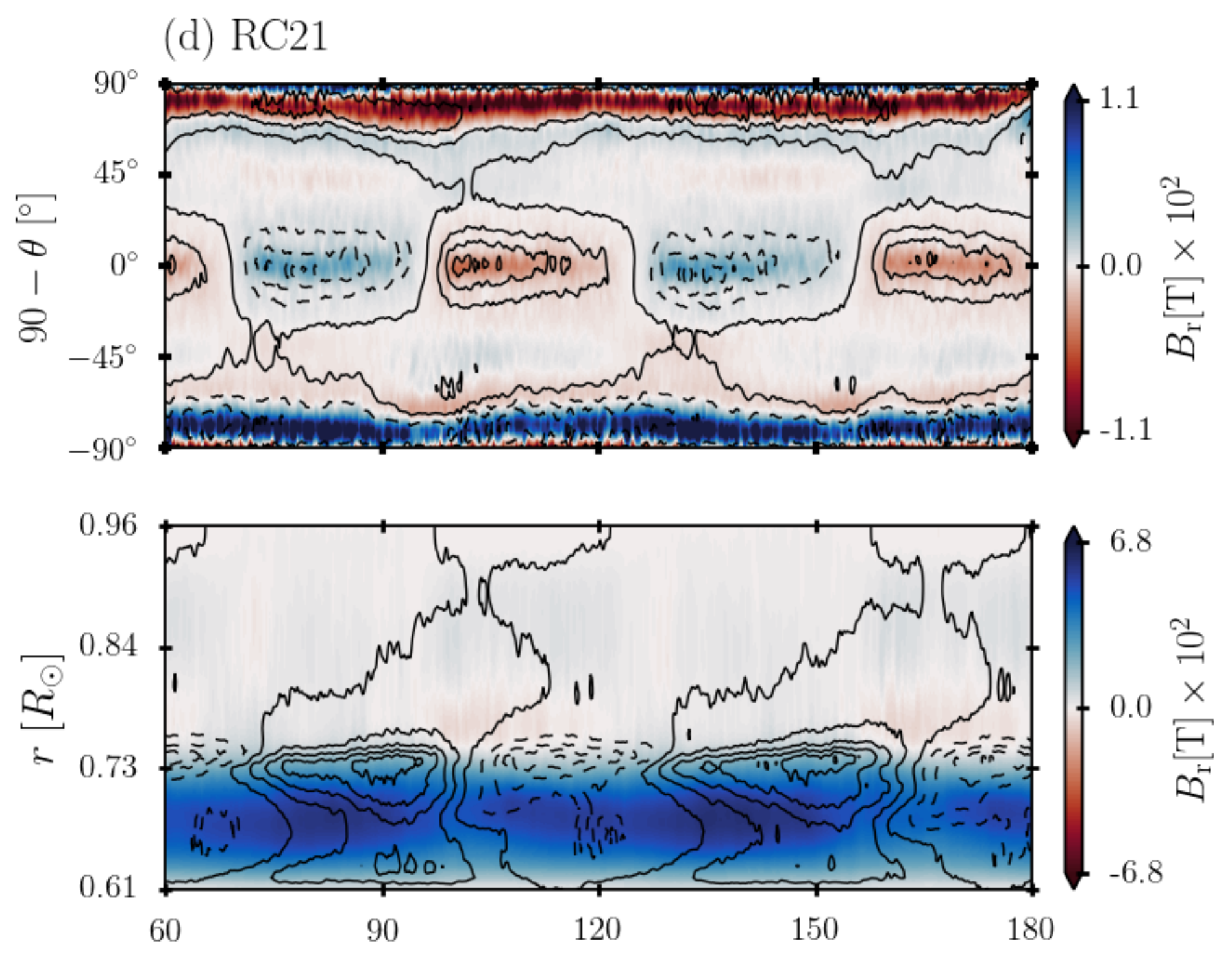}\\
\includegraphics[width=0.4\columnwidth]{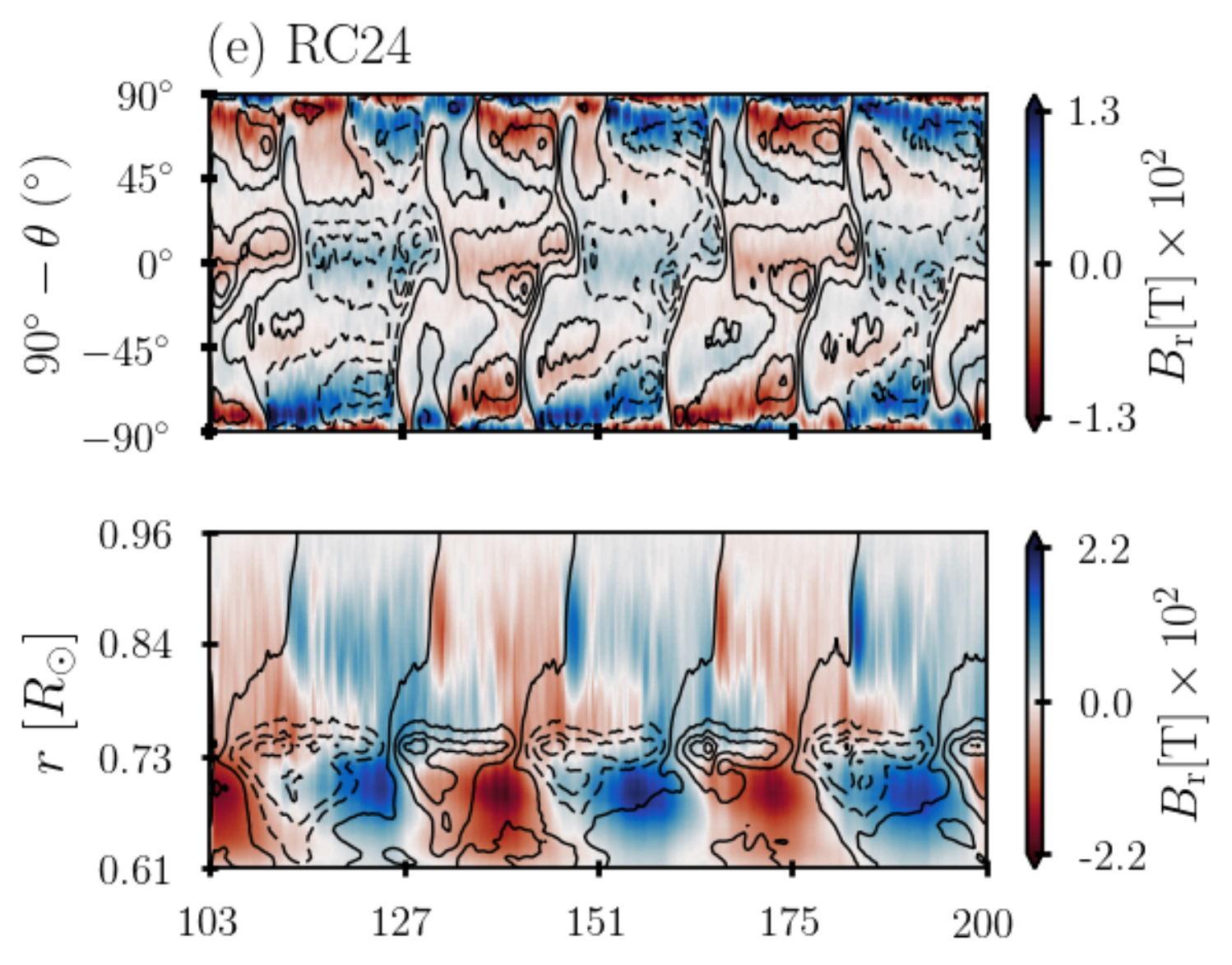}
\hspace{0.4cm}
\includegraphics[width=0.4\columnwidth]{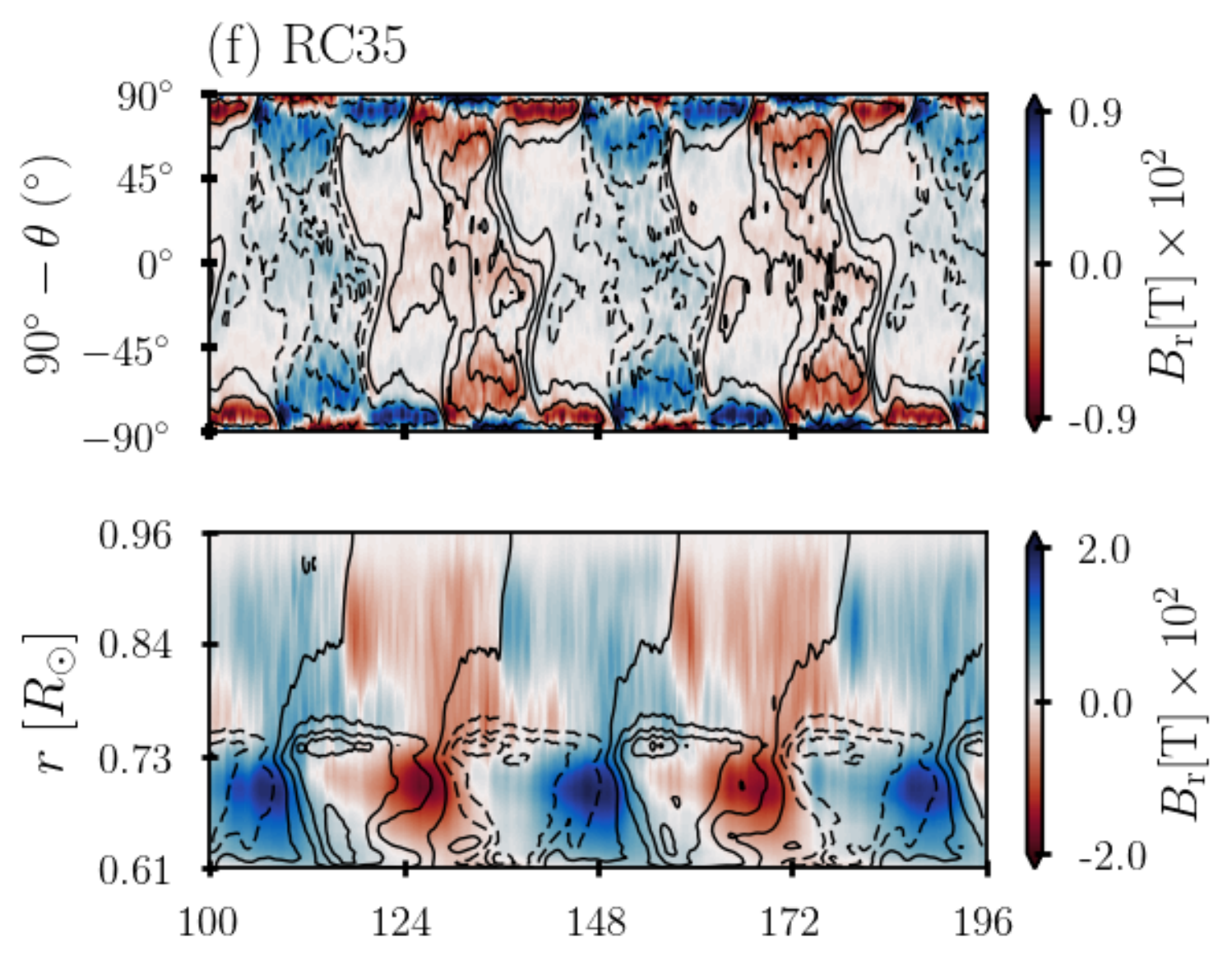}\\
\includegraphics[width=0.4\columnwidth]{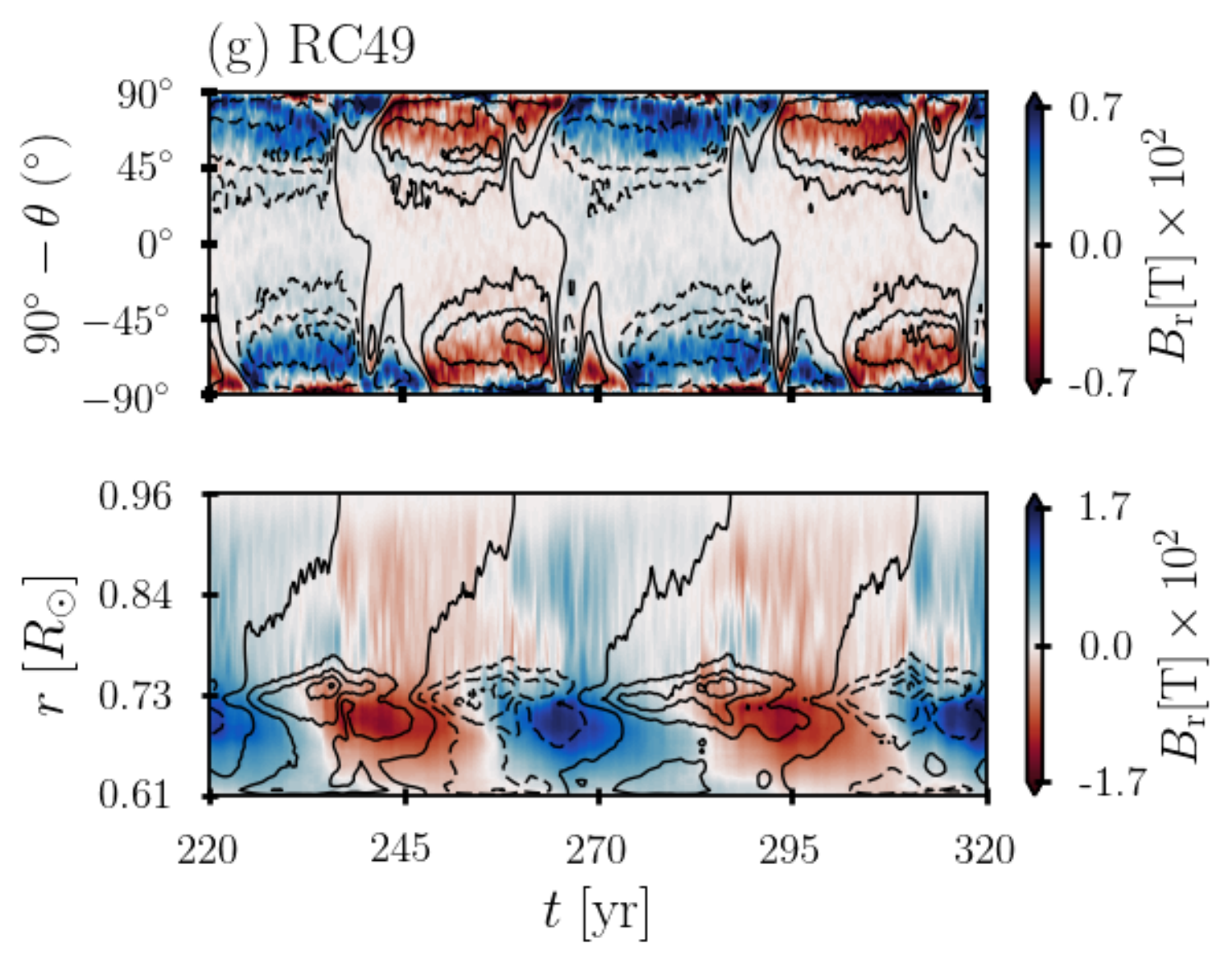}
\hspace{0.4cm}
\includegraphics[width=0.4\columnwidth]{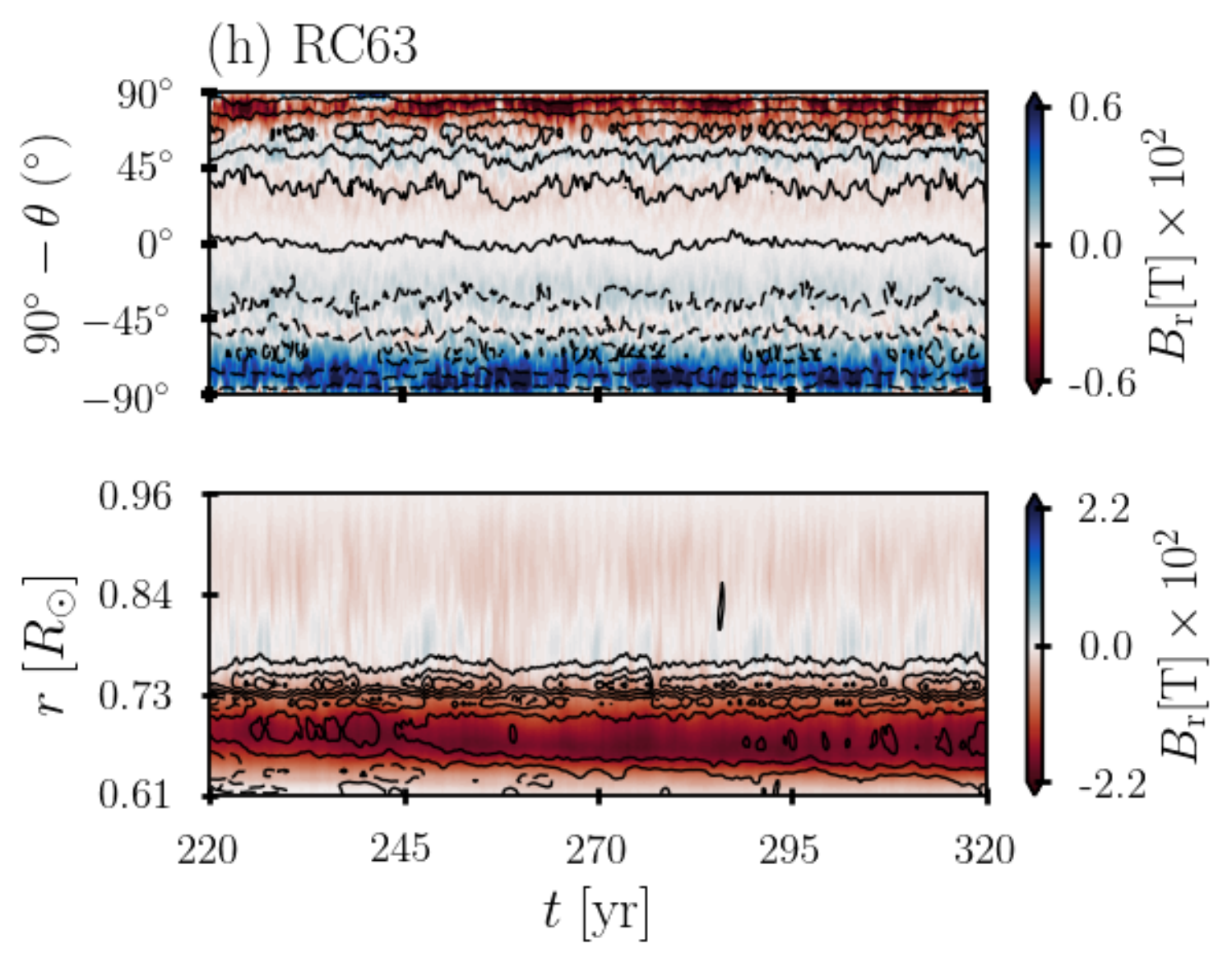}
\caption{Time-latitude (at $r = 0.95\Rs$)  and 
time-radius (at $\theta = 24^{\circ}$) butterfly diagrams showing the evolution 
of models between (a) RC07 and (h) RC63. The coulored contours 
show the radial magnetic field, with its amplitude, in Tesla, depicted in the 
color tables. The continuous (dashed) contour lines show the positive (negative) 
toroidal field.}
\label{fig.bd}
\end{center} 
\end{figure*}

Figure~\ref{fig.bd} shows the time-latitude (at $r=0.95\Rs$) 
and time-radius (at $\theta=24^{\circ}$ latitude) evolution of the 
mean magnetic field associated with the mean-flows described above 
($\mean{B}_{r}$ by colored contours and $\mean{B}_{\phi}$ by contour lines).  
Several dynamo modes  can be distinguished for different values of $\Ro$.  
For the fastest rotating
model (smallest $\Ro$) RC07, the rotation profile is 
almost homogeneous in the whole domain and the shear is small (Fig.~\ref{fig.dr}a). 
Even though there is a steady magnetic 
field in the radiative zone, an oscillatory dynamo with a short period of 2.8 years
develops predominantly in the upper convection zone. 
In model~RC14 (b) the radial shear at the tachocline starts to develop leading 
to the formation of an antisymmetric steady dynamo.  
In this model the magnetic field is oscillatory in the convection zone, however, 
it does not show hemispheric polarity reversals.  
In model~RC18 (panel c), the dynamo also develops mainly at the tachocline.
The field is steady and no cyclic variations of the field are observed in the 
convection zone.  In model~RC21 (panel d) the solution exhibits bimodality, 
i.e., two dynamo modes are simultaneously excited with the magnetic field
periodic in the equatorial region but steady at the poles.
In the radius-time diagram, periodic reversals of the toroidal field (see 
continuous and dashed lines) can be observed while the poloidal field remains 
steady.   Models RC24-RC49 
(e-g) are periodic 
with well defined magnetic field polarity reversals. 
The magnetic field is generated mainly at the tachocline, but the dynamo 
action occurs in the entire convection zone. The radial magnetic field 
is also observed in the convectively stable layer, and reverses polarity
during the toroidal field maxima.   
The parity of these models varies in time and is not well defined. The slowly rotating 
models RC56 (the butterfly diagram of which is not shown in
Fig.~\ref{fig.bd}, but it is presented in figure 6 (c) of \cite{GSDKM16a}) and RC63 
(h) are all antisymmetric steady 
dynamos. We have also performed simulations (not shown here) for longer 
rotational periods, i.e., 112 and 140 days. In these cases there is still dynamo 
action but the magnetic field is weak, with no back reaction on the convection zone 
dynamics. For models with the rotation period longer than 224 days 
the dynamo instability does not develop.

\begin{figure*}[h]
\begin{center}
\includegraphics[width=0.32\columnwidth]{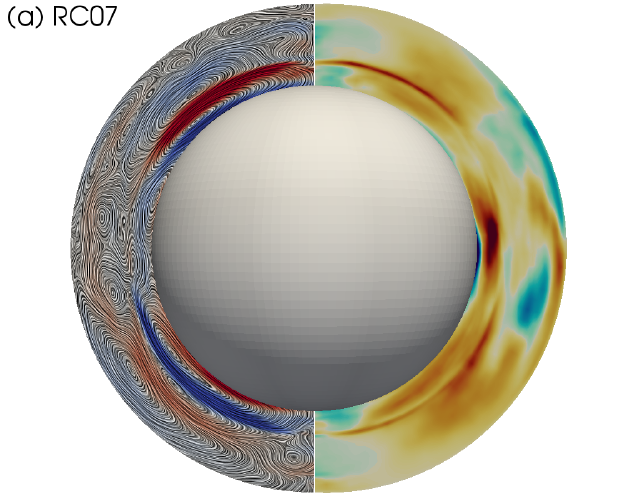} \hspace{0cm}
\includegraphics[width=0.32\columnwidth]{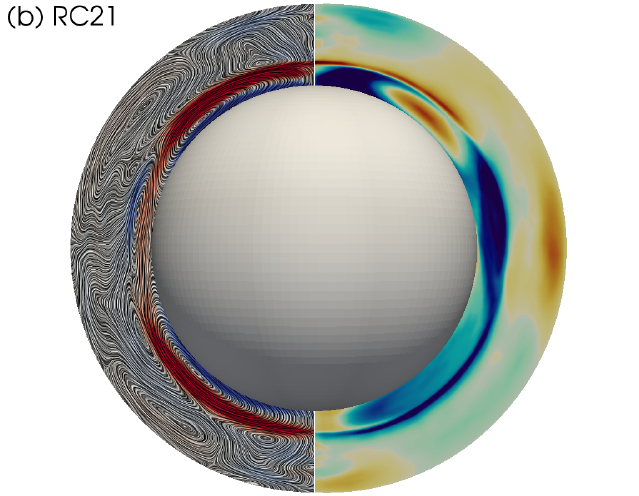} \hspace{0cm}
\includegraphics[width=0.32\columnwidth]{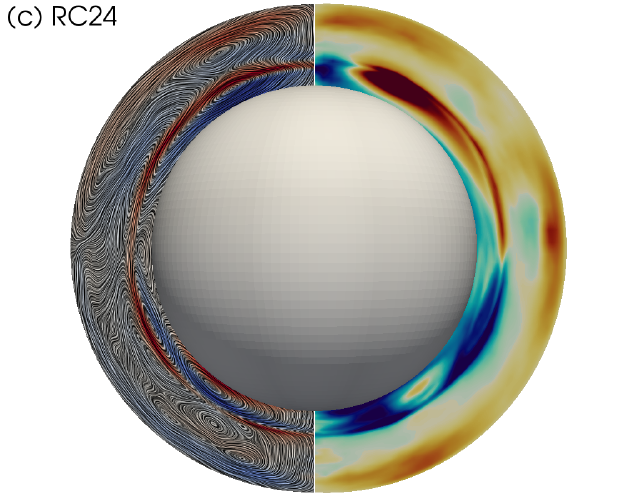}\\
\vspace{0.4cm}
\includegraphics[width=0.32\columnwidth]{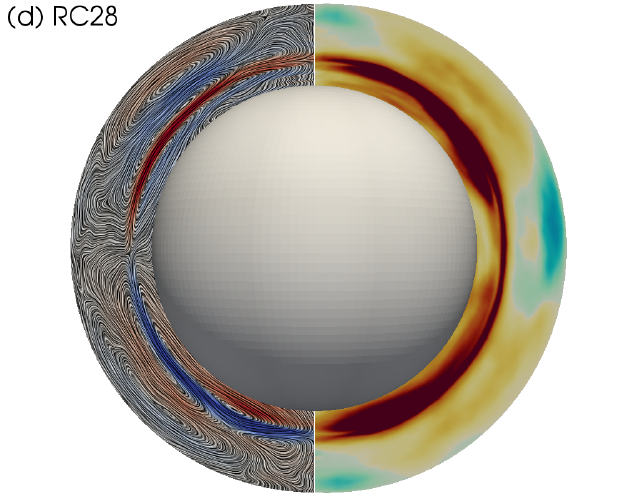} \hspace{0cm}
\includegraphics[width=0.32\columnwidth]{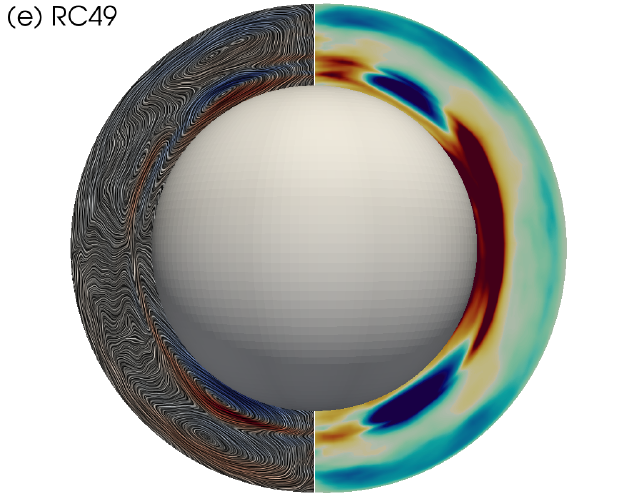} \hspace{0cm}
\includegraphics[width=0.32\columnwidth]{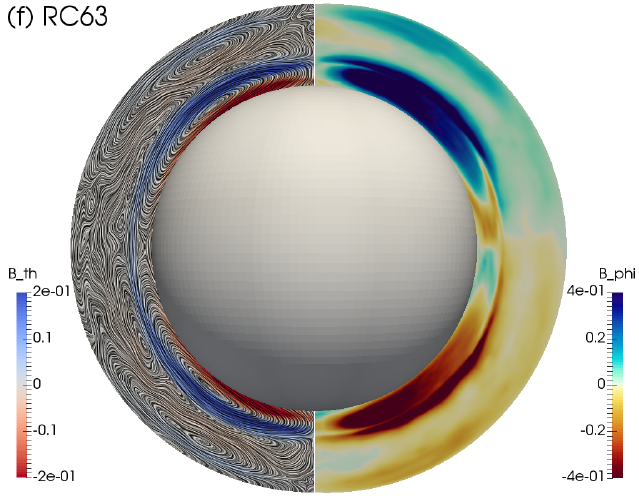}\\
\caption{Snapshots indicating the distribution of the mean magnetic field 
components for some characteristic simulations. On the left, the LIC representation depicts the 
distribution of the meridional field lines in the convection zone with the color indicating the magnitude 
of the mean latitudinal field, $\bar{B}_{\theta}(t,\theta,r)$.  The colored contours on the right quadrants 
correspond to the azimuthal mean magnetic field, $\bar{B}_{\phi}(t,\theta,r)$. 
Because the fields at the tachocline are about one order of magnitude larger than the fields
at the convection zone, the contours are highly saturated. For a better comprehension of these 
results and behavior, animations corresponding to some of these
cases are available as supplementary material or in the link
\url{http://lilith.fisica.ufmg.br/~guerrero/cycle_global.html}.
In these movies, specially for the upper regions of
simulations with period 
larger than 28 days, we note that the LIC part seems to  evolve faster. This is due to the way 
by which  the LIC is generated from the convolution between the vector field and a random white 
noise background, which causes an artificial advance with respect to the rapid time scale of the 
magnetic flux emergence in the frames.}
\label{fig.top} 
\end{center}
\end{figure*}

Complementary to Fig.~\ref{fig.bd}, Figure~\ref{fig.top} depicts
the latitudinal distribution of the fields in the meridional plane ($r-\theta$).
It highlights the series of animations available as supplementary material.
On the left of each panel the line integral convolution (LIC) representation
depicts the distribution of the poloidal field lines 
with the color indicating the magnitude and direction of the mean latitudinal 
field, $\mean{B}_{\theta}$. On the right of each panel the colored contours show  
the distribution of the mean toroidal magnetic field, $\mean{B}_{\phi}$. They
make clear that a layer of strong toroidal field is formed at and below the 
tachocline, specially for models with $\prot \gtrsim 21$ days.

Figure \ref{fig.bd} and \ref{fig.top} evidence the complexity of large scale 
dynamos. Different dynamo modes can be excited depending on local conditions. 
Examples of this are the cases RC07-RC21, where the steady and oscillatory 
modes are mixed. In the time-latitude diagrams of models RC24-RC49 it can 
be seen that the field generated 
at the base of the convection zone is superposed with the field generated near
the surface forming irregular branches that end up mixing  the parity of the 
dynamo solutions. Because the simulated magnetic field is generated  by sources at 
different locations, in the next section we explore the magnetic 
field amplitude and dynamo coefficients by performing
volume averaging over three radial regions: the tachocline (TAC), from  
$0.63\Rs \le r < 0.74\Rs$, the bulk of the convection zone (CZ), from 
$0.74\Rs \le r < 0.89\Rs$, and the near-surface layer (NSL), from 
$0.89\Rs \le r < 0.96\Rs$.  In latitude, we 
separate the domain into polar (POL) and equatorial (EQU) regions. 
The averages over these regions are denoted by angle brackets, $\brac{}$.  

\subsection{Mean-field analysis:  magnetic field and dynamo coefficients}
\label{sec.mfa}

For the understanding of the dynamo solutions as a function
of the Rossby number we perform a systematic analysis of the 
simulations results in terms of the mean-field dynamo framework. 
Upon the condition that the results are axisymmetric, the magnetic field as well
as the velocity field may be decomposed in their large-scale and turbulent components.
This analysis leads to the governing equation for a mean-field $\alpha^2 \Omega$ dynamo 
\citep{Moffatt78}:
\begin{equation}
\frac{\partial \mean{\bm{B}}}{\partial t} = \nabla \times (\mean{\bm{u}} \times \mean{\bm{B}})
                                          + \nabla \times (\alpha \mean{\bm{B}}) 
                                          - \nabla \times (\eta \nabla \times \mean{\bm{B}}),
\label{eq.mfie}
\end{equation}
where $\mean{\bm {B}}=(\mean{B}_r, \mean{B}_{\theta},\mean{B}_{\phi})$ 
is the magnetic field averaged over longitude and 
$\mean{\bm {u}} =  (\mean{u}_r, \mean{u}_{\theta},\mean{u}_{\phi})$ is the
velocity field  averaged over longitude and time in an interval during the dynamo 
saturated phase (see Fig.~\ref{fig.dr}). 
In the second term on the right hand side (rhs), the $\alpha$ term
stems for the $\alpha$-effect which has kinetic and magnetic contributions,
namely $\alpha=\alpha_{\rm k}+\alpha_{\rm m}$.  These terms 
generate the large-scale magnetic 
field from small scale helical motions and currents, respectively. In the third 
term, $\eta = \eta_m + \etat$, is the 
sum of the molecular and the turbulent magnetic diffusivities.  

If the mean-velocity field is expressed as 
$\mean{\bm u} = r \sin\theta \Omega {\bm {\hat e}}_{\phi} + \mean{\bm u}_p$, with 
$\mean{\bm u}_p = (\mean{u}_r, \mean{u}_{\theta}, 0)$, and the mean magnetic field as 
$\mean{\bm B} = \mean{B}_{\phi} + \mean{\bm B}_p$, with 
$\mean{\bm B}_p = (\mean{B}_r, \mean{B}_{\theta}, 0)$, then
Eq.~(\ref{eq.mfie}) can be written as

\begin{equation}
\frac{\partial \mean{\bm{B}}}{\partial t} = [ r \sin\theta {\bm B}_p \cdot \nabla \Omega]
                                          + \nabla \times (\mean{{\bm u}_p} \times \mean{\bm{B}})
                                          + \nabla \times (\alpha \mean{\bm{B}}) 
                                          - \nabla \times (\eta \nabla \times \mean{\bm{B}}).
\label{eq.mfie2}
\end{equation}

Here, the first term on the rhs represents the rotational shear which 
generates the toroidal field from the poloidal one. The second term corresponds to 
the advective transport by  the meridional circulation, $\mean{{\bm u}}_p$.
The $\alpha$-effect, third term in the above equation, represented by a
pseudo scalar, $\alpha$, operates on both components of the field.  Rigorously, 
$\alpha$ is a second order tensor which acts as source of the toroidal and poloidal 
fields and also advects them \citep{Moffatt78}. The turbulent magnetic diffusivity, $\etat$, 
comes from a third order tensor, $\beta$, which also might have source terms \citep{BRRK08}.  
For our analysis we
estimate $\alpha$ and $\etat$ by using the first order 
smoothing approximation (FOSA) as detailed in Appendix~\ref{ap.B}. 
This approximation assumes isotropy such that both $\alpha$ and $\beta$ 
become scalars. 
We notice that this approximation is valid only for low magnetic Reynolds 
numbers\footnote{A complete
determination of the dynamo coefficients can be performed via the so-called test 
field method \citep{Warnecke+18}. }.
However, it has been shown that the coefficients profiles obtained
with FOSA are qualitatively compatible to those obtained by directly inverting
the electromotive force \citep{RCGS11}, and to the ones obtained by the test-field
method \citep{Warnecke+18}. In spite of possible differences with their actual values, 
the systematic use of the same technique for all the simulations provides a reliable 
picture of the change of the helicities with the Rossby number. 
Details of the computation of $\alpha_k$ and $\alpha_m$, as well as of the
turbulent diffusion coefficient are presented in Appendix \ref{ap.B}. Meridional profiles of
the kinetic, magnetic and total $\alpha$-effect for representative models between
RC07 and RC63 are shown in Fig.~\ref{fig.a3}(a)-(h).  The radial profile of $\etat$ 
for the same models is presented  in Fig.~\ref{fig.a2}(d). 

Writing Eq.~\ref{eq.mfie2} in a non-dimensional form, and using a characteristic 
dynamo time-scale, $\taud= \Rs^2/\eta_0$, where $\eta_0$ is a suitable 
value of the diffusivity coefficient, we can define non-dimensional dynamo coefficients 
which  compare the inductive, $C_{\alpha} = \alpha \taud / \Rs$ and 
$C_{\Omega} = \Delta \mean{\Omega} \taud$; and advective, 
$C_{u} = \mean{u}_{p} \taud / \Rs$, effects with diffusion.  Here, 
$\mean{u}_{p} = \sqrt{\mean{u}_r^2 + \mean{u}_{\theta}^2}$, is the amplitude of 
the meridional motions. Because both, the radial and latitudinal, derivatives of 
$\mean{\Omega}$  contribute 
to the generation of $\mean{B}_{\phi}$ (by stretching poloidal field lines 
in the azimuthal direction),  the parameter $C_{\Omega}$ has  
two components, $C_{\Omega_r} = \Rs \partial_r \mean{\Omega} \taud$ 
and $C_{\Omega_{\theta}} = \Rs (\partial_{\theta} \mean{\Omega}/r) \taud $.

\begin{figure*}
\begin{center}
\includegraphics[width=0.8\columnwidth]{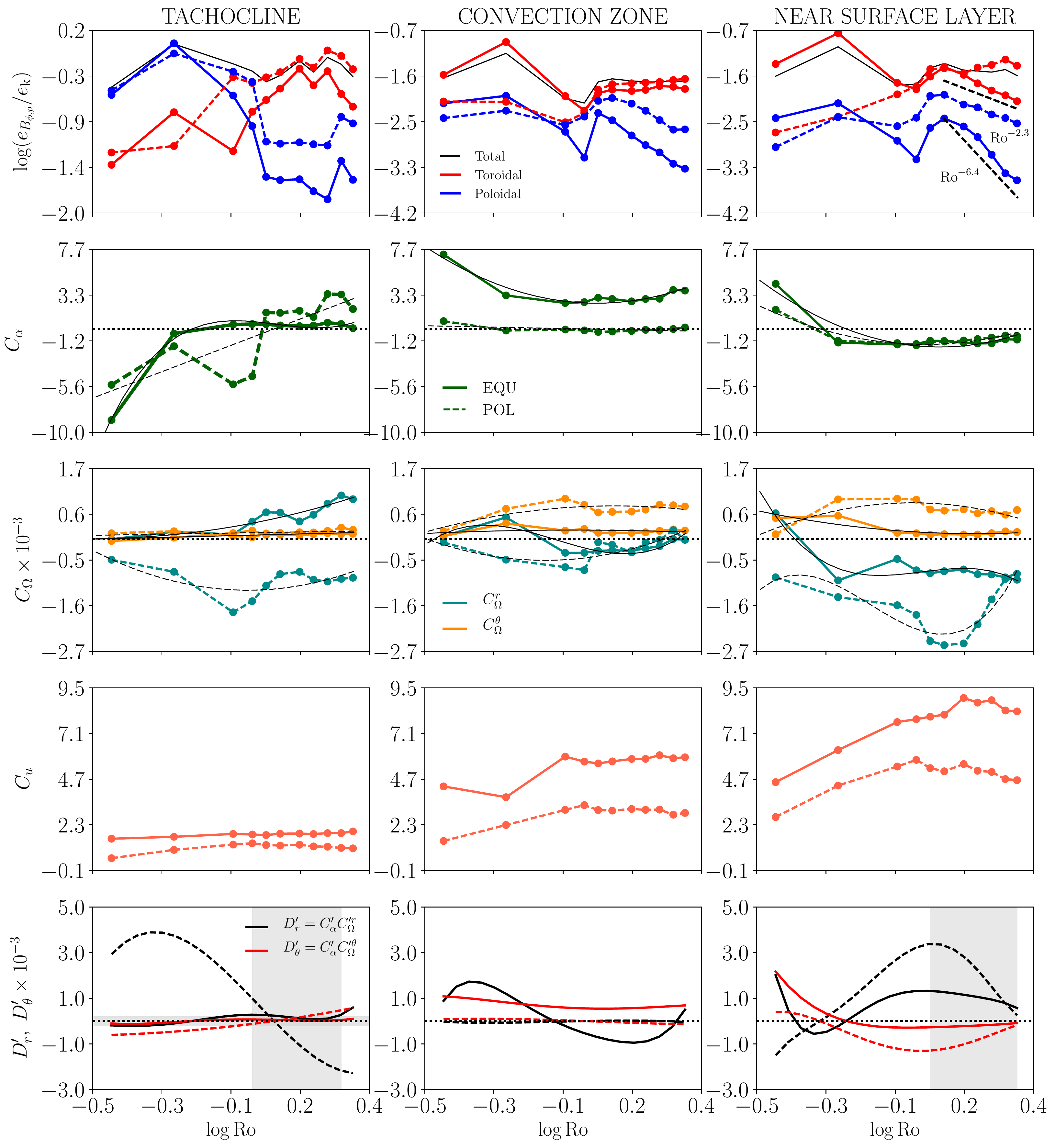}\\
\caption{Top panels: toroidal (blue) and poloidal (red) magnetic field 
energy densities as a function of $\Ro$. The left, middle and 
right columns show averages over the TAC, CZ and NSL regions, respectively. 
In all panels the continuous (dashed) lines depict the values at the equatorial
(polar) latitudes.  The magnetic energy densities are normalized to the 
kinetic energy density of each model averaged over the entire volume, $e_{k}$. 
The black lines in these upper row panels correspond to the magnetic energy density 
averaged over the entire northern hemisphere for the TAC, CZ and NSL regions.
The second, third and fourth rows correspond to the dynamo coefficients,
$C_{\alpha}$, $C_{\Omega}$ and $C_u$, respectively, as a function of $\Ro$.  
The thin continuous black lines in these 
panels show polynomial fits to these coefficients.  The bottom panel shows the dynamo 
numbers, $D_r^{\prime}$ (black) and  $D_{\theta}^{\prime}$ (red). The shades in the 
left panel exhibit the regions in the $\Ro-D_r^{\prime}$ space where the dynamos are 
oscillatory at the tachocline.  In the right panel the shades show the range of $\Ro$ for 
which the magnetic energy densities decay. Finally, the black dotted lines indicate 
the zero level of each quantity.}
\label{fig.dyn}
\end{center}
\end{figure*}

The energy density of the magnetic field components as well as the dynamo 
coefficients as a function of the Rossby number, $\Ro$, are presented in Fig.~\ref{fig.dyn}. 
The left, middle and right columns correspond to the TAC, CZ, and NSL regions, respectively. 
In each panel the continuous and dashed lines depict the volume averages over the 
polar (POL) and equatorial (EQU) regions and the corresponding 
radial extent.  The top row shows the poloidal 
($e_{B_{p}} = \mean{B}_{p}^2/2 \mu_0$, 
blue lines and symbols) and toroidal ($e_{B_{\phi}} = \mean{B}_{\phi}^2/2 \mu_0$, red) magnetic 
energy densities. They are normalized to the 
kinetic energy density, $e_k = \brac{\rho_s} \urms^2 /2$, where $\brac{\rho_s}$ is the mean
value of the isentropic density, and $\urms$ is the time-volume averaged rms velocity.
The second, third and fourth rows show, respectively, $C_{\alpha}$, $C_{\Omega}$ and $C_u$.
The bottom row depicts the radial and latitudinal dynamo numbers, 
$D_r^{\prime} = C_{\alpha}^{\prime} C_{\Omega}^{\prime r}$ and 
$D_{\theta}^{\prime} = C_{\alpha}^{\prime} C_{\Omega}^{\prime \theta}$. 
The primes in these definitions come from the fact that the dynamo numbers
are computed from polynomial fits to the dynamo coefficients. We describe next 
the main characteristics of these quantities from the TAC to the NSL 
regions.

\subsubsection{Tachocline region, TAC}
In the tachocline region (TAC), 
at both, polar and equatorial altitudes, the magnetic 
energy shows two separate branches. For  $\Ro \lesssim 1$, 
the poloidal field energy is larger than the toroidal one. Both 
trends cross at  $\Ro \sim 1$,
and the toroidal field energy becomes larger for $\Ro \ge 1$.  The total 
magnetic energy averaged
in the entire hemisphere (see the continuous black line) is,  thus, 
roughly constant for all the Rossby numbers.   It reaches roughly 50\% of the
kinetic energy. Note that most of the magnetic 
energy is in the tachocline magnetic fields, therefore, the normalized
magnetic energy density, averaged over the entire domain, is independent 
of $\Ro$. For $\Ro \lesssim 1$ the $\alpha$-effect is negative and its amplitude 
decreases as $\Ro$ increases. The meridional profiles in Fig.~\ref{fig.a3}(a)-(h) 
indicate that this term has magnetic origin. In the  EQU region, $C_{\alpha}$ 
reaches zero at $\Ro \sim 0.5$, while in the POL region it does this at $\Ro \sim 1$,
roughly at the same point where the toroidal and magnetic energy densities
have similar values. Note that in Fig.~\ref{fig.bd}(a)-(c), corresponding to the same 
range of $\Ro$, the radial field in the stable layer is positive. 
For $\Ro > 1$,  $C_{\alpha}$ increases in POL and is roughly null in EQU. 
Regarding $C_{\Omega}^r$, it is negative in POL and positive in EQU. 
It means that the rotation 
goes from faster to slower at polar latitudes and from slower to faster at 
equatorial latitudes. Such as it is observed in the Sun, in all simulations the 
largest shear occurs in the POL region. In both latitudinal zones the radial shear 
increases with $\Ro$.  The latitudinal shear, $C_{\Omega}^{\theta}$, is 
rather similar in both
latitudinal zones.  It is positive and also increases slowly with $\Ro$. 
The meridional flow coefficient, $C_u$,  has small values at all latitudes
and is roughly independent of with $\Ro$.

Although the trends are clear, the curves of the dynamo coefficients fluctuate.
We found useful and more clean to plot the dynamo numbers, a multiplication of 
$C_{\alpha}$ and $C_{\Omega}$, by using polynomial fits 
to these coefficients. The radial (black) and latitudinal (red) dynamo numbers are
presented in the last row of Fig.~\ref{fig.dyn}. In the TAC region, it is evident that
the most relevant dynamo number is $D_r^{\prime}$ at the polar latitudes 
(dashed black line).
It is positive for the fast rotating cases and decreases with the Rossby number 
until $\Ro\sim 1$ where it becomes negative. Its amplitude increases with $\Ro$ for 
the slow rotating cases. In the equatorial zone (continuous black line) the radial 
dynamo number is positive
(except for $\Ro \sim 0.5$). It has small values because of small $C_{\alpha}^{EQU}$ 
and is roughly independent of $\Ro$. The latitudinal dynamo number, $D_{\theta}^{\prime}$,
at POL is negative for the fast rotating cases and positive for the models with 
$\Ro \gtrsim 1$.  In the EQU region it is roughly null. The shaded region shows the interval
of Rossby numbers for oscillatory dynamos, $1 \lesssim \Ro \lesssim 1.7$ 
(cases RC21, which exhibits bimodality,  to RC49).

\begin{figure*}[ht]
\begin{center}
\includegraphics[width=0.3\columnwidth]{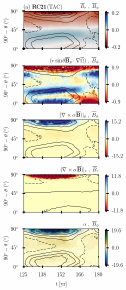}
\includegraphics[width=0.3\columnwidth]{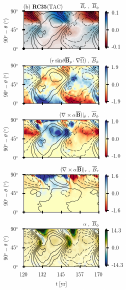}
\includegraphics[width=0.3\columnwidth]{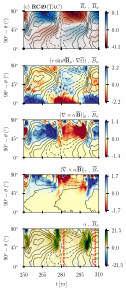}
\caption{Mean-field dynamo sources in the tachocline region (TAC) compared 
with the toroidal and radial 
mean magnetic fields for models (a) RC21, (b) RC35 and (c) RC49.
All the quantities are averaged in longitude and over the radial TAC extent.
From top to bottom the panels display the radial, $\bar{B}_r$ (colored contours) and toroidal,
$\bar{B}_{\phi}$ (contour lines) mean magnetic fields; the sources of the toroidal field, 
$r \sin \theta (\bar{B}_p \cdot \nabla) \bar{\Omega}$ and 
$(\nabla \times \alpha \bar{\bf B})|_{\phi}$ (colored contours), 
both compared with $\bar{B}_{\phi}$; 
the source of the radial field,  $(\nabla \times \alpha \bar{\bf B})|_{r}$ (color), 
compared with $\bar{B}_r$ (contour lines); and $\alpha = \alpha_k + \alpha_m$ 
(colored contours),  compared with $\bar{B}_{\phi}$. 
In the color map the dimensions are [T] for the magnetic field, 
$10^{-8}$ [T/s] for the source terms, and [m/s] for the $\alpha$-effect, respectively. 
The red vertical lines in the bottom panel at right indicate the time interval
described in Sec.~\ref{sec.period}.}
\label{fig.sources} 
\end{center}
\end{figure*}

In Fig.~\ref{fig.sources} we explore how the dynamo sources
contribute to the spatio-temporal evolution of the mean magnetic fields.
We compare the source terms in Eq.~(\ref{eq.mfie2}) with the dynamo generated mean magnetic 
fields in the simulations (a) RC21, (b) RC35 and (c) RC49. 
All the quantities are averaged in longitude and over the radial extent of TAC and 
only the northern hemisphere is presented for clarity.  
The upper panel shows a time-latitude butterfly diagram with $\bar{B}_r$ presented in
colored contours and $\bar{B}_{\phi}$ with solid and dashed contour lines as in 
Fig.~\ref{fig.bd}.
The second and third panels compare, respectively, the shear term,  
$r \sin \theta (\bar{B}_p \cdot \nabla) \bar{\Omega}$,  and the azimuthal component of 
the $\alpha$ source term, $(\nabla \times \alpha \bar{\bf B})|_{\phi}$,
with the toroidal field, $\bar{B}_{\phi}$. The fourth panel depicts the colored contours
of the radial component of the $\alpha$ source term, $(\nabla \times \alpha \bar{\bf B})|_{r}$,
and the contour lines of $\bar{B}_r$. The bottom panel shows the time-latitude 
evolution of $\bar{B}_{\phi}$ plotted as contour lines over the colored contours 
of $\alpha=\alpha_k+\alpha_m$.

From the second row of Fig.~\ref{fig.sources}
it can be noticed that positive (negative) values of the shear source terms
correlate well with positive (negative) values of the toroidal field at low latitudes 
in the three presented cases. The amplitude of the shear source increases with
the rotational period (from the left to the right panel). As a matter of fact, 
it is the amplitude of the equatorial shear what makes the dynamo of simulation
RC21 oscillatory.  In the simulations RC35 and RC49, the shear also correlates
with the toroidal field near the pole; above $85^{\circ}$ degrees. As it will
be discussed in Sec.~\ref{sec.period}, for these cases the reversal of the 
toroidal field starts at these latitudes. At intermediate latitudes the signs 
of the shear source and 
the toroidal field are opposite. The panels in the third row from the top
reveal that the $\alpha$  source term is the main responsible for the generation 
of the toroidal field, near the poles for the simulation RC21, and in latitudes 
between $30$ and $80$ degrees for RC35 and RC49. In the latter cases,
representative of the slow rotating simulations, its contribution increases 
in amplitude and spatial extent from the faster to the slower rotational rate. 
The fourth row of panels shows a clear correlation between the radial 
component of the $\alpha$ source term and the radial field. (We have verified
that in simulation RC21, and other simulations with faster rotation rate,
the $\alpha$ source term correlates better with the latitudinal field,
$\mean{B}_{\theta}$. However, since $\mean{B}_{\theta}$ changes sign within
the TAC region,  this correlation is observed by averaging 
over smaller radial extents.  For consistency we have decided to present
only the correlations between the $\alpha$ source term and $\mean{B}_r$.)
In the simulations RC35 and RC49, the quantity 
$(\nabla \times \alpha \bar{\bf B})|_{r}$ is
concentrated at the polar latitudes from where it migrates equatorward. 
It induces a change of polarity of the radial field which follows the 
same pattern of migration until $(\nabla \times \alpha \bar{\bf B})|_{r}$  
reverses sign. 

According to this analysis we can infer that in this region of the domain,
where the strongest magnetic field is generated, the mean-field coefficients
capture well the physics of the dynamo mechanism. Thus, we can
conclude that the dynamos operating in the simulations are of $\alpha^2\Omega$ type with 
the $\alpha$-effect generated in the stable layer.  The amplitude of this quantity is not 
constant in time but varies dynamically with the cycle evolution, as can be seen in the 
bottom panels of Fig.~\ref{fig.sources}. 

\subsubsection{Convection zone, CZ}
\label{sec.cz}

In the convection zone, CZ (middle column of Fig.~\ref{fig.dyn}), the toroidal
magnetic energy is larger than the poloidal one for all values of $\Ro$ and seems
to be independent of it.  On the other hand, for $\Ro \gtrsim 1.2$, the poloidal field
energy decays in all latitudes. The total magnetic energy (black dotted line) in this 
part of the domain is about 3\% of the kinetic energy. $C_{\alpha}$ is positive at 
EQU and consistent with zero at POL. In this case, $\alpha$  is mainly due to the kinetic 
helicity of the convective motions.

The radial shear, $C_{\Omega}^r$ is negative at POL with a minimum for $\Ro \sim 1$ 
(same as in TAC). At lower latitudes, it is positive (negative) for the fast (slow) 
rotating cases. The latitudinal shear, $C_{\Omega}^{\theta}$, is positive for 
all $\Ro$ with larger amplitudes in polar latitudes.  The meridional flow term, 
$C_{u}$, for $\Ro \gtrsim 1$,  has values comparable (or even larger) than 
$C_{\alpha}$, specially at equatorial latitudes.  Therefore, it is likely that 
this motion plays a significant role in transporting the magnetic field inside 
the convection  zone.  Finally, the dynamo numbers, $D_r^{\prime}$ and 
$D_{\theta}^{\prime}$ are roughly zero at POL due to the small values of $C_{\alpha}$. 
At EQU, $D_{\theta}^{\prime}$
follows the profile of $C_{\alpha}$, while  $D_r^{\prime}$ follows  
$C_{\Omega}^r$.

\begin{figure*}[ht]
\begin{center}
\includegraphics[width=1.01\columnwidth]{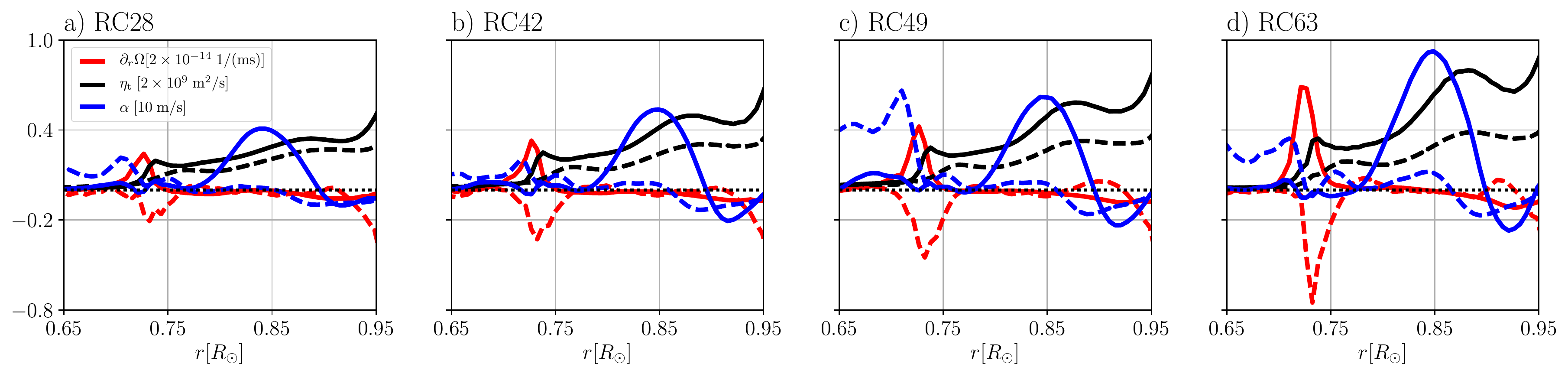}
\caption{Radial profiles of $\alpha$ (blue lines), $\partial_r \Omega$ (red lines) 
and $\etat$ (black lines) averaged over time and over POL (dashed lines) and EQU (solid
lines) for simulations (a) RC28, (b) RC42, (c) RC49 and (d) RC63.  The black dotted 
line depicts zero. For clarity, the profiles of $\alpha$,  $\partial_r \Omega$, and
$\etat$ from all the simulations were normalized to $10$ m/s, $2\times10^{14}$ 1/(ms) and
$2\times10^9$ m$^2$/s, respectively.}
\label{fig.coef} 
\end{center}
\end{figure*}

In contrast to the analysis of the TAC region, the interpretation of 
Fig.~\ref{fig.dyn} in the CZ region (middle top panel)  in 
terms of mean-field theory is less conclusive. For $\Ro \lesssim 1$ an analysis 
similar to that of Fig.~\ref{fig.sources} (not presented here for the sake of
brevity) indicates local dynamo action. However, for $\Ro \gtrsim 1$ 
no clear correlation between the source terms and the magnetic field is observed. 
For instance, Fig.~\ref{fig.dyn} shows that the poloidal field decays 
in the POL and EQU regions, while concomitantly its source term, $C_{\alpha}$, is either zero
or $\sim 3$, respectively. As this appears counterintuitive, we note that the local 
magnetic diffusivity  $\etat$, as specified in Eq. (\ref{eq.etat}), generally varies 
between the pole and the equator.  Because the coefficients in Fig.~\ref{fig.dyn} 
were computed using  $\tau_{\rm dyn} = \Rs/\eta_0$, where $\eta_0$ is the 
average of the $\etat$ over the entire CZ, this variability is not reflected 
in the dynamo coefficients.  For  substantiation, Fig.~\ref{fig.coef} displays 
normalized radial profiles of $\alpha$ 
(blue line),  $\partial_r \Omega$ (red line) and $\etat$ (black line) at POL 
(dashed line) and EQU (solid line) regions for models (a) RC28, (b) RC42, 
(c) RC49, and (d) RC63.  The figure shows orderly difference between the EQU and POL 
magnetic diffusivity increasing with $\Ro$, which, in principle, could be responsible 
for different decay rates of the poloidal field energy 
at POL and EQU. Moreover, the time evolution of the mean-magnetic field 
(see movies in the supplementary material) reveals non-local contributions to the 
local magnetic field, i.e., magnetic buoyancy. The energy transported by non-local
processes is not captured by Eq.~\ref{eq.mfie2}, therefore, it cannot  
be quantified by the dynamo coefficients. The intricacy of the magnetic fields 
in CZ can also undergo the advective action of the $\alpha$-effect, the so 
called turbulent pumping \citep{GDP08}, or the meridional circulation.

\subsubsection{Near-surface layer, NSL}
\label{sec.nsl}

The right column of Fig.~\ref{fig.dyn} presents the magnetic energy, dynamo
coefficients and numbers corresponding to the near-surface layer, NSL. 
The upper panel shows that while the toroidal field energy density at EQU 
increases with $\Ro$, despite the fluctuations, at POL it decreases. 
Since it is larger than the poloidal energy density, the total energy in the 
entire hemisphere is independent of $\Ro$ and about 5\% of the kinetic energy.
For $\Ro \lesssim 1$, the poloidal energy density,  $e_{B_p}/e_k$, has large fluctuations
at POL (specially odd is the case RC21), and increases slowly at EQU.  Starting from the 
case RC28, $\Ro \sim 1.2$, the poloidal energy shows a clear decay. It is fast 
at EQU, $e_{B_p}/e_k \propto \Ro^{-6.4}$ (see black dashed lines), and slow at POL, 
$e_{B_p}/e_k \propto \Ro^{-2.3}$ (a similar trend is observed 
in $e_{B_{\phi}}/e_k$ at POL). 

\begin{figure*}[ht]
\begin{center}
\includegraphics[width=0.3\columnwidth]{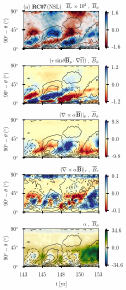}
\includegraphics[width=0.3\columnwidth]{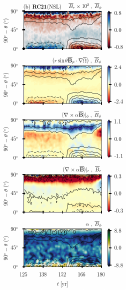}
\includegraphics[width=0.3\columnwidth]{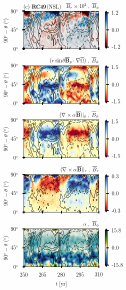}
\caption{Same as in Fig.~\ref{fig.sources} but for simulations
(a) RC07, (b) RC21 and (c) RC24. The average in this case is done over the NSL radial extent.}
\label{fig.sources_nsl} 
\end{center}
\end{figure*}

In the NSL The $\alpha$-effect coefficient, $C_{\alpha}$, is positive in both, POL 
and EQU latitudes for model RC07, and negative for all the other cases. 
The shear coefficient,  $C_{\Omega}^{\theta}$, is positive at both, POL and EQU, with 
amplitudes  similar to those in CZ. On the other hand the radial shear coefficient,
$C_{\Omega}^r$,  as in the solar near-surface shear layer, is negative at 
POL and EQU (except for case RC07 at EQU) . At POL
it reaches its minimum value for the case RC35, $\Ro \sim 1.25$, followed by a
step decrease for larger values of $\Ro$. 

The dynamo coefficients show trends that partially explain the 
behavior of the magnetic field energy density in this region. 
Nevertheless, similarly to the CZ region, non-local processes
seem to be relevant for defining the amplitude of the magnetic fields.

In Fig.~\ref{fig.sources_nsl} we performed the same analysis as in 
Fig.~\ref{fig.sources} but for models (a) RC07, (b) RC21,  and (c) RC49, and with
all the quantities averaged in longitude and over the radial extent of the 
NSL region. 
Together with Fig.~\ref{fig.dyn}, this figure illustrates the dynamo
behavior for different Rossby numbers at the outer layers.  Of particular 
interest  are the following points:

\begin{itemize}
\item
In the POL region the toroidal field energy increases with $\Ro$ (Fig.~\ref{fig.dyn}). 
This trend does not follow 
the scaling of $C_{\Omega}^r$, which at similar latitudes has a minimum at $\Ro \sim 1$
and decreases sharply for the largest $\Ro$.
We note that the scaling of $e_{B_{\phi}}/e_k$ with $\Ro$ in this region reproduces
the behavior of the toroidal energy at TAC, evidently with less energy. 
The second and third row panels, from top to bottom, in Fig.~\ref{fig.sources_nsl} 
show that for model RC07 (a), which is an example of simulations with 
$\Ro < 1$, there are no
dynamo sources at high latitudes. Correspondingly, the toroidal field is 
rather small.  For the simulation RC21 (b), where $\Ro \sim 1$, 
the local shear and $\alpha$-effect are responsible for the generation of a toroidal 
field. For simulation RC49 (c), representing $\Ro > 1$, 
the values of both, the shear and the $\alpha$ effect are significant, 
nevertheless the toroidal 
field shows poor correlation with these quantities.  The movies presented in the 
supplementary material clearly show that in simulations RC24-63, the toroidal field at 
TAC is transported from the tachocline to the NSL. 
\item
In the EQU region, for  $\Ro < 1$ the toroidal magnetic energy, $e_{B_{\phi}}/e_k$, seems 
to be independent from its values at TAC. The second and third row panels of 
Fig.~\ref{fig.sources_nsl} (a) 
suggest $C_{\alpha}$ and $C_{\Omega}^r$ contribute locally to the generation of 
the toroidal field.  The source terms seem to be out of phase with the magnetic 
field because the generation occurs slightly below in CZ (see also the panel (a) of 
Fig.~\ref{fig.dyn} and the movie corresponding to simulation RC07). 
Local toroidal field generation is also observed in simulation RC21 where the field 
correlates well with the shear term (second panel from top to bottom of 
Fig.~\ref{fig.sources_nsl} (b)). 
In model RC49 (representative of $\Ro > 1$) 
there is only a marginal correlation between the toroidal field and its local source terms. 
Thus, the decay of the toroidal energy for $\Ro \gtrsim 1.2$ is likely a consequence of 
the decaying toroidal field at TAC and the enhanced magnetic diffusivity at lower 
latitudes. 
\item
Unlike $\mean{B}_{\phi}$ which is generated by the shear and the $\alpha$-effect, 
the poloidal magnetic field is solely generated by $\alpha$, specifically by the term 
$\nabla \times \alpha \mean{B}_{\phi}$. Since the poloidal source term, $C_{\alpha}$, is
roughly constant with $\Ro$ in the NSL region,  one should expect that the poloidal
magnetic energy roughly follows the trend of the toroidal energy.  This proportionality
is observed at the POL region for $\Ro \lesssim 1$. However, for $\Ro > 1$ while  
$e_{B_{\phi}}/e_k$ increases  $e_{B_{p}}/e_k$ decays. Relevant hints to understand
this behavior can be found by watching the supplementary movies for simulations
RC24-RC49. We find that when the toroidal field (on the left side quadrants) quickly 
rises from TAC to NSL, the existent poloidal flux is rapidly redistributed in the 
bulk of the convection zone, and a new poloidal field of opposite polarity is generated. 
This is a clear evidence of the non-local effects present in the simulations.  
Although it is not easy to make a quantitative analysis, we suggest that the 
decay of $e_{B_{p}}/e_k$ with $\Ro$ is due to the fact that the total 
magnetic energy is independent of the rotation rate. Thus,
the more toroidal field is rapidly deposited into the upper layers of the domain, 
the less poloidal field might reside there.
\item
At the EQU region, $e_{B_{p}}/e_k$ roughly follows $e_{B_{\phi}}/e_k$ for all values of 
$\Ro$. Note in Fig.~\ref{fig.sources_nsl} (fourth row panels from the top) how 
the field is locally generated for simulation RC07 (a), has a minimum for simulation 
RC21 (b) and shows diffusive values in model RC49 (c), in agreement to what is observed in 
Fig.~\ref{fig.dyn}.  In simulations RC56 and RC63 the toroidal field at TAC is steady 
at polar latitudes, therefore a weak poloidal field develops at EQU.
\end{itemize}

\subsection{Comparison with the observations}

The top boundary of our model is placed at $r_t = 0.96$ of the stellar radius.
However, this does not preclude the relevance of simulated NSL properties to observations.
Unfortunately it is not yet clear how magnetic fields erupt to the surface to form star spots 
and how this emergence process depends on fluid properties such as rotation and convective 
motions. For instance, it is not clear what is the correspondence between the $\rhk$ flux 
and the magnetic field in stellar interiors. Also, there is not a complete interpretation 
of the magnetic fields inferred by the ZDI technique  and a few shortcomings 
of this method have been recently identified \citep{LHJMV19}.

To stablish some connection between the results presented above and the observations, 
we remind that the results of \cite{Vidotto+14} correspond to solar-like stars
with Rossby numbers spanning from ~0.3 to ~3.  All the stars in this sample follow the 
relation $\bar{B}  \propto  \Ro^{-1.38}$.  In addition, \cite{See+2015} reported a power law 
relation between the toroidal and poloidal magnetic energies 
with $\bar{B}_{\phi}^2 \propto (\bar{B}_{p}^2)^{1.25}$,
for stars with masses between $0.5$ and $1.5$ solar masses in the same range of $\Ro$. 

\begin{figure*}[ht]
\begin{center}
\includegraphics[width=0.96\columnwidth]{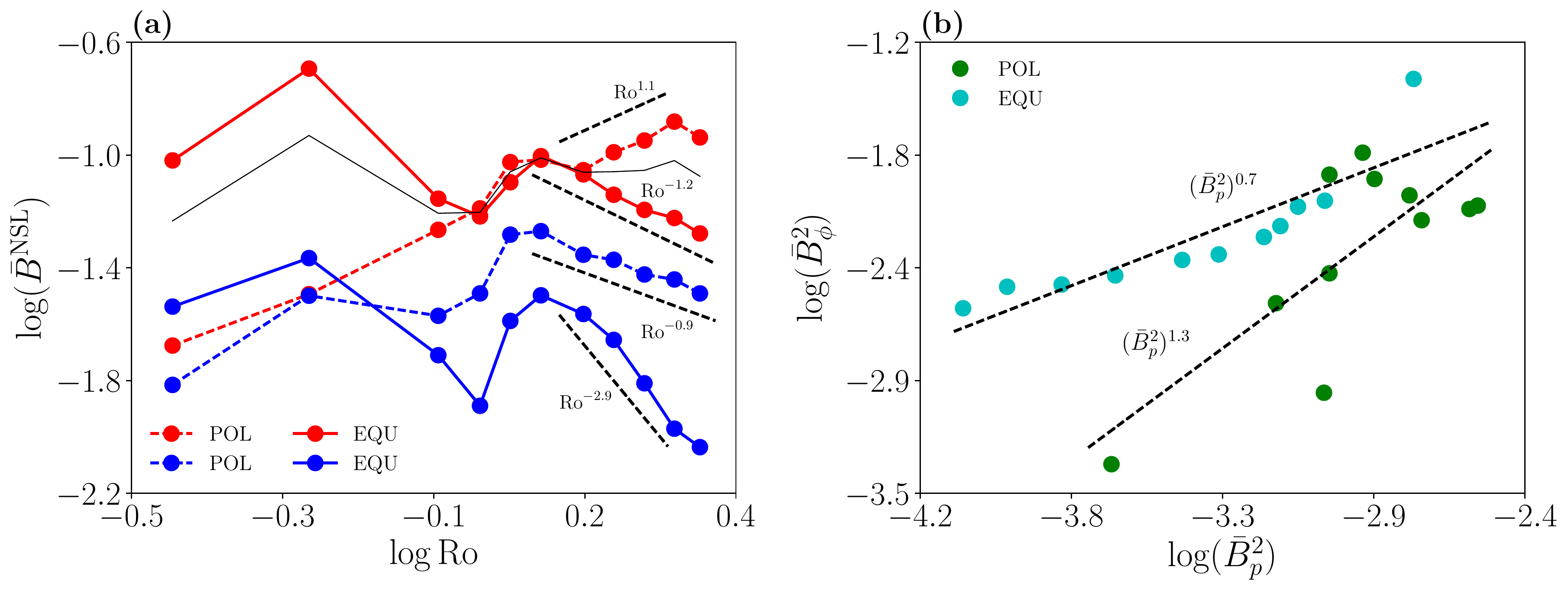}
\caption{(a) Scaling of the NSL magnetic field components with $\Ro$, the conventions
are the same as in Fig. \ref{fig.dyn}.  (b) correlation
between the toroidal and the poloidal magnetic energies at the NSL. The light blue (green) 
dots correspond to the EQU (POL) region. The dashed lines show the power law.}
\label{fig.bfs} 
\end{center}
\end{figure*}

As depicted in Fig.~\ref{fig.bfs}(a) our simulations correspond to $0.36 < \Ro < 2.12$ 
($0.26 < \Ro < 3.03$ according to the definition of \cite{Noyes+84b}).  Within this 
interval we found different behaviors at the EQU and POL regions and, unlike the 
observations, two different scalings with $\Ro$.   
In the EQU region, for $\Ro \lesssim 1.2$, both $\mean{B}_{\phi}$ and 
$\mean{B}_p$ seem independent of $\Ro$. For $\Ro \gtrsim 1.2$, our simulations 
predict $\mean{B}_{\phi}^{NSL} \propto \Ro^{-1.2}$
and $\mean{B}_p^{NSL}  \propto  \Ro^{-2.9}$.  In the POL region, the toroidal field
increases with the Rossby number as $\mean{B}_{\phi}^{NSL} \propto \Ro^{1.1}$. Yet,
the poloidal field increases similarly to $\mean{B}_{\phi}$ for 
$\Ro \lesssim 1.2$, and then decreases as $\bar{B}_p^{NSL}  \propto  \Ro^{-0.9}$ for 
large values of $\Ro$.  

As discussed in the previous section, there is not a straightforward interpretation 
of these scaling relations, however, we can summarize our findings as follows. 
For $\Ro \lesssim 1$ there 
is local dynamo action which occurs mostly at the EQU region. On the other
hand, for $\Ro \gtrsim 1$, there is a rapid transport of 
$\mean{B}_{\phi}$ from the 
bottom to the top of the convection zone. These effects, however, are more
pronounced and effectively change the scaling relations for $\Ro \gtrsim 1.2$.  
Thus, the toroidal field at surface levels
scales with $\Ro$ in the same way as it does at TAC, i.e., it increases and
is located closer and closer to the poles with increasing $\Ro$. Consequently, the toroidal 
field at the equator diminishes as $\Ro$ increases. Furthermore, large fractions of poloidal 
magnetic flux are quickly removed from the NSL and redistributed in the convection zone. 
We believe that the faster decay of the poloidal field at EQU compared to POL is due to 
the latitudinal variation of the turbulent magnetic diffusivity. Also, it is worth noticing 
that the contribution of other transport mechanisms like meridional circulation or turbulent 
pumping cannot be ruled out.

The correlation between the toroidal and the poloidal magnetic field energies is 
presented in Fig.~\ref{fig.bfs}(b). The light blue and green points correspond to the 
EQU  and POL regions, respectively.  It can be seen that the poloidal field energy is an 
increasing power law function of the toroidal energy with coefficients $0.7$ at EQU 
and $1.3$ at POL. However, the correlation 
is better defined at the equator than at the poles confirming that at lower latitudes  
the field strength depends mostly on local dynamo contributions.  At the pole, the 
relation shows more dispersion, specially for stronger magnetic fields.
This is understandable if non-local sources are contributing to the toroidal field but not to
the poloidal one,  which, moreover, is expelled from the places where the toroidal flux increases.  

\subsection{The dynamo period}
\label{sec.period}

The magnetic cycles in Fig.~\ref{fig.bd} are clear and well defined for most 
of the models. We 
compute the period by using a Fourier transform of the data. The technical
details are presented in Appendix~\ref{ap.C}.
In Fig.~\ref{fig.per} we show the (a) $\pcyc$ vs $\prot$, and (b)  
log($\prot/\pcyc$) vs. log($1/\Ro$) representations of  the
dynamo cycle period for the simulations ~RC07~-~RC49
(black solid stars).
In both panels we have plotted the observational data as reported by \cite{BMM17}. 
The blue and red 
stars correspond to the active and inactive branches, respectively.  The
11-yr cycle of the Sun is represented with a yellow star. 
In panel (a) we notice that, similar to the observations, the magnetic cycle 
period increases with the period of rotation.  If the simulation RC21 
(i.e., the third black star from left to right in panel (a)) is 
discarded, the trend appears linear and nearly separates the active (blue)
from the inactive (red) branch.

\begin{figure*}[ht]
\begin{center}
\includegraphics[width=0.46\columnwidth]{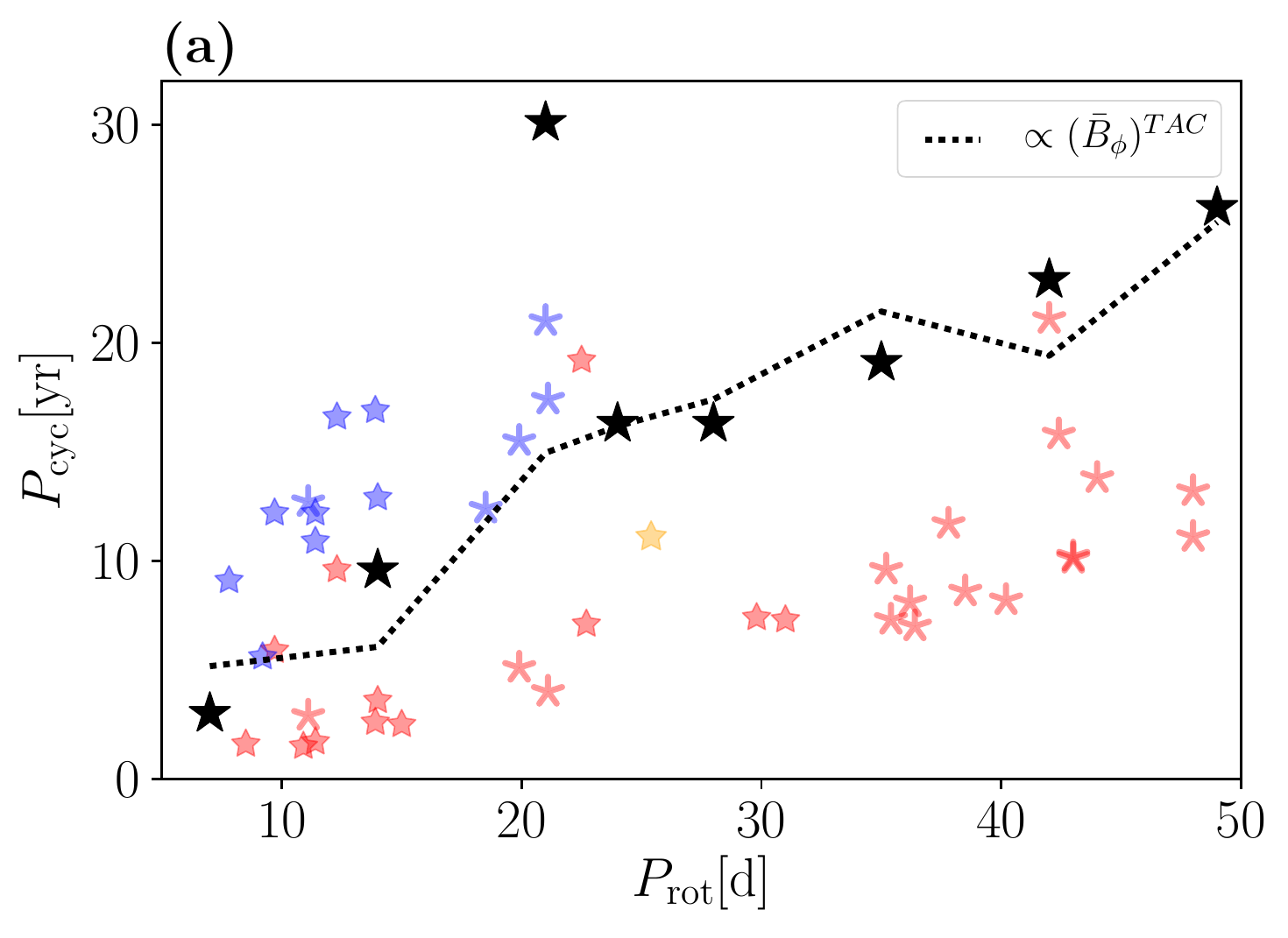}\hspace{0.4cm}
\includegraphics[width=0.46\columnwidth]{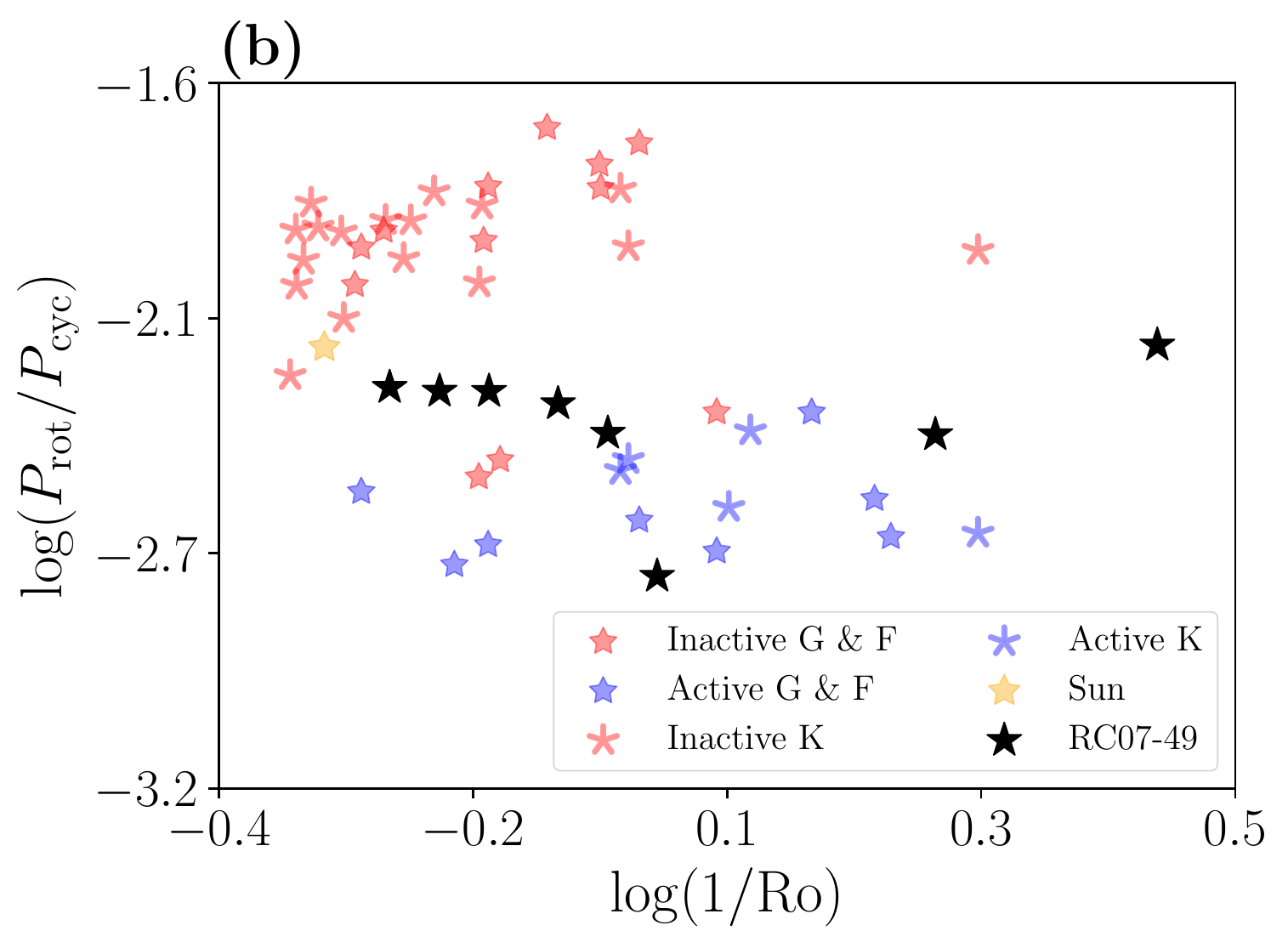}\\
\caption{(a) $\pcyc$ vs $\prot$, and (b) log($\prot/\pcyc$) vs. log($1/\Ro$)
representations of the magnetic cycle period against the rotation period.
The simulations results are shown with black solid stars. 
The blue and red stars symbols
correspond, respectively, to active and inactive branches of activity with data taken 
from \cite{BMM17}. Filled stars correspond to F and G stars, * symbols correspond 
to K type stars. The yellow filled stars show the 2 and 11 years solar
activity cycles. In the lhs panel, the green, orange and black dashed lines 
compare the rotation period with 
$(D_r^{\prime})^{-1/2}$, $\tau_{\rm dyn}$ and $B_{\phi}^{TAC}$, respectively.
These quantities are normalized to the cycle period of model RC28.}
\label{fig.per} 
\end{center}
\end{figure*}

In the right panel, Fig.~\ref{fig.per} (b), all the values of $\prot/\pcyc$  
fall closer to the active branch.  However two different trends can be identified, 
one with positive inclination for the cases RC07-RC21  (see black stars for 
$\log(1/\Ro) \gtrsim 0$),  
and another one, with a negative inclination, for cases RC21-RC49
($\log(1/\Ro) \lesssim -0.1$).   The simulation RC18, which shows no signals of periodicity 
in the convection zone, indicates that the simulation RC21 actually is close to
the transition between bifurcated regimes, one with the oscillatory
dynamo operating mainly in the convection zone, and the other with the tachocline having a 
predominant role.  We recall that in model RC21 a relevant oscillatory toroidal field 
developed at the tachocline coexists with a steady poloidal field in the stable layer.  
Thus, the model RC21 effectively belongs to the branch with negative inclination together 
with models RC24-RC49. 

In \cite{GSDKM16a} we discussed the instabilities that may occur in the
tachoclines and radiative zones.  We showed that the so-called magneto-shear 
instability is the most likely source of the magnetic $\alpha$-effect. 
This instability belongs to the Tayler instability family modified by 
the presence of shear. The Tayler instability is related to the decay of a 
large-scale toroidal field in a stable stratified layer \citep{Tayler73}. 
In the non-linear 
phase it results in a saturated state with non-zero helicity \citep{BU12}, 
which, in turn, might develop large-scale magnetic fields. The growth 
rate of this instability is inversely related to the ratio 
$\Omega_0/B_{\phi}$ \citep{BU13}. 
This means that it is inhibited by fast rotation or enhanced by strong toroidal 
magnetic fields. On the other hand, the shear contributes in this complex process by 
replenishing the toroidal field \citep{Miesch07b,SA13}.  However, its relevance 
still needs to be quantified.

The analysis of dynamo results presented in \cite{LSC15,GSDKM16a} suggests
that there is an exchange of energy between the development of a large-scale 
toroidal field and turbulent motions and currents, i.e., the interaction
between the dynamo and magneto-shear. The time-scale of 
this exchange can set the activity cycle period at least in dynamos where 
reversals of both magnetic field components take place in the stable layer 
(RC24~-~RC49). This process also allows for the existence of the bimodality observed 
in RC21 as reported by \cite{Rogers18} for axisymmetric dynamo simulations.  

\begin{figure*}[ht]
\begin{center}
\includegraphics[width=1.01\columnwidth]{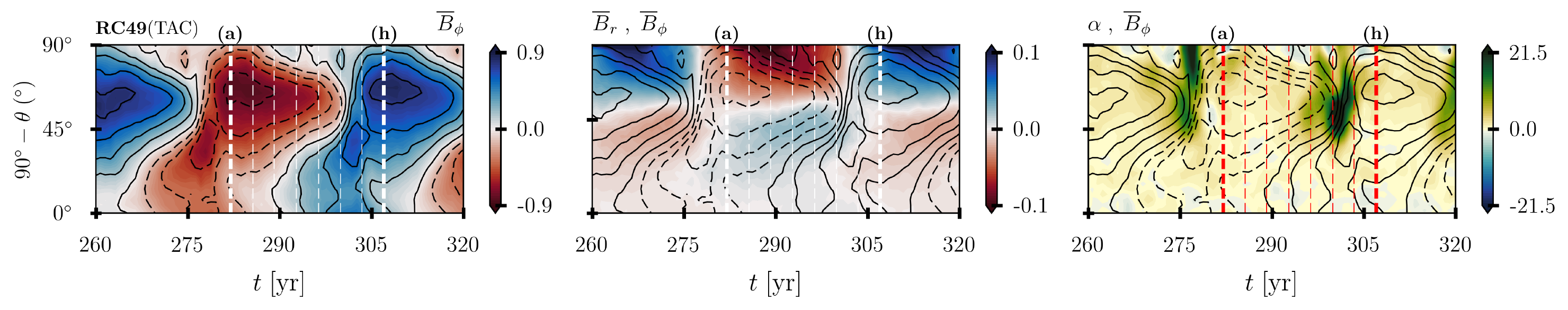}\\\vspace{0.2cm}
\includegraphics[width=0.24\columnwidth]{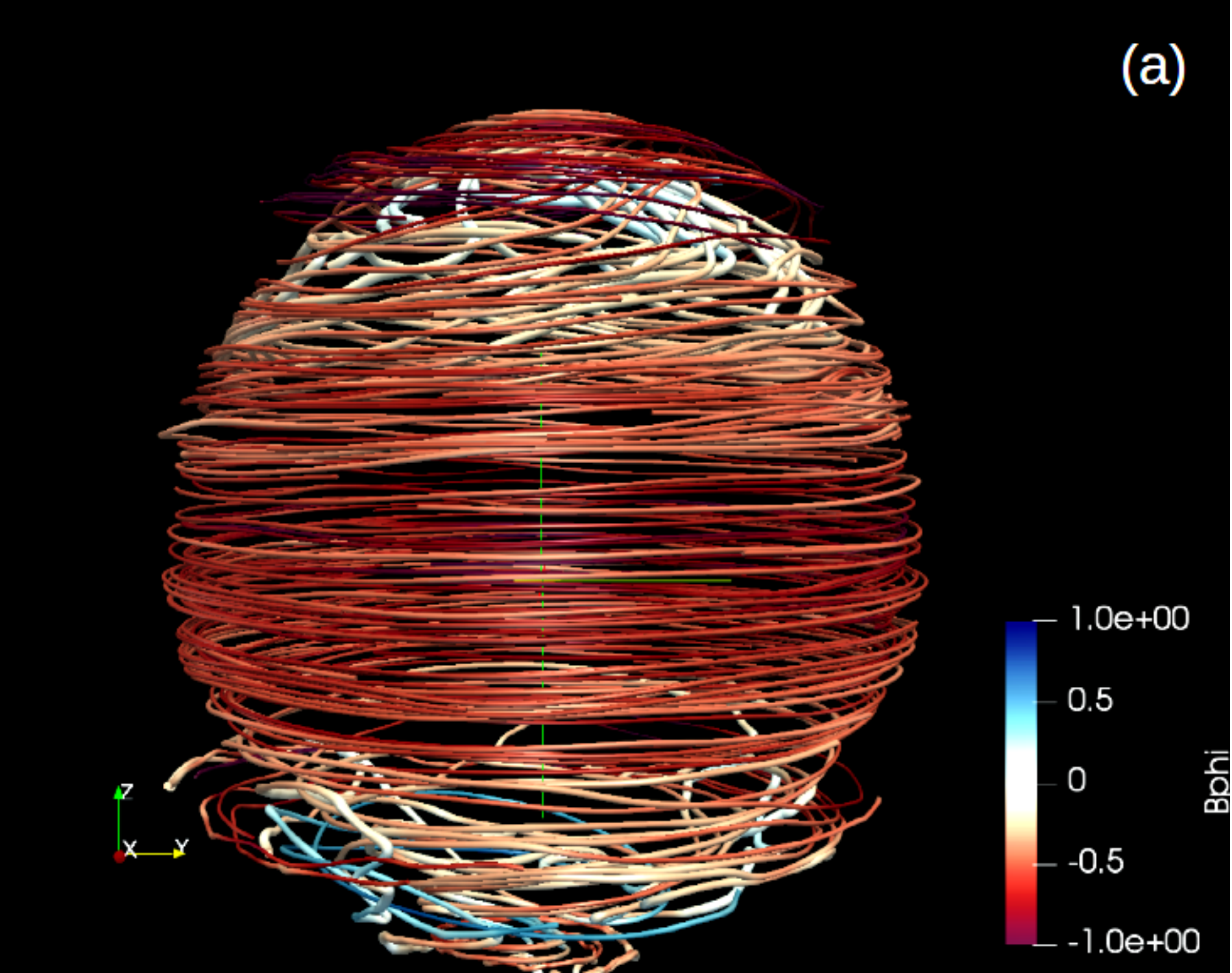}\hspace{0.1cm}
\includegraphics[width=0.24\columnwidth]{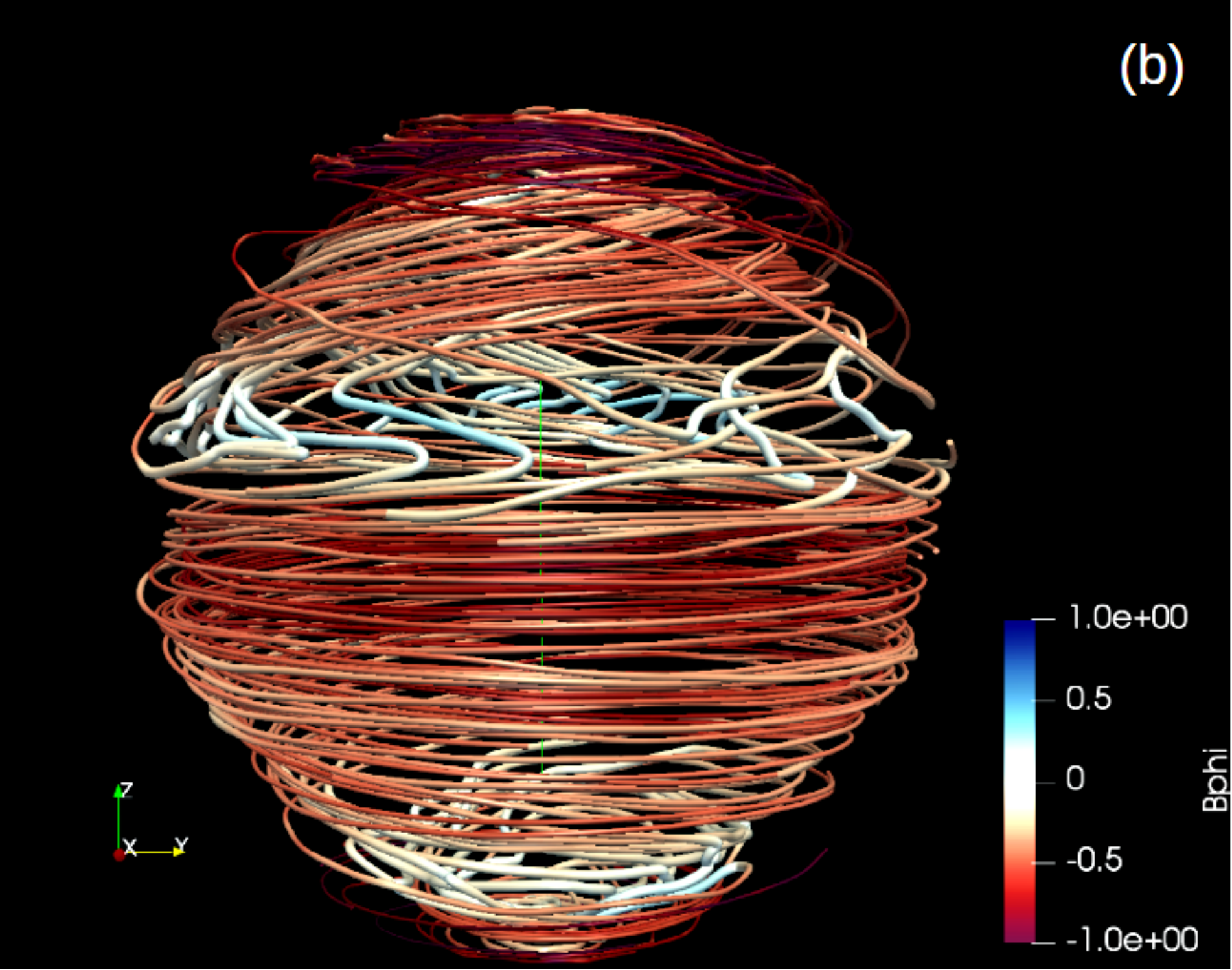}\hspace{0.1cm}
\includegraphics[width=0.24\columnwidth]{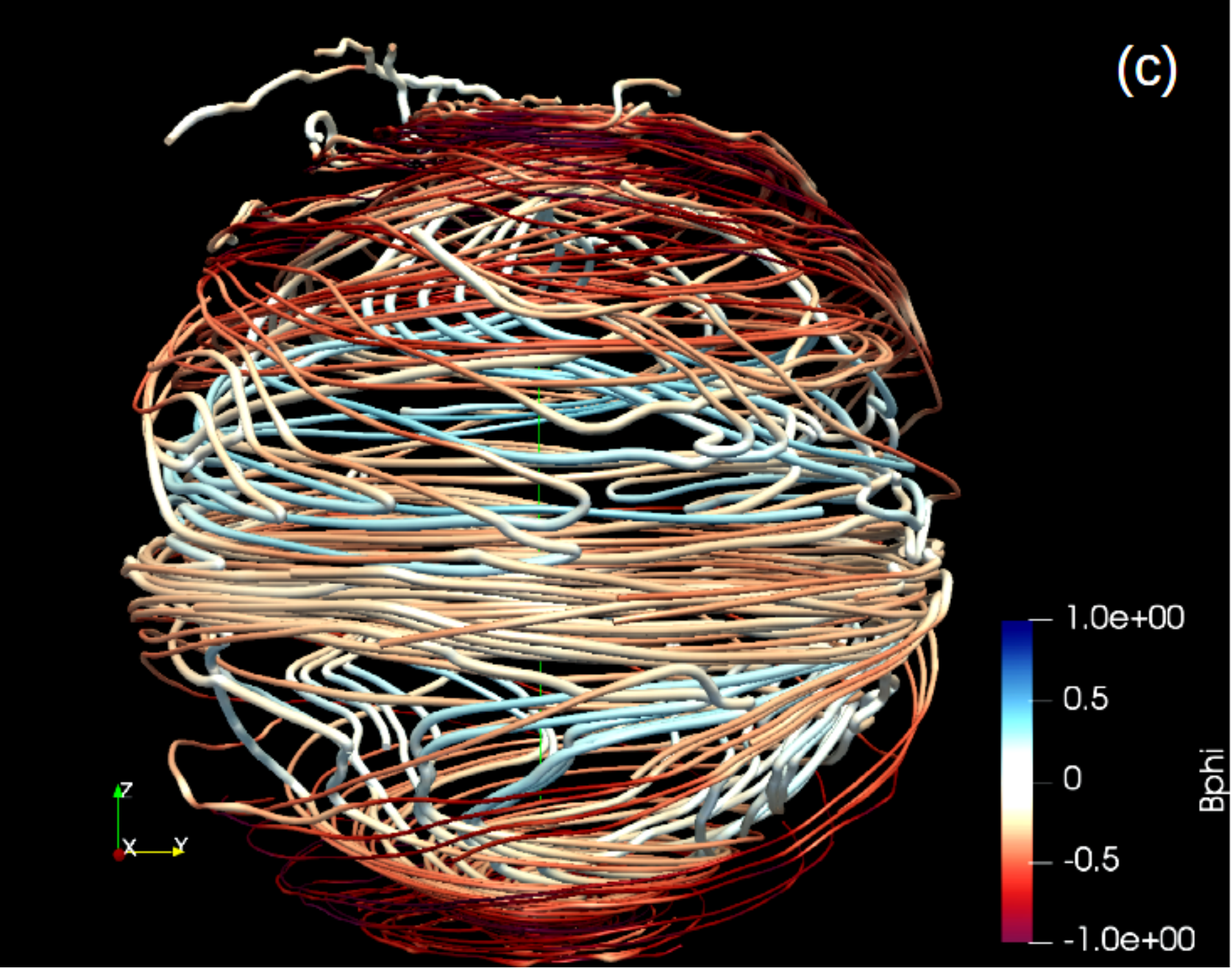}\hspace{0.1cm}
\includegraphics[width=0.24\columnwidth]{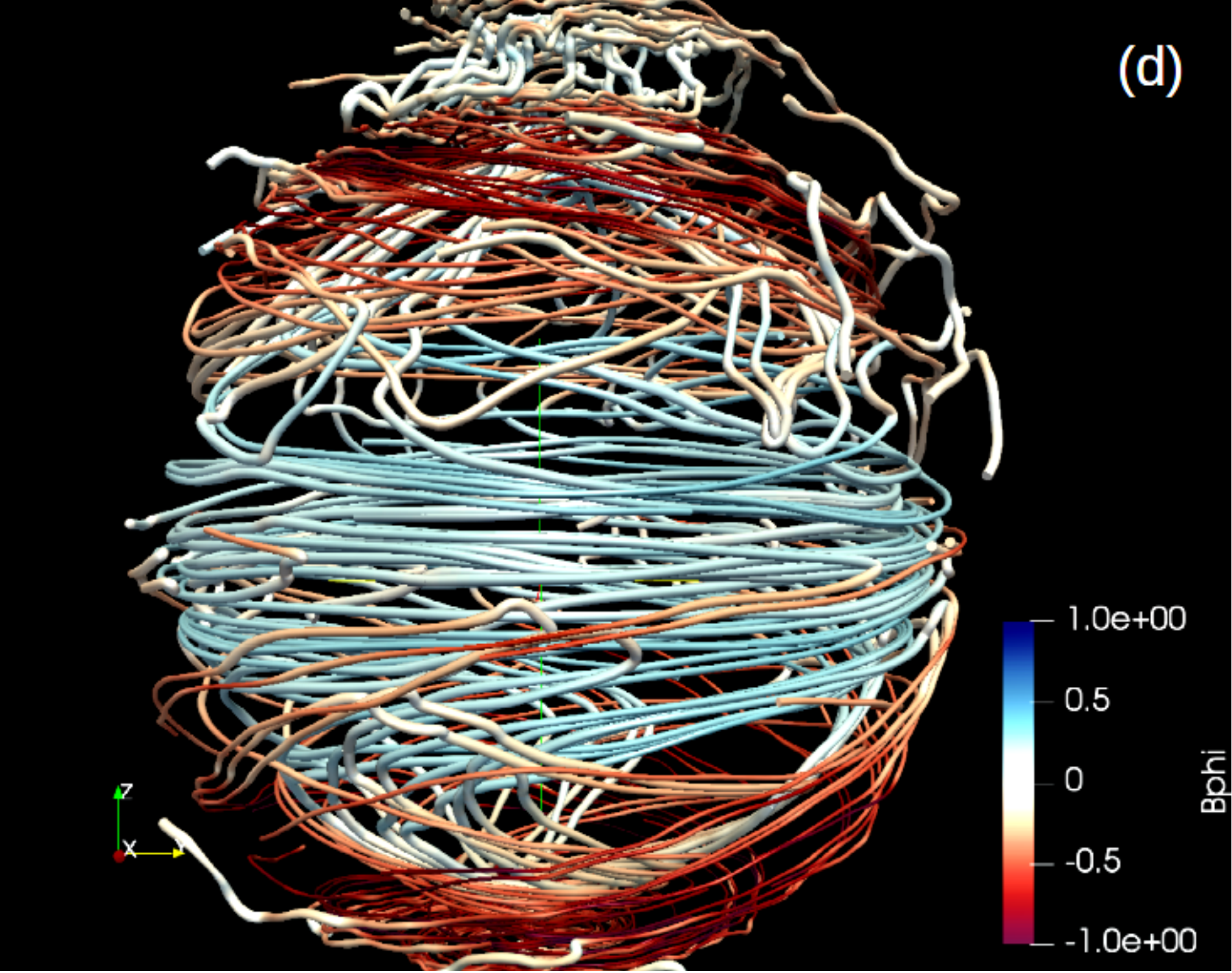}\vspace{0.2cm}
\includegraphics[width=0.24\columnwidth]{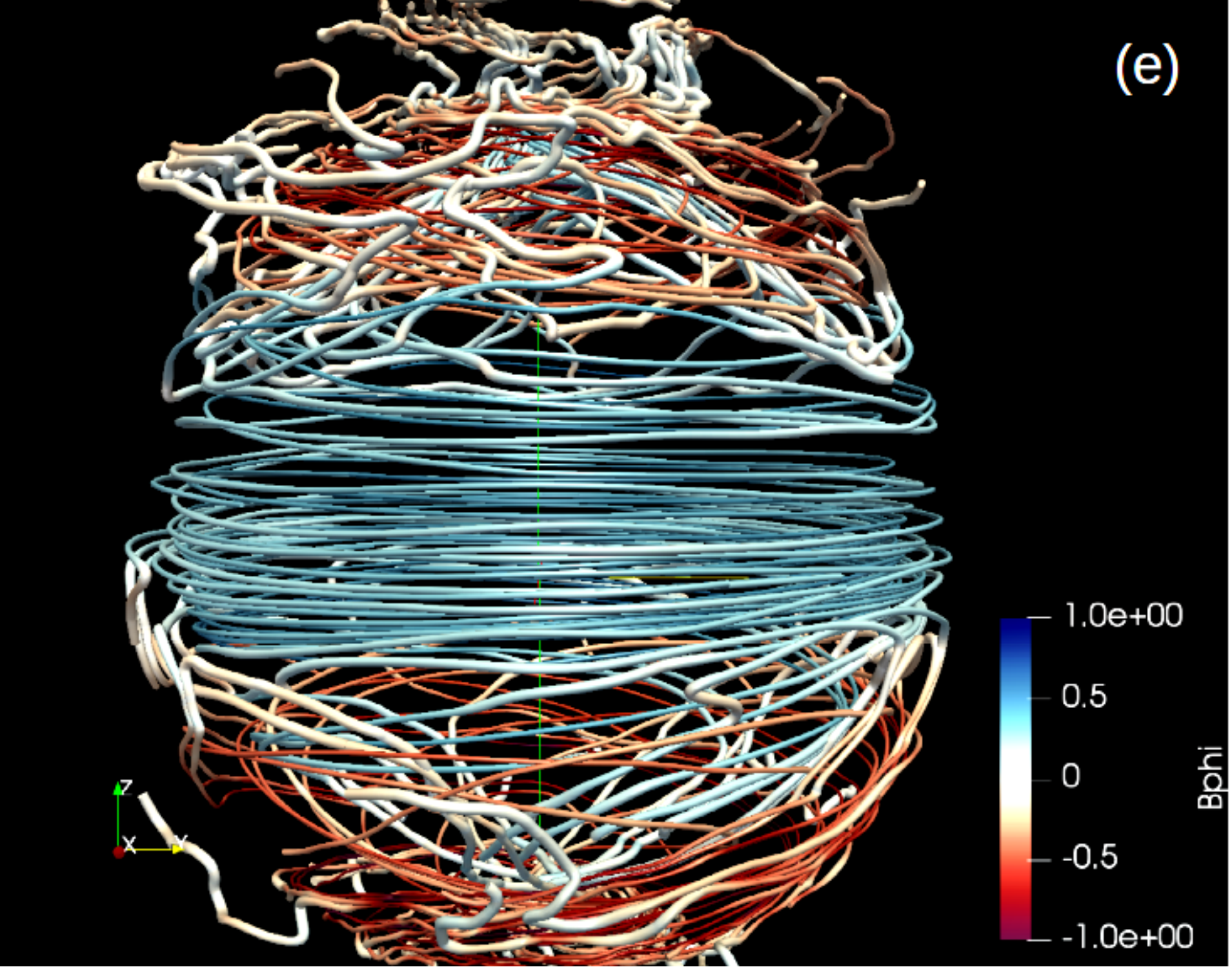}\hspace{0.1cm}
\includegraphics[width=0.24\columnwidth]{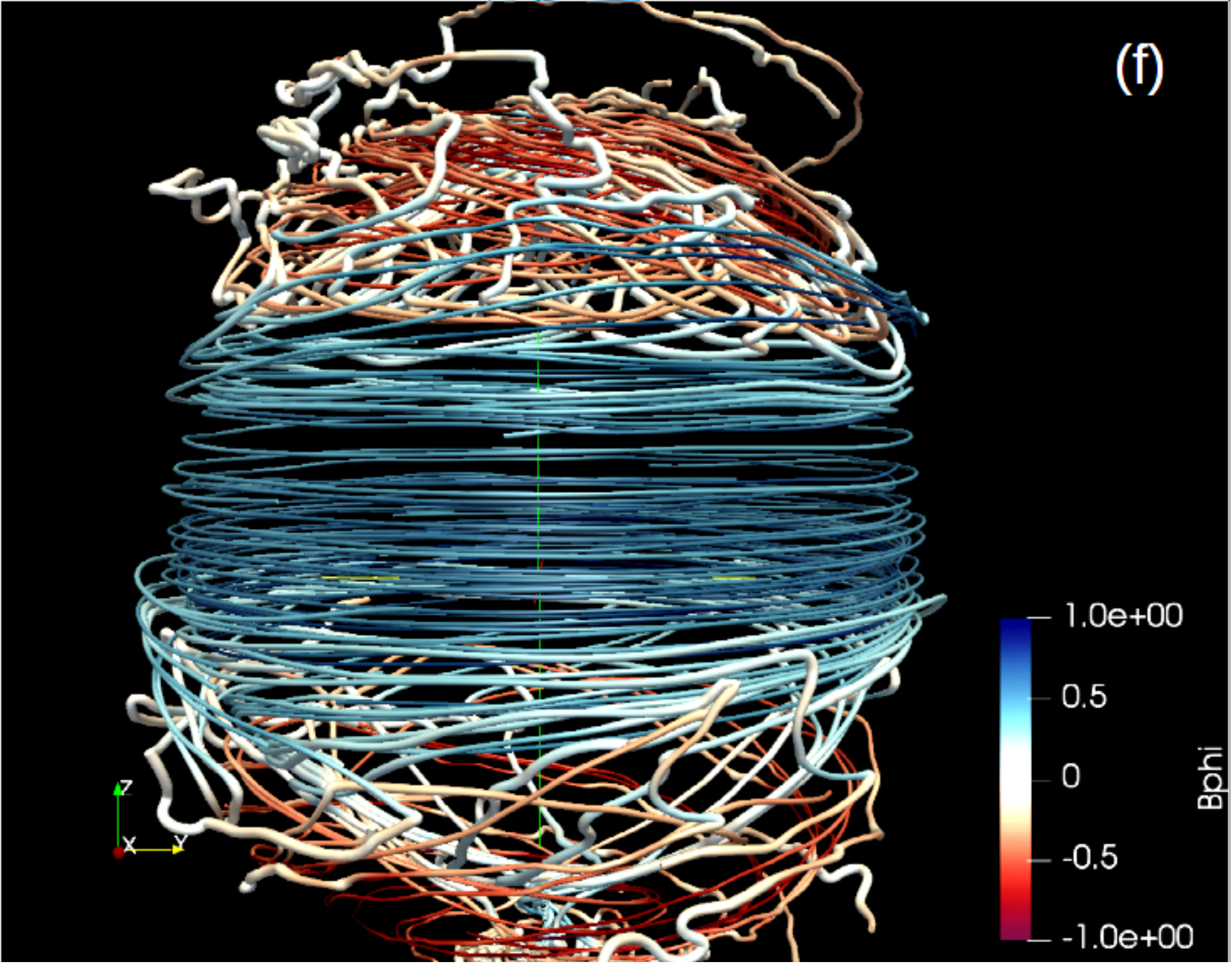}\hspace{0.1cm}
\includegraphics[width=0.24\columnwidth]{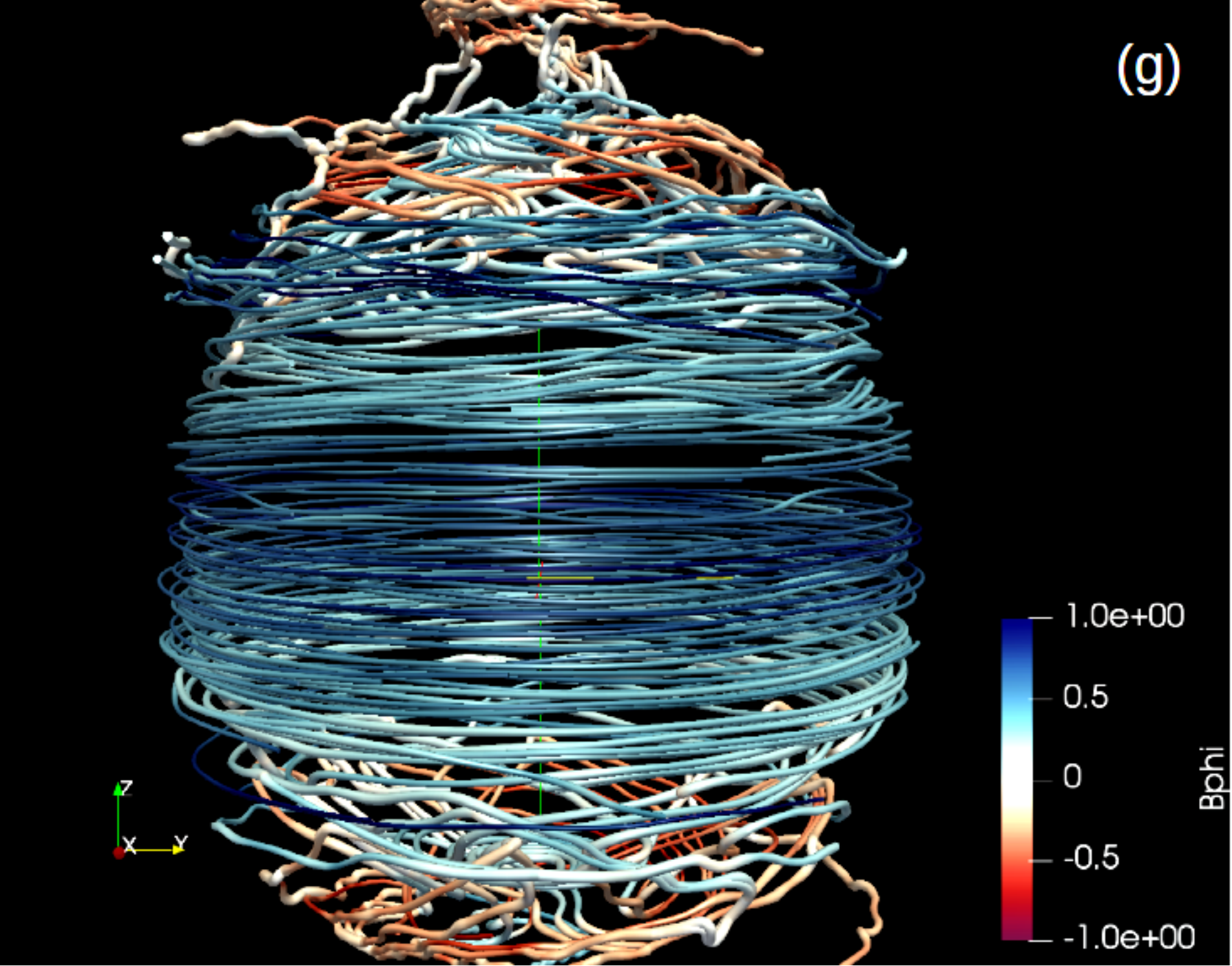}\hspace{0.1cm}
\includegraphics[width=0.24\columnwidth]{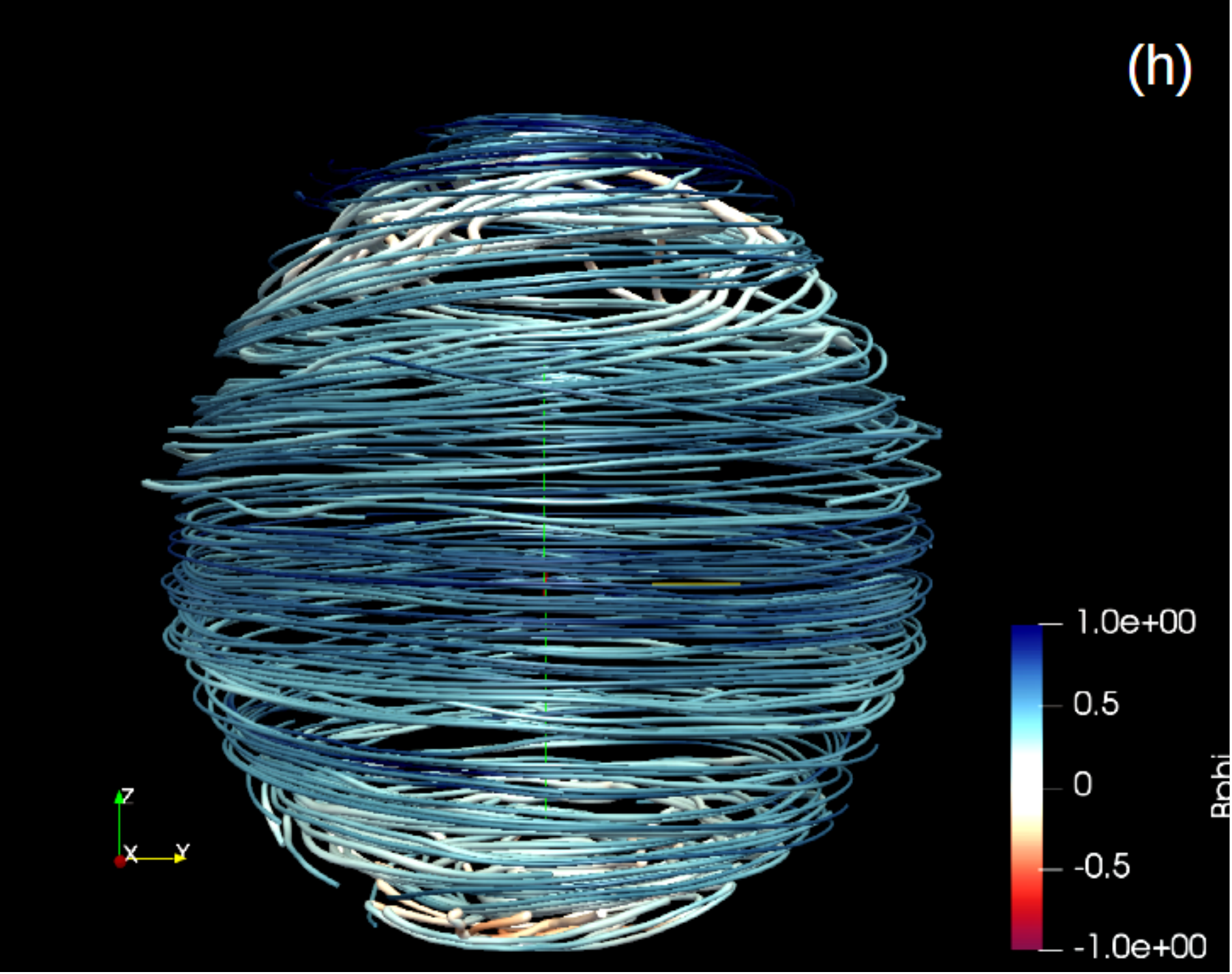}
\end{center}
\caption{Snapshots of the magnetic field lines around $r = 0.66\Rs$ covering one polarity
reversal (half cycle period) of model RC49.  The blue (red) colors correspond to
toroidal field pointing eastward (westward), the thickness of the lines is proportional
to the magnitude of the magnetic field. 
The upper panel of the  figure repeats butterfly diagrams after
Fig.~\ref{fig.sources}(c) presenting the evolution of $\mean{B}_{\phi}$, 
$\mean{B}_{r}$ and $\alpha$ (from left to right) in the TAC region. 
The dashed vertical lines show to the time of corresponding snapshots.
Animations of the field lines evolution of simulations RC28
and RC49 are available as supplement material or in the link
\url{http://lilith.fisica.ufmg.br/~guerrero/cycle_global.html}.}
\label{fig.flines} 
\end{figure*}

The sequence of images in Fig.~\ref{fig.flines}
presents in more detail the reversal process for simulation RC49. 
The figure shows snapshots of the 
magnetic field lines around $r=0.66\Rs$ during half cycle period, the colors 
represent the direction of the toroidal field and the thickness of the lines the 
magnitude of the magnetic field. For the reader convenience, 
the upper panel of the  figure repeats butterfly diagrams after 
Fig.~\ref{fig.sources}(c) presenting the evolution of $\mean{B}_{\phi}$,  
$\mean{B}_{r}$ and $\alpha$ in the TAC region.  The vertical dashed lines 
indicate the time of the corresponding snapshots.

The first vertical dashed line in the bottom-right panel of Fig.~\ref{fig.sources} 
corresponds to a time where a negative toroidal field covers almost the entire 
northern hemisphere. At this moment the radial field is negative and in growing phase. 
Panel (a) of 
Fig.~\ref{fig.flines} corresponds 
to this stage. Few years after,  Fig.~\ref{fig.sources} shows that 
while a new positive toroidal field is originated at polar latitudes,  
a negative $\alpha$ develops at the same locations.
This can be seen in panels (a) and (b) of Fig.~\ref{fig.flines}
where the magnetic field, mostly toroidal and negative, has a
positive inclination near the pole with respect to the equator.  This tilt, characteristic
of the so-called clamshell instability \citep{Cally_03},  generates poloidal field components 
that are winded up by the differential rotation at the poles first 
and then at the equator, where the shear is stronger;  see panels (a)-(d) of
Fig.~\ref{fig.flines}. This generates a new positive toroidal field which 
migrates polewards. In panel (d) the radial field reaches its maximum and the remaining
negative toroidal field at polar latitudes continues to be unstable.  However, 
this time it shows a negative tilt with respect to the horizontal direction, giving rise 
to positive values of $\alpha$; see bottom right panel of Fig.~\ref{fig.sources}.  
This $\alpha$-effect seems to be responsible for the generation of both, a positive radial 
field and a positive toroidal field 
at intermediate latitudes. Note in panels (c)-(f) that although the field lines 
are erratic, some of them are oriented in the latitudinal direction first, and 
in the azimuthal direction later. The positive toroidal field 
finally covers almost the entire hemisphere. A new positive tilt is observed 
near the poles together with the remainings of the negative field. This is the initial
configuration for the second part of the full cycle.

We have described the reversal process for model RC49. For other 
oscillatory simulations with fast rotation the process is fundamentally the same (see the 
movie of the field reversal for simulation RC28 in the supplementary material). However, 
according to Fig.~\ref{fig.dyn}, increasing the rotation leads to larger poloidal field
and smaller toroidal fields. 
Thus, for a field configuration analogous to that of Fig.~\ref{fig.flines}(a), the toroidal
field is less coherent and decays faster giving rise to well organized field lines in the 
latitudinal direction.  These, in turn, are rapidly stretched by the equatorial shear resulting
in a faster cycle. 
 
In Fig.~\ref{fig.per}~(a) the black dotted line depicts the amplitude of the 
toroidal field as a function of the rotational period. Its trend agrees with that of the 
cycle period suggesting proportionality between the two quantities. As discussed in the
third paragraph of this subsection,
this is at odds with the linear theory of the Tayler instability which predicts fast 
development of the unstable modes for strongest fields. Because rotation
also stabilizes toroidal fields, the fact that the toroidal field decays faster for 
simulations  with rapid rotation is also against the linear theory.  Nevertheless, 
our results correspond to a steady non-linear state of the simulations, which is hardly 
comparable  to the linear phase of the instability. 
\cite{Miesch07} studied the development of shear-current instabilities in
the tachocline through non-linear MHD simulations. Even though the unstable modes 
reported by them have similarities with those presented in Fig.~\ref{fig.flines}, 
in their simulations a latitudinal shear is imposed as initial condition.
In our case the shear is mostly radial and it is continuously replenished by the 
rotation and the convective motions in the convection zone. Therefore, the mechanisms 
driving the instabilities are different and the results cannot be compared.
The stability of magnetic fields 
in stable stratified atmospheres under these circumstances has not yet been explored 
in detail. A dynamo operating in these radiative zones has been envisaged by e.g., 
\cite{spruit02,ZB07,Bonanno13},
nevertheless, the results presented here \citep[perhaps also the simulations of][and 
subsequent papers]{GCS10}  are the first to capture the 
entire process from first principles. Idealized simulations, where these 
processes are studied separately, are still necessary to clarify the contribution of 
each one (Guerrero \& Bonanno, 2019, {\it in preparation}).

\newpage
\section{Summary and conclusions}
\label{s.con}

We analyzed the results of global dynamo simulations in models which
have a solar-like stratification and span a wide range of rotation rates. 
In all the simulated cases the resulting mean-flows exhibit a fast equator
and slower poles and the formation of radial shear layers in two locations,
the interface between a convectively stable layer at the bottom of the domain 
and the convection zone (the tachocline), and the uppermost layer of the domain 
(the near-surface shear layer).
The magnetic fields evolve in a variety of dynamo modes, from oscillatory 
dynamos with short period occurring mostly in the convection zone, to dynamos
mostly happening at the tachocline with periodic, steady, and also
mixed modes. The MHD properties of some of the simulations have been studied
in detail in previous works \citep{GSDKM16a,GSDKM16b}.
In this work we focussed our analysis on the magnetic activity, i.e., the
magnetic field strength and cycle period, and compared our results with
recent observational findings. 

The butterfly diagrams of the simulated stellar dynamos demonstrate the 
complexity of the systems (Fig.~\ref{fig.bd}).  One of the characteristics is the
existence of dynamo sources in different places of the domain which add features to
the spatio-temporal evolution.  For instance, oscillatory dynamos for 
rotational periods between 24 and 49 days have their magnetic time scale
governed by the tachocline dynamics, however, due to the dynamo action in the 
near-surface layer new branches of activity appear modifying the solution. 

In view of these intricacies we have performed an analysis considering
three different shells within the domain (TAC, covering the shear region and the
stable stratified layer; CZ, enclosing the bulk of the convection zone; and NSL,
covering the uppermost fraction of the model), and two latitudinal
zones (EQU, form the equator to $45^{\circ}$;  and POL, from
$45^{\circ}$, to the north pole). The volume rms values of the normalized 
magnetic field energy density, the non-dimensional dynamo coefficients 
$C_{\alpha}$, $C_{\Omega}$ and $C_u$, as well as the dynamo numbers, 
$D_r^{\prime}$ and $D_{\theta}^{\prime}$ , were computed for these
regions.  These quantities appear in Fig.~\ref{fig.dyn} as function of the
Rossby number.  This figure summarizes and quantifies what can be also  
observed in Figs.~\ref{fig.dr}, \ref{fig.bd}, and \ref{fig.a3}. 
Our most relevant findings are: 

\begin{itemize}
\item 
The total magnetic energy density is independent of the rotational period, 
yet the energy density in the toroidal and poloidal components of the
magnetic field is determined by $\Ro$ and reaches different amplitudes 
at different depths. Although the dynamo numbers in the three regions
have similar values, the magnetic energy in TAC is one order of magnitude 
larger than in CZ and NSL. This is a consequence of the stable stratified layer
which allows a longer storage and further amplification of the magnetic fields.
\item 
In the TAC region, for  $\Ro \lesssim 1$, the poloidal field is larger
than the toroidal one. They have similar energies at $\Ro \sim 1$,
and for $\Ro > 1$ the toroidal field energy dominates.  This is an outcome
of the scaling of $C_{\alpha}$ (at higher latitudes) and 
$C_{\Omega}^r$ (in the entire hemisphere) with the Rossby number. For $\Ro \lesssim 1$,
$C_{\alpha}$ has larger amplitude and negative values, while $C_{\Omega}^r$ is
smaller. For $\Ro \sim 1$, $C_{\alpha}$ changes sign while $C_{\Omega}^r$ 
reaches larger values that enhance the generation of toroidal field. 
\item
At $\Ro \sim 1$, when $C_{\alpha}$ becomes positive, and the radial shear
at the tachocline
is sufficiently strong to generate large toroidal fields, the dynamo simulations 
start to develop deep seated cyclic dynamos (models RC21 - RC49). For $\Ro \gtrsim 1.7$
(models RC56-RC63),
$C_{\alpha}$ and $C_{\Omega}^r$ are still significant, however, 
Fig~\ref{fig.dr}(h-i) and Fig~\ref{fig.a3}(h) reveal that these coefficients
are prominent in latitudes closer to the poles.
The magnetic field in these slow rotating cases is steady and concentrated
at higher latitudes (Fig.~\ref{fig.bd}(h)). 
\item
The case RC21 exhibits bimodality at the tachocline, i.e., 
a steady radial field is mixed with an oscillatory toroidal field. 
This transitional case is confirmed by the simulationss RC18, with a
steady dynamo in the tachocline and no cyclic behavior in the convection zone,
and RC24, where an oscillatory dynamo is observed in the tachocline.
\item 
In the NSL region, for $\Ro \gtrsim 1.2$, the normalized magnetic energy 
densities, $e_{B_{\phi}}$ and $e_{B_{p}}$, decay at equatorial latitudes 
as $\Ro^{-2.8}$ and $\Ro^{-6.4}$, respectively. 
At higher latitudes $e_{B_{\phi}}$ increases as $\Ro^{1.8}$ and $e_{B_{p}}$ decays 
as $\Ro^{-2.3}$; see the scaling laws for the magnetic field components in 
Fig.~\ref{fig.bfs}. 
In \S\ref{sec.nsl} we have argued that finding an explanation for these scaling
laws is not straightforward. In some parts of the domain the local generation is
relevant, in others there are non-local sources of magnetic flux which cannot be
easily quantified.   Furthermore, we demonstrated that the turbulent magnetic diffusivity
varies in the latitudinal direction affecting differently the decay of 
the field in the POL and EQU regions.  It is important to bear in
mind that we have computed the dynamo turbulent coefficients by considering the FOSA 
approximation. The results of \cite{Warnecke+18} indicate that the turbulent 
diffusivity may be even anisotropic, i.e., it may have different values for the different 
components of the field.   

\item
The scaling of the magnetic field energy with $\Ro$, and also the relation between
the toroidal and poloidal field energy densities in the most external layers
of the domain are reminiscent of the observational results 
\citep{pizzolato+03,Wright+11,Vidotto+14,See+2015,Wright+16}.  
Because the magnetic field may have different local and non-local sources for 
different regimes and regions within the domain, these observations cannot be
explained by a simple model. Before attaining a satisfactory explanation it is 
necessary to understand better the correlation between the different observational 
signatures and the magnetic field in the interior of the stars.

\end{itemize}

In Fig.~\ref{fig.per} we show two different representations, often used in the 
literature, correlating the rotation period, $\prot$, with the magnetic cycle period, 
$\pcyc$.  The simulations results clearly show that $\pcyc$ increases with $\prot$, 
in agreement with the observations of stars of types G to K \citep{BV07, BMM17}.
Most of the cycle periods obtained in this work are consistent with the 
$A$ branch described in the literature. However, the bimodal dynamo observed in case 
RC21 appears out of the curve. We conclude 
that this case is in the transition between cyclic dynamos operating in the 
convection zone and cyclic dynamos in the tachocline.  As mentioned
above, this transition is characterized by the enhanced generation of deep seated 
toroidal field due to an equatorial radial shear that increases monotonically with 
the Rossby number.
The non-linear effects resulting from this strong field (one order of magnitude larger
than in the rest of the domain) are substantial, they can even modify the character of 
of the convective motions \citep[e.g.,][]{PMCG16,Beaudoin+18} and/or give rise to other 
instabilities different from the dynamo \citep{LSC15,GSDKM16a}.

We suggest that the resulting cycle period may be explained from the energy exchange 
between the dynamo and shear-current instabilities occurring below the tachocline.
In the dynamical phase of evolution of the velocity and magnetic fields, both 
instabilities reach an equilibrium state of energy exchange which behaves like 
an $\alpha^2\Omega$ dynamo and determines the cycle period.  A similar oscillatory 
exchange of energy in a shear layer, between the magnetic field and the 
non-axisymmetric kinetic energy was reported by \cite{Miesch07b}.  However, it 
is still necessary to quantify how the period of these oscillations depends
on the strength of the toroidal field and the amount of shear. Since the entire process 
is hard to disentangle in convection simulations, a step by step analysis of these 
instabilities is left to a forthcoming work (Guerrero \& Bonanno, 2019, 
{\it in preparation}).

The resulting magnetic fields are a direct consequence of the self consistent 
development of tachoclines in our simulations. 
This characteristic increase of $\pcyc$ with $\prot$ is precisely what distinguishes 
them from other convective dynamo simulations where $\pcyc$ decreases with the 
increase of $\prot$  \citep{SBCBdN17,Warnecke17}. It is worth mentioning, however, 
that the observations
still are inconclusive in this regard. Nevertheless, 
the estimation of stellar magnetic fields is currently
an exiting and quite active field of research. Luckily in the near future the observations 
will provide a clear picture about the relation between stellar cycle and rotational
periods.  From the numerical point of view, current dynamo models are able to
reproduce both scenarios. 

\acknowledgments
We thank an anonymous referee for insightful comments that
helped to improve the paper.
This work was partly funded by FAPEMIG grant APQ-01168/14 (GG),
FAPESP grant 2013/10559-5 (EMGDP), CNPq grant 306598/2009-4 (EMGDP),
NASA grants NNX09AJ85G, NNX17AE76A and NNX14AB70G.
The simulations
were performed in the NASA cluster Pleiades and the Brazilian supercomputer 
SDumont of the National Laboratory of Scientific Computation (LNCC).


\appendix
\section{The convective turnover time and the Rossby number}
\label{ap.A}

One of the bottlenecks in the interpretation of stellar activity is the 
determination of the convective turnover time, $\tau_c$. It is useful for
computing the Rossby number as well as the turbulent dynamo coefficients. 
Observationally, there is a correlation between $\tau_c$ and the chromospheric
emission $\brac{R'_{HK}}$ \citep{Noyes+84a}.  In simulations this
quantity can be computed through the energy spectrum. Since we are interested
in the scales associated with the turbulent quantities, the spectrum is computed 
from the transformation of, for example,  the non-axisymmetric velocity,  
${\bm u}' = {\bm u} - \meanv{u}$,
into spherical harmonics, $Y_l^m(\theta,\phi) = P_l^m(\cos\theta) \exp( i m \phi)$, by  
\begin{equation}
{\bm u}'(l,m) = \sum_{l,m} Q_l^m Y_l^m(\theta,\phi) + S_l^m r \nabla Y_l^m(\theta,\phi) - T_l^m {\bm r} \times \nabla Y_l^m(\theta,\phi)\;,
\end{equation}
where the expansion coefficients, $Q_l^m$, $S_l^m$ and $T_l^m$, 
are computed with the optimized library SHTns \citep{shtns}.  

The kinetic energy spectra is computed by:
\begin{equation}
\tilde{E}_k = 4 \pi \sum_{m = -l}^l  \left[ |Q_l^m|^2 + l(l+1) ( |S_l^m|^2 + |T_l^m|^2)\right]
\end{equation}

\begin{figure*}[ht]
\begin{center}
\includegraphics[width=0.96\columnwidth]{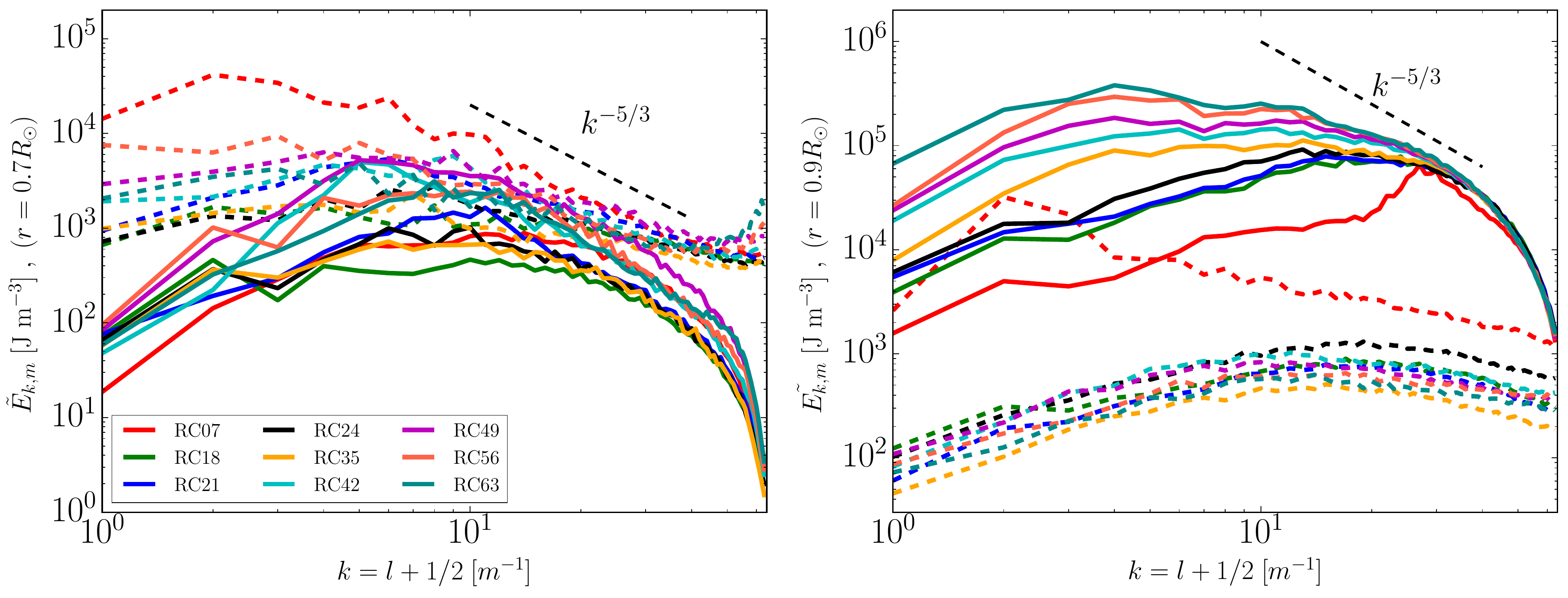}
\caption{Energy spectra of the kinetic (solid line) and magnetic (dashed lines) energies 
computed for $r=0.7\Rs$ (left) and $r=0.9\Rs$ (right) for some representative models. The
black dashed line shows the Kolmogorov $k^{-5/3}$ energy scaling.}
\label{fig.a1} 
\end{center}
\end{figure*}

\begin{figure*}[ht]
\begin{center}
\includegraphics[width=0.96\columnwidth]{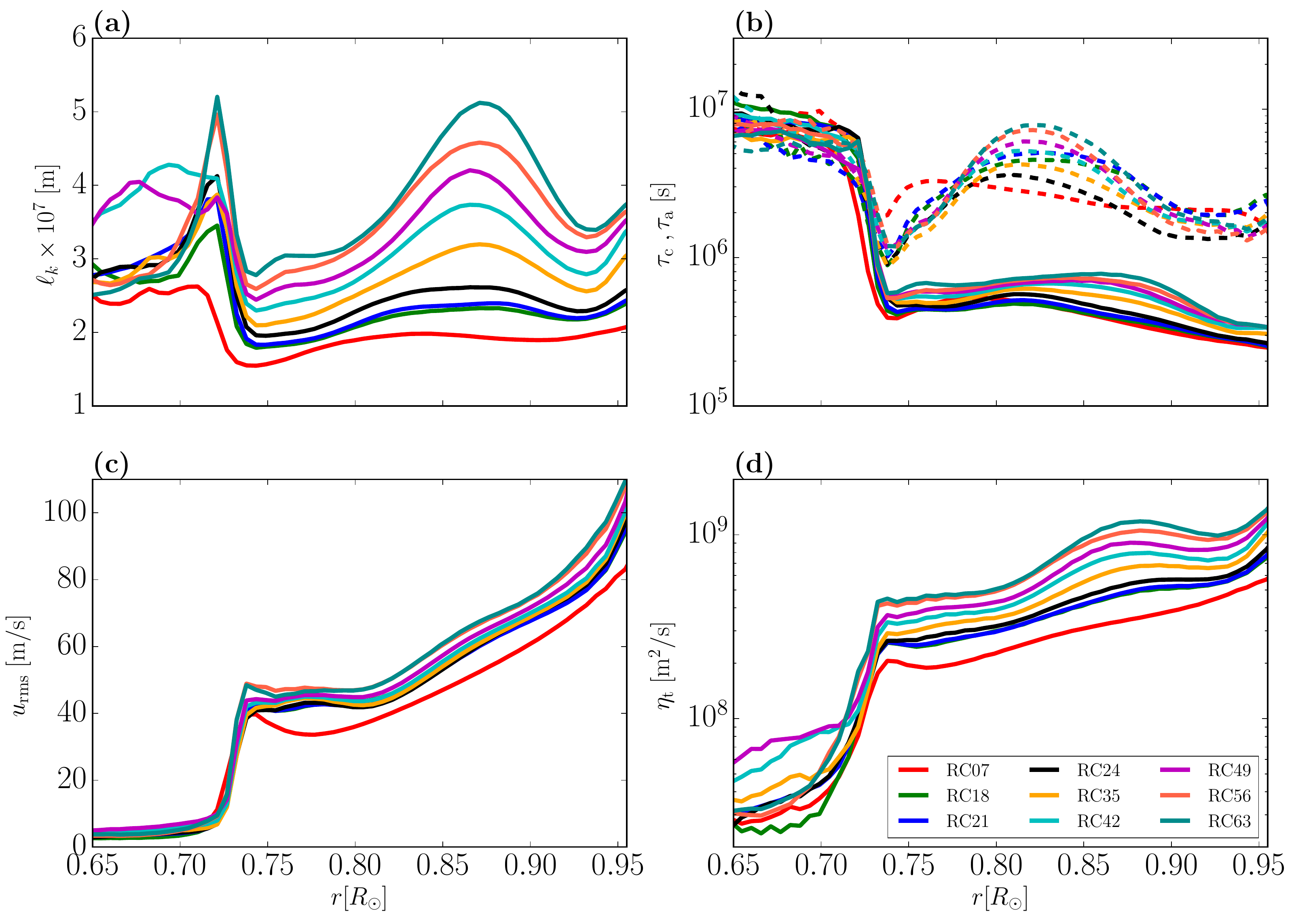}
\caption{(a) Length scale of the energy carrying eddies, $\ell(r)$; (b)  
convective turnover times, $\tau_c$; (c) radial profile of the $\urms$ velocity and
(d) turbulent magnetic diffusivity. Different colors are assigned to 
some representative simulations  between RC07 and RC63 (see annotations in panel (d)). 
The dashed lines in panel (b) depict the Alfven time scale, $\tau_{\rm A}$.
See the text for the definition.}
\label{fig.a2} 
\end{center}
\end{figure*}

With a similar decomposition, we obtain the spectral energy of the non-axisymmetric 
magnetic field, $\tilde{E}_m$.
In Figure~\ref{fig.a1} we present the kinetic (solid line) and magnetic (dashed line)
energy spectra for the radial level $r=0.7\Rs$ (left), and $r=0.9\Rs$ (right). 
Note that in the convection zone the kinetic energy dominates over the magnetic energy.
For the faster rotating simulations the peak of the spectra is at largest $k$.
With the decrease of the rotation (large period), it moves towards smaller values 
of $k$ and reaches large values for the energy. This behaviour is expected 
from simulations dominated by rotation towards simulations dominated by convection. 
In the radiative zone the magnetic energy is dominant and 
the spectra peak at small wave numbers.   Since
we are interested in the scales that carry most of the energy as a function of 
depth, we compute a similar spectra for each radial level of the simulation,
and compute such length scale as
\begin{equation}
\ell_{k,m}(r) = \frac{r\; \int_k \frac{\tilde{E}_{k,m}(k,r)}{k} dk}{\int_k \tilde{E}_{k,m}(k,r) dk}\;,
\end{equation}
where $k=l+1/2$, according to the Jeans' rule.  The convective turnover time
is computed by $\tau_c(r) = {\ell}_k(r) / \urms(r)$.  The radial profiles of 
$\ell$ and $\tau_c$ are depicted in Figure~\ref{fig.a2} (a) and (b). Finally, 
the Rossby number is computed as $\Ro = \prot/(2 \pi \tau_c^{NSL})$ (see Table~\ref{table.1}). 
 
\section{Mean-field turbulent coefficients}
\label{ap.B}

We use the first order smoothing approximation (FOSA) to compute the 
dynamo turbulent coefficients. Under this formalism the $\alpha$-effect is 
given by
\begin{equation}
\alpha = \alpha_{\rm k} + \alpha_{\rm m} = -\frac{\tau_c}{3}\brac{{\bm \omega'} \cdot {\bm u'}} 
                  + \frac{\tau_c}{3}\brac{{\bm j'} \cdot {\bm B'}}/\rho_e \;,
\label{equ:alpha}
\end{equation}
where ${\bm \omega}'= \nabla \times {\bm u}'$ and ${\bm j}'= \nabla \times {\bm B}'$ are
the small scale vorticity and current, respectively, and $\tau_c$ is the convective 
turnover time defined in the previous appendix.  Note that the convective motions 
spread along the unstable region therefore $\tau_c$ in Eq. \ref{equ:alpha} is 
valid only down to $r\sim 0.74\Rs$. This is not much important for $\alpha_{\rm k}$ 
since the velocity drops to small values in the radiative zone. However, it 
presents a problem for $\alpha_{\rm m}$ 
as a significant amount of current helicity, $\brac{{\bm j'} \cdot {\bm b'}}$,
develops below this radius. 

As can be noticed in the lhs panel of  Fig.~\ref{fig.a1}, 
at $r = 0.7\Rs$ the non-axisymmetric magnetic energy dominates over the kinetic energy.
We believe that the magnetic field develops at and below the tachocline because
of shear-current instabilities. Therefore, we associate the time scale of the 
magnetic $\alpha$-effect below the tachocline to one of the time-scales associated 
with these instabilities, namely the  Alfven time: $\tau_{\rm A} = \ell_m/v_{\rm A}$ 
(see dashed lines in the upper rhs panel of Fig.~\ref{fig.a2}), 
where $v_{\rm A} = \Brms^{\prime}/(\mu_0 \rho_e)^{1/2}$. Here, 
$\Brms^{\prime} = (\brac{B_r'^2 + B_{\theta}'^2  + B_{\phi}'^2}_{\phi,\theta,t})^{1/2}$ is the 
non-axisymmetric magnetic field averaged over $\phi$, $\theta$ and time.  

\begin{figure*}[ht]
\begin{center}
\includegraphics[width=0.42\columnwidth]{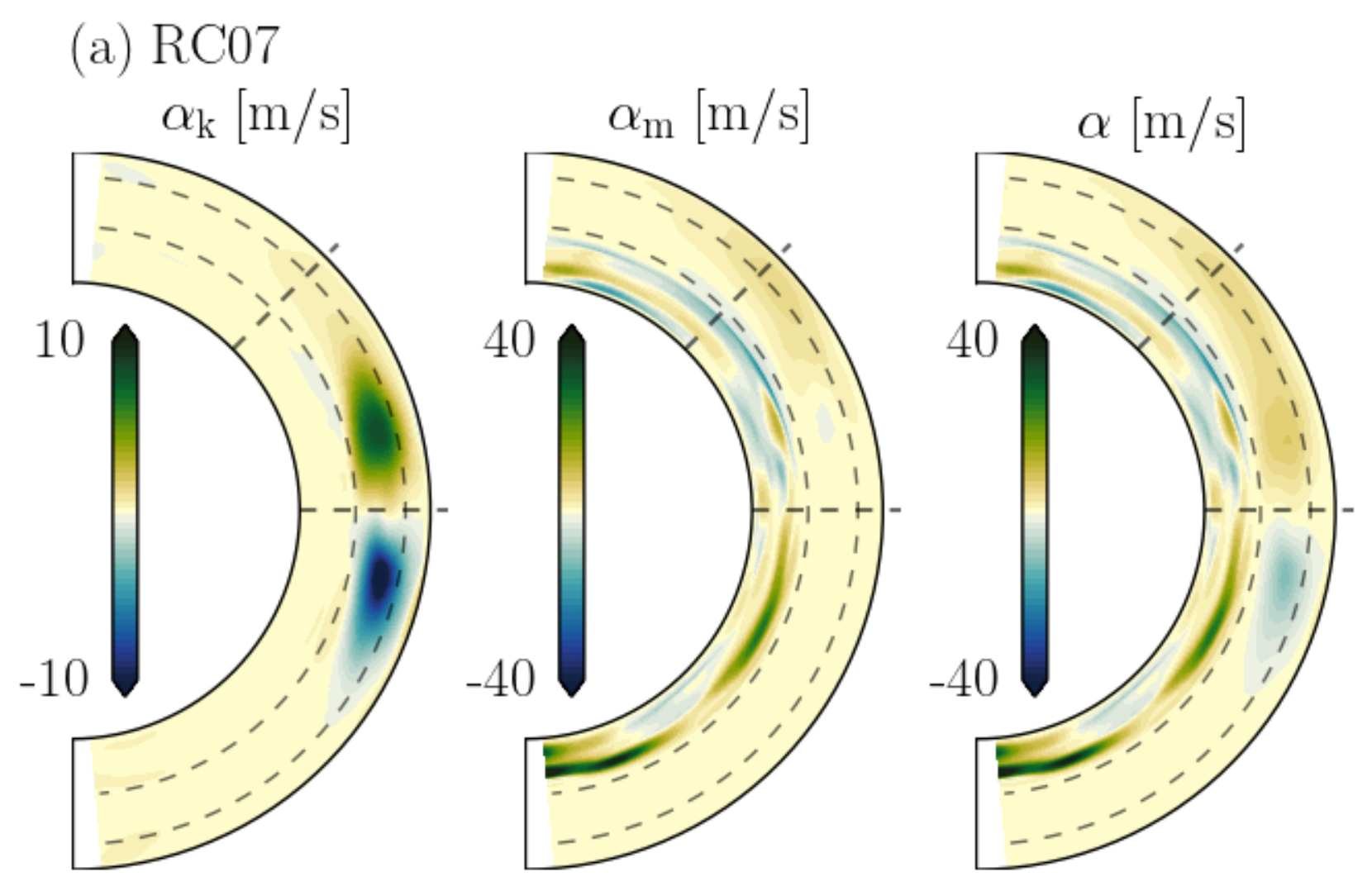}
\hspace{0.2cm}
\includegraphics[width=0.42\columnwidth]{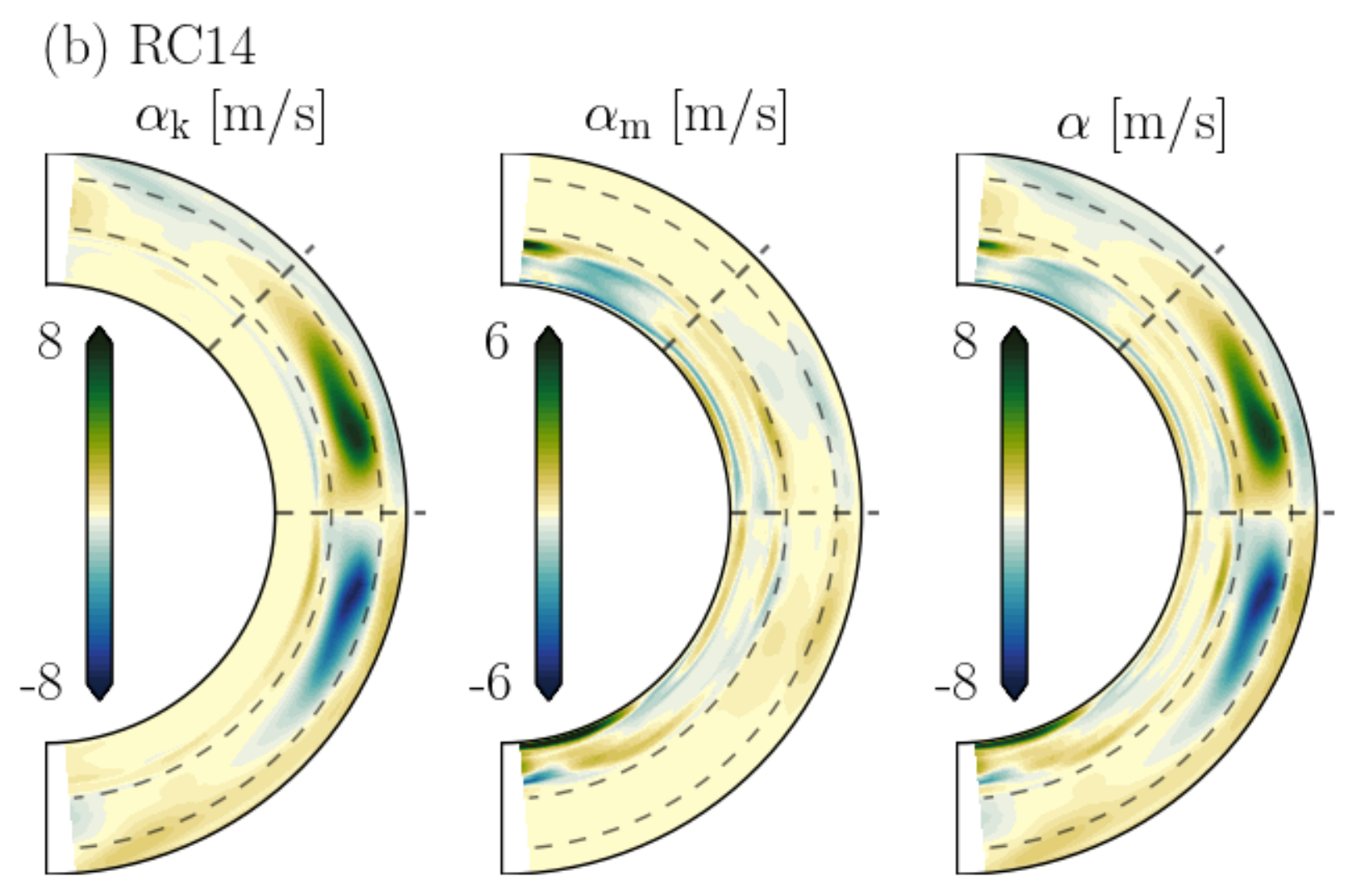}\\
\includegraphics[width=0.42\columnwidth]{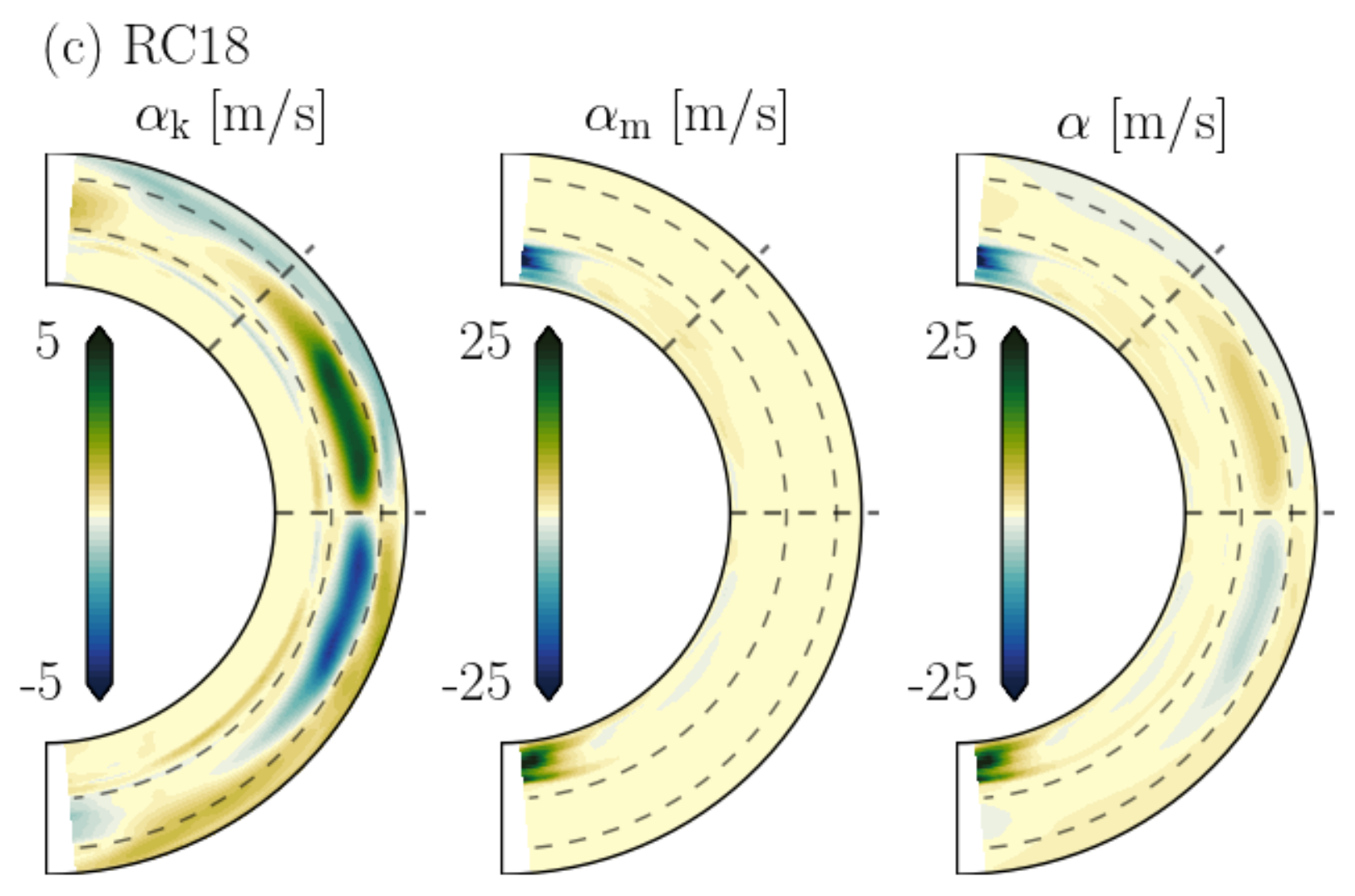}
\hspace{0.2cm}
\includegraphics[width=0.42\columnwidth]{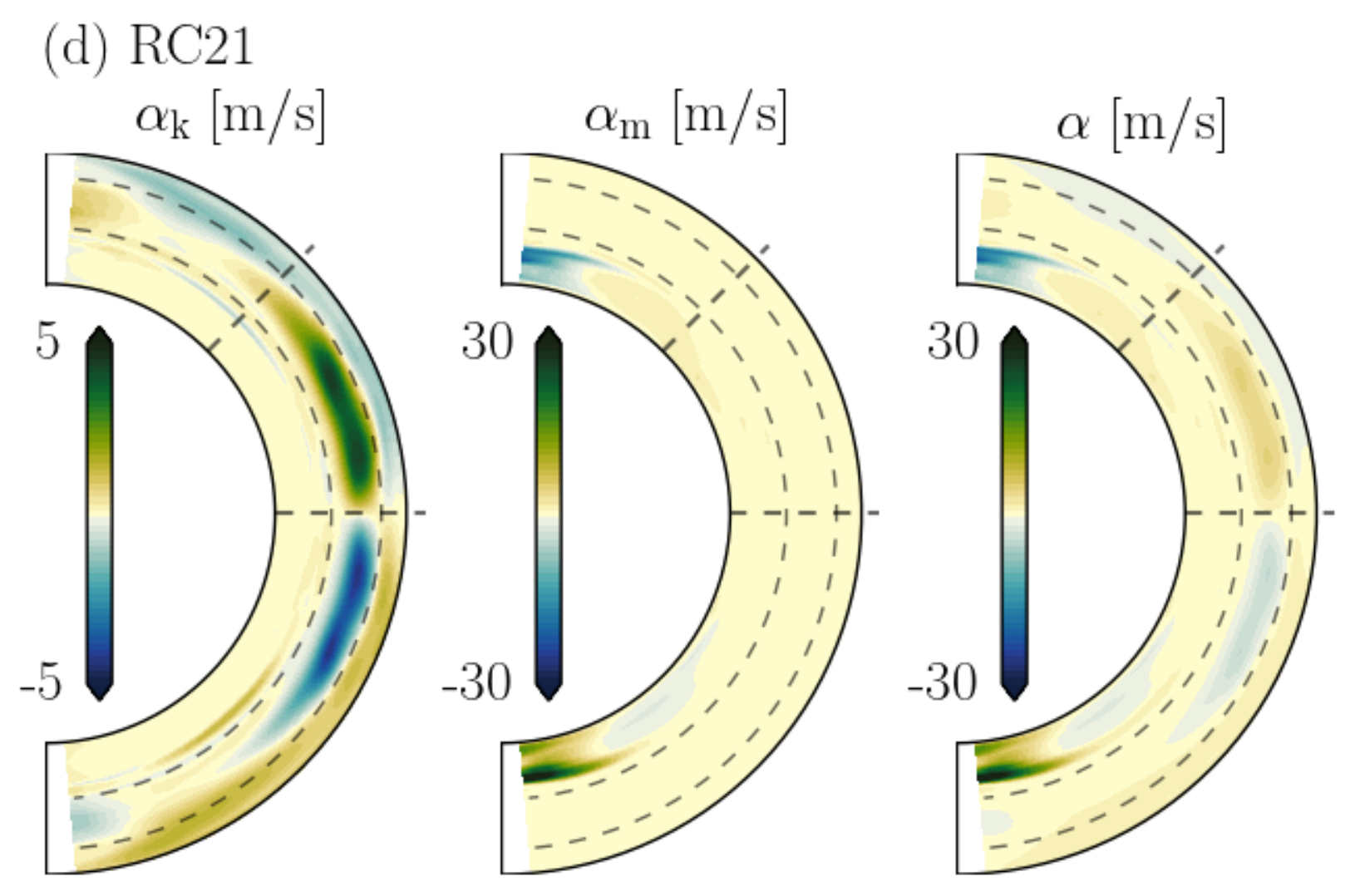}\\
\includegraphics[width=0.42\columnwidth]{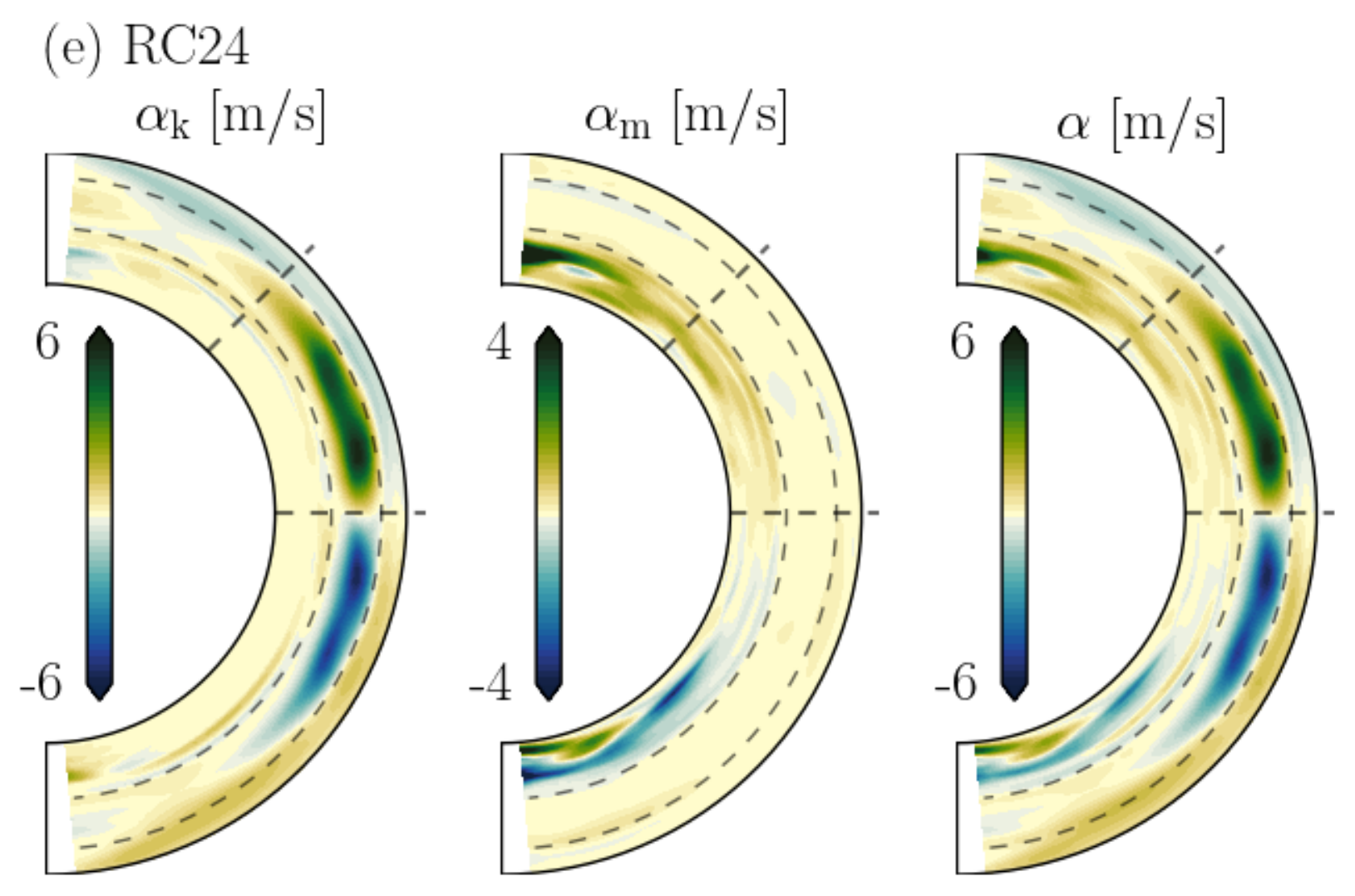}
\hspace{0.2cm}
\includegraphics[width=0.42\columnwidth]{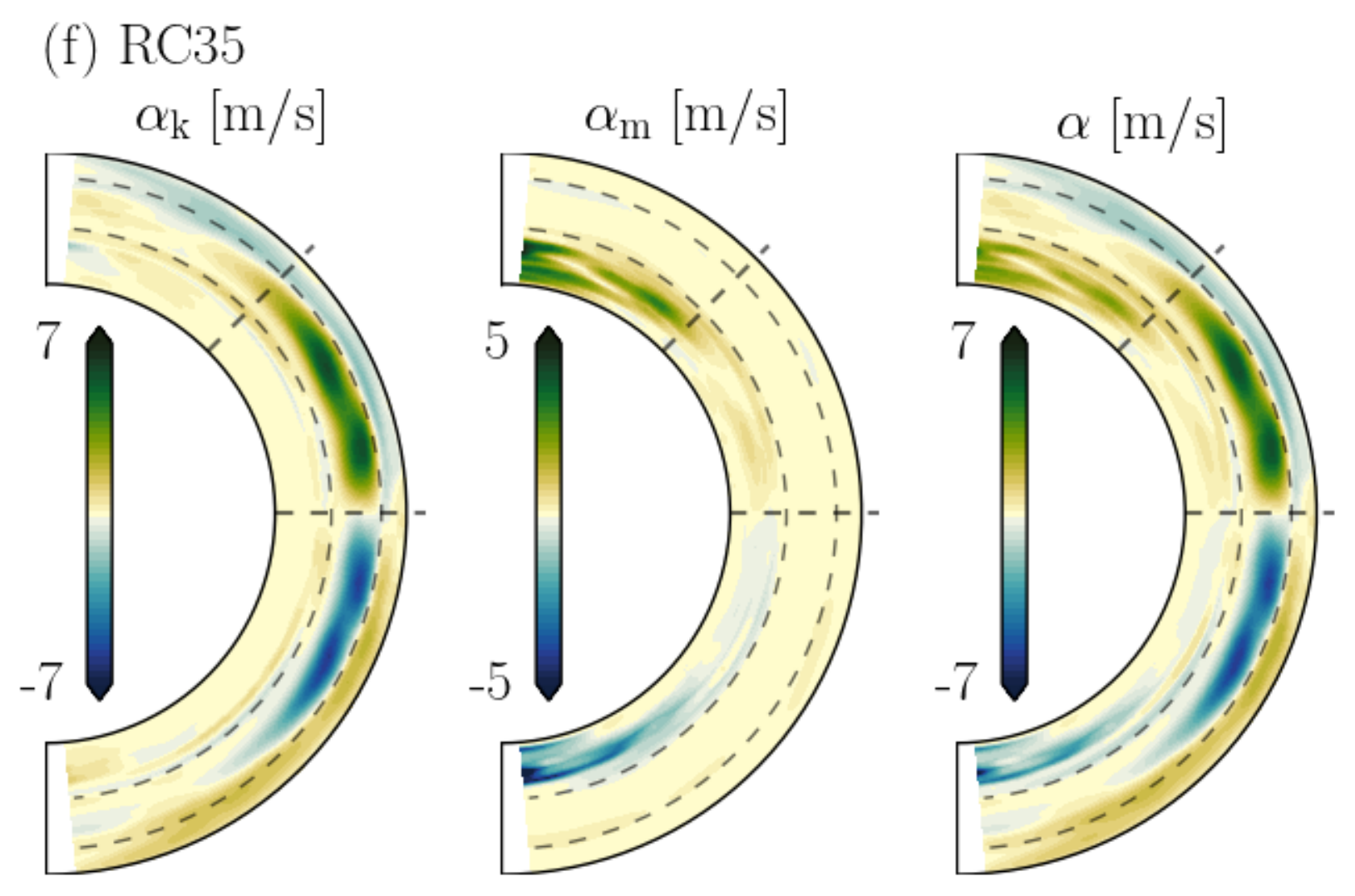}\\
\includegraphics[width=0.42\columnwidth]{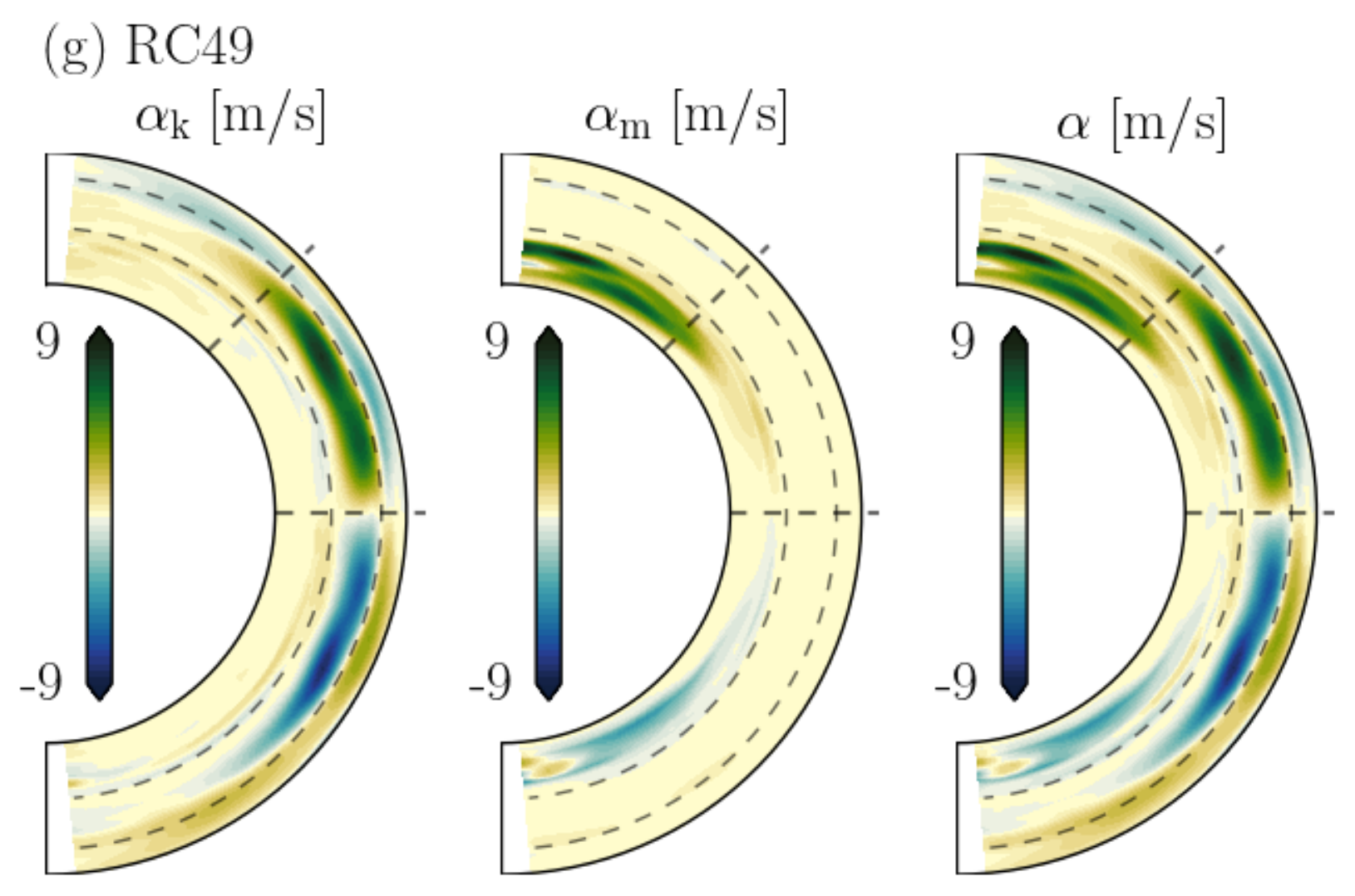}
\hspace{0.2cm}
\includegraphics[width=0.42\columnwidth]{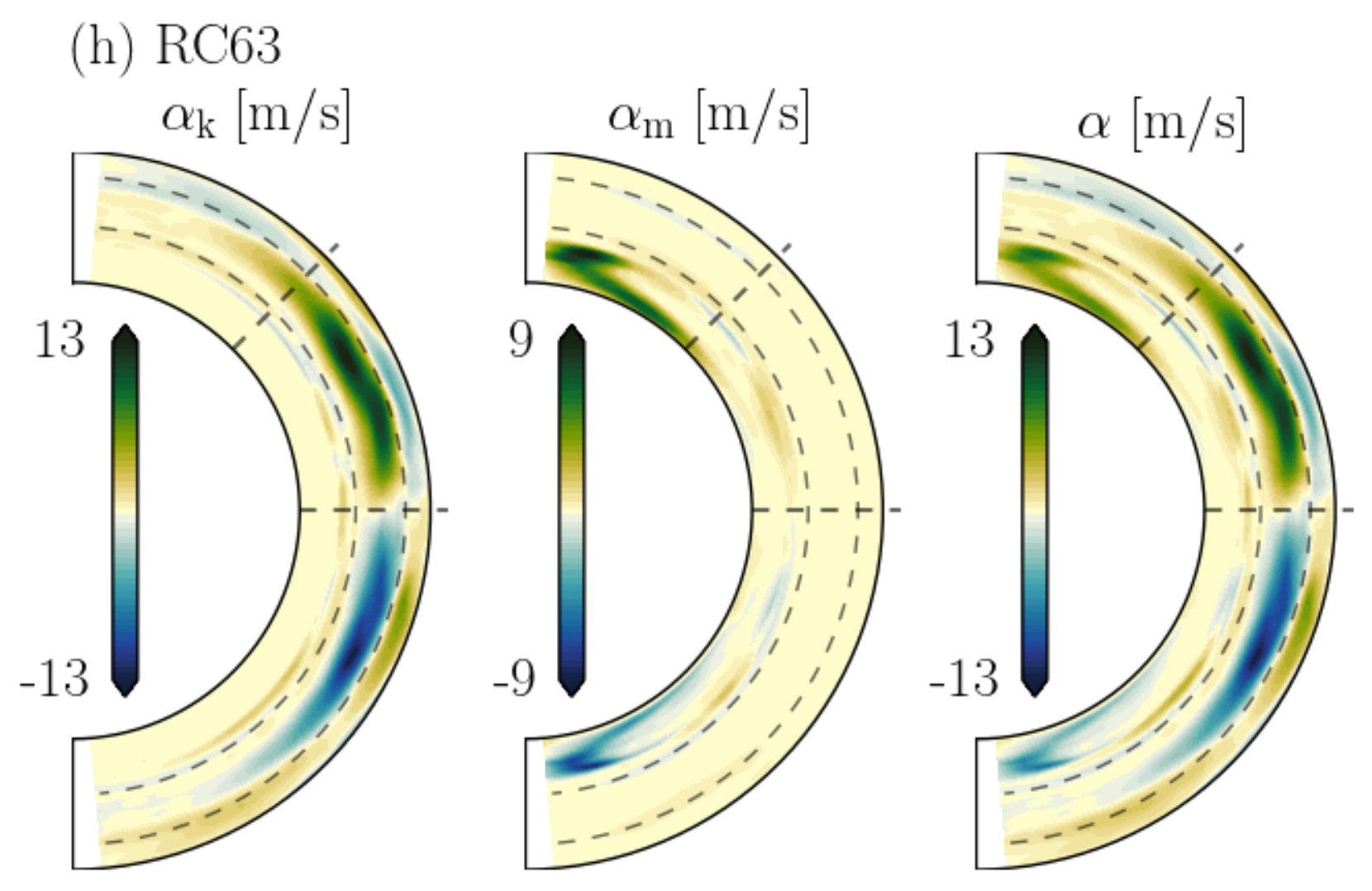}\\
\caption{Kinetic, magnetic and total $\alpha$-effect, from left to rigth, for 
representative models between RC07 and RC63, in panels (a)-(h), respectively.}
\label{fig.a3} 
\end{center}
\end{figure*}

The profiles of the kinetic (left), magnetic (middle) and total (right) $\alpha$-effect,
for the simulations RC07~-RC63 are presented in panels (a) to (h) of Fig.~\ref{fig.a3}. 
A quantitative analysis of these profiles is presented in \S\ref{sec.mfa}. Finally,
the turbulent diffusivity coefficient presented in Fig.~\ref{fig.a2}(d) is evaluated as
\begin{equation}
\etat = \frac{1}{3} \tau_c u^{\prime 2} \;.
\label{eq.etat}
\end{equation}

\section{Cycle period evaluation}
\label{ap.C}
\begin{figure*}[ht]
\begin{center}
\includegraphics[width=0.96\columnwidth]{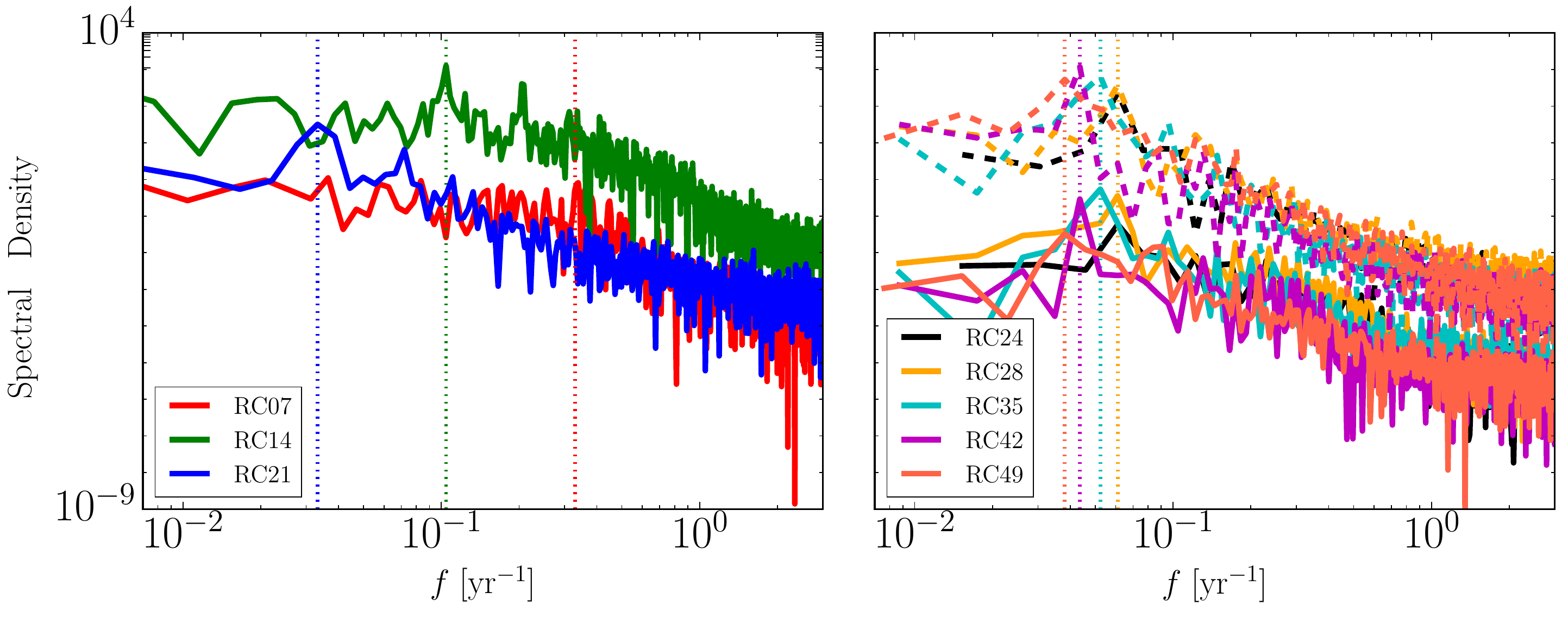}
\caption{Spectral density computed from the volume averaged time series
of the poloidal field $\brac{B_p}$. The dotted lines show the frequency
where the spectral density peaks.  For the models RC24~-~RC49 the period was
computed in the regions NSL (continuos lines) and TAC (dashed lines). 
For a clear visualization the spectral density in the region TAC was multiplied 
by a factor $10^2$.}
\label{fig.a4} 
\end{center}
\end{figure*}

The magnetic cycle period, $\pcyc$, is evaluated through the Fourier transformation 
of the rms vertical magnetic field (see Fig.~\ref{fig.a4}).  
Since the oscillatory behavior observed in models RC07~-~RC21 (left panel) is
prominent close 
to the equator, to obtain a better estimation of the period we consider a 
latitude range between $\pm 10^{\circ}$ latitude. In radius these models are
oscillatory only in the CZ, therefore,  the radial average is made in the layer 
NSL.  While for the models  RC14 and RC21 the frequency with maximal spectral 
density is clearly defined, the model RC07 seems to have multiple periodicities.
We have chosen the peak in the spectral density that matches better with the 
periodicity observed in the butterfly diagram ($\pcyc=2.9$ days). 
For the models  RC24~-~RC49 (right panel of Fig.~\ref{fig.a4}) we consider 
latitudes between $0^{\circ}$ and $90^{\circ}$. We compute the period in 
the shells NSL (continuous lines) and TAC (dashed lines).  For all these
models the periods agree in both regions, they are presented in Table~\ref{table.1}.

\bibliographystyle{aasjournal}
\bibliography{bib}

\begin{thebibliography}{}
\expandafter\ifx\csname natexlab\endcsname\relax\def\natexlab#1{#1}\fi
\providecommand{\url}[1]{\href{#1}{#1}}

\bibitem[{{Baliunas} {et~al.}(1995){Baliunas}, {Donahue}, {Soon}, {Horne},
  {Frazer}, {Woodard-Eklund}, {Bradford}, {Rao}, {Wilson}, {Zhang}, {Bennett},
  {Briggs}, {Carroll}, {Duncan}, {Figueroa}, {Lanning}, {Misch}, {Mueller},
  {Noyes}, {Poppe}, {Porter}, {Robinson}, {Russell}, {Shelton}, {Soyumer},
  {Vaughan}, \& {Whitney}}]{Baliunas+95}
{Baliunas}, S.~L., {Donahue}, R.~A., {Soon}, W.~H., {et~al.} 1995, \apj, 438,
  269

\bibitem[{{Beaudoin} {et~al.}(2018){Beaudoin}, {Strugarek}, \&
  {Charbonneau}}]{Beaudoin+18}
{Beaudoin}, P., {Strugarek}, A., \& {Charbonneau}, P. 2018, \apj, 859, 61

\bibitem[{{Blackman} \& {Thomas}(2015)}]{BT15}
{Blackman}, E.~G., \& {Thomas}, J.~H. 2015, \mnras, 446, L51

\bibitem[{{B{\"o}hm-Vitense}(2007)}]{BV07}
{B{\"o}hm-Vitense}, E. 2007, \apj, 657, 486

\bibitem[{{Bonanno}(2013)}]{Bonanno13}
{Bonanno}, A. 2013, \solphys, 287, 185

\bibitem[{{Bonanno} \& {Urpin}(2012)}]{BU12}
{Bonanno}, A., \& {Urpin}, V. 2012, \apj, 747, 137

\bibitem[{{Bonanno} \& {Urpin}(2013)}]{BU13}
---. 2013, \mnras, 431, 3663

\bibitem[{{Brandenburg} {et~al.}(2017){Brandenburg}, {Mathur}, \&
  {Metcalfe}}]{BMM17}
{Brandenburg}, A., {Mathur}, S., \& {Metcalfe}, T.~S. 2017, \apj, 845, 79

\bibitem[{{Brandenburg} {et~al.}(2008){Brandenburg}, {R{\"a}dler},
  {Rheinhardt}, \& {K{\"a}pyl{\"a}}}]{BRRK08}
{Brandenburg}, A., {R{\"a}dler}, K.-H., {Rheinhardt}, M., \& {K{\"a}pyl{\"a}},
  P.~J. 2008, \apj, 676, 740

\bibitem[{{Brandenburg} {et~al.}(1998){Brandenburg}, {Saar}, \&
  {Turpin}}]{BST98}
{Brandenburg}, A., {Saar}, S.~H., \& {Turpin}, C.~R. 1998, \apjl, 498, L51

\bibitem[{{Cally}(2003)}]{Cally_03}
{Cally}, P.~S. 2003, \mnras, 339, 957

\bibitem[{{Cossette} {et~al.}(2017){Cossette}, {Charbonneau}, {Smolarkiewicz},
  \& {Rast}}]{Cossette+17}
{Cossette}, J.-F., {Charbonneau}, P., {Smolarkiewicz}, P.~K., \& {Rast}, M.~P.
  2017, \apj, 841, 65

\bibitem[{{do Nascimento} {et~al.}(2016){do Nascimento}, {Vidotto}, {Petit},
  {Folsom}, {Castro}, {Marsden}, {Morin}, {Porto de Mello}, {Meibom},
  {Jeffers}, {Guinan}, \& {Ribas}}]{dN+16}
{do Nascimento}, Jr., J.-D., {Vidotto}, A.~A., {Petit}, P., {et~al.} 2016,
  \apjl, 820, L15

\bibitem[{{Domaradzki} {et~al.}(2003){Domaradzki}, {Xiao}, \&
  {Smolarkiewicz}}]{DXS03}
{Domaradzki}, J.~A., {Xiao}, Z., \& {Smolarkiewicz}, P.~K. 2003, Physics of
  Fluids, 15, 3890

\bibitem[{{Donati}(2001)}]{Donati01}
{Donati}, J.-F. 2001, in Lecture Notes in Physics, Berlin Springer Verlag, Vol.
  573, Astrotomography, Indirect Imaging Methods in Observational Astronomy,
  ed. H.~M.~J. {Boffin}, D.~{Steeghs}, \& J.~{Cuypers}, 207

\bibitem[{{Donati} \& {Brown}(1997)}]{Donati+Brown_97}
{Donati}, J.-F., \& {Brown}, S.~F. 1997, \aap, 326, 1135

\bibitem[{{Donati} \& {Landstreet}(2009)}]{Donati+Landstreet_09}
{Donati}, J.-F., \& {Landstreet}, J.~D. 2009, \araa, 47, 333

\bibitem[{{Elliott} \& {Smolarkiewicz}(2002)}]{ES02}
{Elliott}, J.~R., \& {Smolarkiewicz}, P.~K. 2002, International Journal for
  Numerical Methods in Fluids, 39, 855

\bibitem[{{Fletcher} {et~al.}(2010){Fletcher}, {Broomhall}, {Salabert}, {Basu},
  {Chaplin}, {Elsworth}, {Garcia}, \& {New}}]{Fea10}
{Fletcher}, S.~T., {Broomhall}, A.-M., {Salabert}, D., {et~al.} 2010, ApJL,
  718, L19

\bibitem[{{Ghizaru} {et~al.}(2010){Ghizaru}, {Charbonneau}, \&
  {Smolarkiewicz}}]{GCS10}
{Ghizaru}, M., {Charbonneau}, P., \& {Smolarkiewicz}, P.~K. 2010, ApJL, 715,
  L133

\bibitem[{{Gleissberg}(1971)}]{Gl71}
{Gleissberg}, W. 1971, \solphys, 21, 240

\bibitem[{{Gray}(1984)}]{Gray84}
{Gray}, D.~F. 1984, \apj, 277, 640

\bibitem[{{Gregory} {et~al.}(2012){Gregory}, {Donati}, {Morin}, {Hussain},
  {Mayne}, {Hillenbrand}, \& {Jardine}}]{GDMHNHJ12}
{Gregory}, S.~G., {Donati}, J.-F., {Morin}, J., {et~al.} 2012, \apj, 755, 97

\bibitem[{{Guerrero} \& {de Gouveia Dal Pino}(2008)}]{GDP08}
{Guerrero}, G., \& {de Gouveia Dal Pino}, E.~M. 2008, \aap, 485, 267

\bibitem[{{Guerrero} {et~al.}(2016{\natexlab{a}}){Guerrero}, {Smolarkiewicz},
  {de Gouveia Dal Pino}, {Kosovichev}, \& {Mansour}}]{GSDKM16a}
{Guerrero}, G., {Smolarkiewicz}, P.~K., {de Gouveia Dal Pino}, E.~M.,
  {Kosovichev}, A.~G., \& {Mansour}, N.~N. 2016{\natexlab{a}}, \apj, 819, 104

\bibitem[{{Guerrero} {et~al.}(2016{\natexlab{b}}){Guerrero}, {Smolarkiewicz},
  {de Gouveia Dal Pino}, {Kosovichev}, \& {Mansour}}]{GSDKM16b}
---. 2016{\natexlab{b}}, \apjl, 828, L3

\bibitem[{{Guerrero} {et~al.}(2013){Guerrero}, {Smolarkiewicz}, {Kosovichev},
  \& {Mansour}}]{GSKM13b}
{Guerrero}, G., {Smolarkiewicz}, P.~K., {Kosovichev}, A.~G., \& {Mansour},
  N.~N. 2013, \apj, 779, 176

\bibitem[{{Jouve} {et~al.}(2010){Jouve}, {Brown}, \& {Brun}}]{JBB10}
{Jouve}, L., {Brown}, B.~P., \& {Brun}, A.~S. 2010, \aap, 509, A32

\bibitem[{{Karak} {et~al.}(2014){Karak}, {Kitchatinov}, \& {Choudhuri}}]{KKC14}
{Karak}, B.~B., {Kitchatinov}, L.~L., \& {Choudhuri}, A.~R. 2014, \apj, 791, 59

\bibitem[{{Kochukhov} \& {Piskunov}(2002)}]{KK02}
{Kochukhov}, O., \& {Piskunov}, N. 2002, \aap, 388, 868

\bibitem[{{Lawson} {et~al.}(2015){Lawson}, {Strugarek}, \&
  {Charbonneau}}]{LSC15}
{Lawson}, N., {Strugarek}, A., \& {Charbonneau}, P. 2015, ArXiv e-prints,
  arXiv:1509.07447

\bibitem[{{Lehmann} {et~al.}(2019){Lehmann}, {Hussain}, {Jardine}, {Mackay}, \&
  {Vidotto}}]{LHJMV19}
{Lehmann}, L.~T., {Hussain}, G.~A.~J., {Jardine}, M.~M., {Mackay}, D.~H., \&
  {Vidotto}, A.~A. 2019, \mnras, 483, 5246

\bibitem[{{Lehtinen} {et~al.}(2016){Lehtinen}, {Jetsu}, {Hackman}, {Kajatkari},
  \& {Henry}}]{Lehtinen+16}
{Lehtinen}, J., {Jetsu}, L., {Hackman}, T., {Kajatkari}, P., \& {Henry}, G.~W.
  2016, \aap, 588, A38

\bibitem[{{Margolin} {et~al.}(2006){Margolin}, {Smolarkiewicz}, \&
  {Wyszogradzki}}]{MSW06}
{Margolin}, L.~G., {Smolarkiewicz}, P.~K., \& {Wyszogradzki}, A.~A. 2006,
  Journal of Applied Mechanics, 73, 469

\bibitem[{{Mestel} \& {Landstreet}(2005)}]{Mestel+Landstreet_05}
{Mestel}, L., \& {Landstreet}, J.~D. 2005, in Lecture Notes in Physics, Berlin
  Springer Verlag, Vol. 664, Cosmic Magnetic Fields, ed. R.~{Wielebinski} \&
  R.~{Beck}, 183

\bibitem[{{Miesch}(2007)}]{Miesch07b}
{Miesch}, M.~S. 2007, \apjl, 658, L131

\bibitem[{{Miesch} {et~al.}(2007){Miesch}, {Gilman}, \& {Dikpati}}]{Miesch07}
{Miesch}, M.~S., {Gilman}, P.~A., \& {Dikpati}, M. 2007, \apjs, 168, 337

\bibitem[{{Moffatt}(1978)}]{Moffatt78}
{Moffatt}, H.~K. 1978, {Magnetic field generation in electrically conducting
  fluids}

\bibitem[{{Noyes} {et~al.}(1984{\natexlab{a}}){Noyes}, {Hartmann}, {Baliunas},
  {Duncan}, \& {Vaughan}}]{Noyes+84a}
{Noyes}, R.~W., {Hartmann}, L.~W., {Baliunas}, S.~L., {Duncan}, D.~K., \&
  {Vaughan}, A.~H. 1984{\natexlab{a}}, \apj, 279, 763

\bibitem[{{Noyes} {et~al.}(1984{\natexlab{b}}){Noyes}, {Weiss}, \&
  {Vaughan}}]{Noyes+84b}
{Noyes}, R.~W., {Weiss}, N.~O., \& {Vaughan}, A.~H. 1984{\natexlab{b}}, \apj,
  287, 769

\bibitem[{{Olspert} {et~al.}(2017){Olspert}, {Lehtinen}, {K{\"a}pyl{\"a}},
  {Pelt}, \& {Grigorievskiy}}]{Olspert+17}
{Olspert}, N., {Lehtinen}, J.~J., {K{\"a}pyl{\"a}}, M.~J., {Pelt}, J., \&
  {Grigorievskiy}, A. 2017, ArXiv e-prints, arXiv:1712.08240

\bibitem[{{Parker}(1955)}]{Pa55}
{Parker}, E.~N. 1955, \apj, 122, 293

\bibitem[{{Passos} {et~al.}(2017){Passos}, {Miesch}, {Guerrero}, \&
  {Charbonneau}}]{PMCG16}
{Passos}, D., {Miesch}, M., {Guerrero}, G., \& {Charbonneau}, P. 2017, \aap,
  607, A120

\bibitem[{{Petit} {et~al.}(2008){Petit}, {Dintrans}, {Solanki}, {Donati},
  {Auri{\`e}re}, {Ligni{\`e}res}, {Morin}, {Paletou}, {Ramirez Velez},
  {Catala}, \& {Fares}}]{Petit+08}
{Petit}, P., {Dintrans}, B., {Solanki}, S.~K., {et~al.} 2008, \mnras, 388, 80

\bibitem[{{Pevtsov} {et~al.}(2003){Pevtsov}, {Fisher}, {Acton}, {Longcope},
  {Johns-Krull}, {Kankelborg}, \& {Metcalf}}]{Pevtsov+03}
{Pevtsov}, A.~A., {Fisher}, G.~H., {Acton}, L.~W., {et~al.} 2003, \apj, 598,
  1387

\bibitem[{{Pipin} \& {Kosovichev}(2016)}]{PK16}
{Pipin}, V.~V., \& {Kosovichev}, A.~G. 2016, \apj, 823, 133

\bibitem[{{Pizzolato} {et~al.}(2003){Pizzolato}, {Maggio}, {Micela},
  {Sciortino}, \& {Ventura}}]{pizzolato+03}
{Pizzolato}, N., {Maggio}, A., {Micela}, G., {Sciortino}, S., \& {Ventura}, P.
  2003, \aap, 397, 147

\bibitem[{{Prusa} {et~al.}(2008){Prusa}, {Smolarkiewicz}, \&
  {Wyszogrodzki}}]{PSW08}
{Prusa}, J.~M., {Smolarkiewicz}, P.~K., \& {Wyszogrodzki}, A.~A. 2008, Comput.
  Fluids, 37, 1193

\bibitem[{{Racine} {et~al.}(2011){Racine}, {Charbonneau}, {Ghizaru}, {Bouchat},
  \& {Smolarkiewicz}}]{RCGS11}
{Racine}, {\'E}., {Charbonneau}, P., {Ghizaru}, M., {Bouchat}, A., \&
  {Smolarkiewicz}, P.~K. 2011, \apj, 735, 46

\bibitem[{{Rogers}(2011)}]{Rogers18}
{Rogers}, T.~M. 2011, \apj, 735, 100

\bibitem[{{Saar}(1988)}]{Saar88}
{Saar}, S.~H. 1988, \apj, 324, 441

\bibitem[{{Saar} \& {Brandenburg}(1999)}]{SB99}
{Saar}, S.~H., \& {Brandenburg}, A. 1999, \apj, 524, 295

\bibitem[{Schaeffer(2013)}]{shtns}
Schaeffer, N. 2013, Geochemistry, Geophysics, Geosystems, 14, 751

\bibitem[{{Schrijver} {et~al.}(1989){Schrijver}, {Cote}, {Zwaan}, \&
  {Saar}}]{Schrijver+89}
{Schrijver}, C.~J., {Cote}, J., {Zwaan}, C., \& {Saar}, S.~H. 1989, \apj, 337,
  964

\bibitem[{{See} {et~al.}(2015){See}, {Jardine}, {Vidotto}, {Donati}, {Folsom},
  {Boro Saikia}, {Bouvier}, {Fares}, {Gregory}, {Hussain}, {Jeffers},
  {Marsden}, {Morin}, {Moutou}, {do Nascimento}, {Petit}, {Ros{\'e}n}, \&
  {Waite}}]{See+2015}
{See}, V., {Jardine}, M., {Vidotto}, A.~A., {et~al.} 2015, \mnras, 453, 4301

\bibitem[{{Semel}(1989)}]{Semel89}
{Semel}, M. 1989, \aap, 225, 456

\bibitem[{{Smolarkiewicz}(2006)}]{S06}
{Smolarkiewicz}, P.~K. 2006, International Journal for Numerical Methods in
  Fluids, 50, 1123

\bibitem[{Smolarkiewicz \& Charbonneau(2013)}]{SC13}
Smolarkiewicz, P.~K., \& Charbonneau, P. 2013, J. Comput. Phys., 236, 608.
\newblock \url{http://dx.doi.org/10.1016/j.jcp.2012.11.008}

\bibitem[{{Smolarkiewicz} \& {Margolin}(2007)}]{GMR07}
{Smolarkiewicz}, P.~K., \& {Margolin}, L.~G. 2007 (Cambridge University Press),
  413

\bibitem[{{Spruit}(2002)}]{spruit02}
{Spruit}, H.~C. 2002, \aap, 381, 923

\bibitem[{{Steenbeck} {et~al.}(1966){Steenbeck}, {Krause}, \&
  {R{\"a}dler}}]{SKR66}
{Steenbeck}, M., {Krause}, F., \& {R{\"a}dler}, K.-H. 1966, Zeitschrift
  Naturforschung Teil A, 21, 369

\bibitem[{{Stix}(1976)}]{Stix76}
{Stix}, M. 1976, in IAU Symposium, Vol.~71, Basic Mechanisms of Solar Activity,
  ed. V.~{Bumba} \& J.~{Kleczek}, 367

\bibitem[{{Strugarek} {et~al.}(2017){Strugarek}, {Beaudoin}, {Charbonneau},
  {Brun}, \& {do Nascimento}}]{SBCBdN17}
{Strugarek}, A., {Beaudoin}, P., {Charbonneau}, P., {Brun}, A.~S., \& {do
  Nascimento}, J.-D. 2017, Science, 357, 185

\bibitem[{{Szklarski} \& {Arlt}(2013)}]{SA13}
{Szklarski}, J., \& {Arlt}, R. 2013, \aap, 550, A94

\bibitem[{{Tayler}(1973)}]{Tayler73}
{Tayler}, R.~J. 1973, \mnras, 161, 365

\bibitem[{{Vidotto} {et~al.}(2014){Vidotto}, {Gregory}, {Jardine}, {Donati},
  {Petit}, {Morin}, {Folsom}, {Bouvier}, {Cameron}, {Hussain}, {Marsden},
  {Waite}, {Fares}, {Jeffers}, \& {do Nascimento}}]{Vidotto+14}
{Vidotto}, A.~A., {Gregory}, S.~G., {Jardine}, M., {et~al.} 2014, \mnras, 441,
  2361

\bibitem[{{Viviani} {et~al.}(2018){Viviani}, {Warnecke}, {K{\"a}pyl{\"a}},
  {K{\"a}pyl{\"a}}, {Olspert}, {Cole-Kodikara}, {Lehtinen}, \&
  {Brandenburg}}]{Viviani+18}
{Viviani}, M., {Warnecke}, J., {K{\"a}pyl{\"a}}, M.~J., {et~al.} 2018, \aap,
  616, A160

\bibitem[{{Warnecke}(2017)}]{Warnecke17}
{Warnecke}, J. 2017, ArXiv e-prints, arXiv:1712.01248

\bibitem[{{Warnecke} {et~al.}(2018){Warnecke}, {Rheinhardt}, {Tuomisto},
  {K{\"a}pyl{\"a}}, {K{\"a}pyl{\"a}}, \& {Brandenburg}}]{Warnecke+18}
{Warnecke}, J., {Rheinhardt}, M., {Tuomisto}, S., {et~al.} 2018, \aap, 609, A51

\bibitem[{{Waruszewski} {et~al.}(2018){Waruszewski}, {K{\"u}hnlein},
  {Pawlowska}, \& {Smolarkiewicz}}]{WKPS18}
{Waruszewski}, M., {K{\"u}hnlein}, C., {Pawlowska}, H., \& {Smolarkiewicz},
  P.~K. 2018, Journal of Computational Physics, 359, 361

\bibitem[{{Wright} \& {Drake}(2016)}]{Wright+16}
{Wright}, N.~J., \& {Drake}, J.~J. 2016, \nat, 535, 526

\bibitem[{{Wright} {et~al.}(2011){Wright}, {Drake}, {Mamajek}, \&
  {Henry}}]{Wright+11}
{Wright}, N.~J., {Drake}, J.~J., {Mamajek}, E.~E., \& {Henry}, G.~W. 2011,
  \apj, 743, 48

\bibitem[{{Zahn} {et~al.}(2007){Zahn}, {Brun}, \& {Mathis}}]{ZB07}
{Zahn}, J.-P., {Brun}, A.~S., \& {Mathis}, S. 2007, \aap, 474, 145

\end{thebibliography}

\end{document}